# Marco de teoría de categorías para el modelado macroeconómico: El caso de la economía bimonetaria de Argentina

Autor: Luciano Pollicino

Afiliación: Universidad Nacional de Cuyo

Email: lucianopollicinoc3@gmail.com

Fecha: 17 de agosto, 2025

Resumen

Los modelos macroeconómicos tradicionales, basados en sistemas algebraicos estáticos, no logran capturar la dinámica de una economía bimonetaria como la de Argentina. Este trabajo propone un marco basado en la teoría de categorías para desarrollar un modelo más flexible y estructurado que represente las relaciones evolutivas entre variables clave como las expectativas de inflación, las tasas de interés y la demanda de divisas. Utilizando conceptos como objetos, morfismos, functores de aprendizaje/olvido, límites y colimites, el modelo se aplica a datos empíricos de 2018-2023. Los hallazgos revelan un desajuste estructural significativo entre el tipo de cambio de equilibrio y el observado, y proponen un nuevo indicador agregado para medir el riesgo de devaluación. El marco demuestra una gran sinergia con herramientas computacionales modernas como el machine learning, ofreciendo un enfoque más robusto para el análisis de políticas y la previsión en economías complejas.

Palabras clave: Teoría de Categorías, Modelado Macroeconómico, Economía Bimonetaria, Argentina, Functores, Equilibrio Económico, Expectativas de Devaluación.

Abstract

Traditional macroeconomic models, based on static algebraic systems, fail to capture the dynamics of a bimonetary economy like Argentina's. This paper proposes a framework based on category theory to develop a more flexible and structured model that represents the evolving relationships between key variables such as inflation expectations, interest rates, and currency demand. Using concepts like objects, morphisms, learning/forgetful functors, limits, and colimits, the model is applied to empirical data from 2018-2023. The findings reveal a significant structural misalignment between the equilibrium and observed exchange rates and propose a new aggregate indicator to measure devaluation risk. The framework demonstrates a strong synergy with modern computational tools like machine learning, offering a more robust approach to policy analysis and forecasting in complex economies.

# Índice

# **Introducción**

## Problema

Los modelos tradicionales de macroeconomía no logran capturar adecuadamente la dinámica de una economía bimonetaria como la argentina, donde las decisiones económicas de los agentes están fuertemente influenciadas por expectativas de inflación, variaciones en las tasas de interés en pesos y dólares, relación histórica peso/dólar, expectativas de devaluación, variables monetarias, entre otras. Estos modelos, basados en sistemas algebraicos estáticos, carecen de la flexibilidad necesaria para describir cómo las interacciones entre estas variables evolucionan en el tiempo y responden a cambios en la política monetaria y shocks externos.

El problema central radica en que la economía es inherentemente dinámica, mientras que los enfoques algebraicos tradicionales son estructuras rígidas y estáticas, diseñadas para modelar relaciones fijas en lugar de sistemas en constante transformación. En el caso argentino, donde las expectativas juegan un rol determinante en la toma de decisiones económicas, resulta insuficiente modelar la oferta y demanda de dinero con ecuaciones cerradas que no capturan las interdependencias y efectos de retroalimentación del sistema.

## Antecedentes

La aplicación de teoría de categorías es emergente en el estudio dinámico de la economía, no existe mucho material al respecto; no obstante, Tohmé (2023) propone su uso para modelar interacciones entre modelos económicos, aunque nosotros partiremos de un enfoque distinto. Este autor particularmente ha explorado en otra ocasión (Tohmé, F., & Auday, M. (2019)) la posible utilización de teoría de categorías a la economía. Otros autores prefieren enfocarse en el aspecto meramente matemático de categorías, más allá de posibles aplicaciones a teoría de juegos, como Abramsky, S., & Winschel, V. (2017). También mencionaré a una autora, Emily Riehl, cuyo trabajo ha sido fundamental para el desarrollo formal de teoría de categorías, así también como de HoTT (teoría de tipos homotópica), y de quien recopilé definiciones, teoremas y explicaciones para facilitar este trabajo. Principalmente, Riehl, E. (2016). Category theory in context.

# Justificación

El presente trabajo es necesario porque propone un enfoque más dinámico y flexible para el análisis macroeconómico de una economía bimonetaria como la argentina, utilizando la teoría de categorías para abordar las limitaciones de los modelos tradicionales. Esta justificación se basa en los siguientes pilares fundamentales:

1. Limitaciones de los modelos tradicionales
   Los modelos económicos convencionales suelen basarse en ecuaciones diferenciales y sistemas algebraicos rígidos, que son útiles para representar relaciones estáticas, pero no capturan la evolución de las variables económicas en el tiempo. En economías como la argentina, donde las expectativas de inflación y devaluación cambian constantemente en respuesta a políticas monetarias y eventos externos, es crucial contar con herramientas que permitan modelar estos cambios de manera estructurada.
   La teoría de categorías permite:
   - Modelar la economía como un sistema dinámico en el que las relaciones entre variables evolucionan con nuevos eventos o políticas.
   - Representar cambios de régimen sin necesidad de redefinir completamente el modelo, simplemente ajustando los morfismos dentro de la categoría.
   - Introducir herramientas como functores de aprendizaje y olvido para captar cómo los agentes económicos incorporan o descartan información en función de su relevancia.
2. Adaptabilidad a nuevas tecnologías
   La economía moderna exige modelos que puedan integrarse con tecnologías de análisis de datos, inteligencia artificial y machine learning. Los enfoques algebraicos tradicionales, aunque robustos, presentan dificultades para adaptarse a estos entornos.
   La teoría de categorías, en cambio, proporciona una estructura que facilita:
   - La representación de modelos económicos en software y su integración con herramientas de aprendizaje profundo basadas en grafos y redes neuronales.
   - La transformación dinámica del modelo en función de nuevos datos, permitiendo su calibración en tiempo real.
   - La generación de características para modelos de machine learning, mejorando la predicción de fenómenos como inflación, tasas de interés y tipo de cambio.
3. Análisis de estabilidad y equilibrio
   En economías volátiles como la argentina, los enfoques tradicionales basados en puntos fijos y optimización estática son insuficientes. La teoría de categorías permite:
   - Definir la estabilidad en términos de conmutatividad de morfismos, asegurando que las transformaciones económicas no generen inconsistencias estructurales.
   - Representar el equilibrio como un límite categórico, permitiendo estudiar su convergencia y cómo ciertos shocks pueden impedir su estabilidad.

- Modelar shocks económicos como perturbaciones en la categoría, analizando su impacto en la estructura general del sistema en lugar de efectos aislados en ecuaciones específicas (análisis de sensibilidad).
4. Medición de expectativas de devaluación
   Las expectativas de devaluación son cruciales en la economía argentina, influenciando la demanda de pesos y dólares. Sin embargo, su medición ha sido tradicionalmente problemática debido a:
   - Limitaciones en los modelos econométricos tradicionales, que dependen de datos históricos y no capturan la evolución dinámica de las expectativas en tiempo real.
   - Falta de integración de múltiples fuentes de información, como mercados financieros, encuestas y análisis de sentimiento en redes sociales.
   - Dificultad para modelar la retroalimentación entre expectativas e inflación, ya que los modelos clásicos suelen tratar estos fenómenos de forma independiente.

El enfoque basado en teoría de categorías introduce una nueva metodología para medir las expectativas de devaluación, proporcionando herramientas más robustas para entender la dinámica de una economía bimonetaria y superar las limitaciones de los enfoques tradicionales.

# Estructura y objetivos

El trabajo se dividirá en una primera parte, donde introduciremos conceptos de teoría de categorías, así también como veremos de modo superficial (de tal forma que nos sirva luego) un análisis de sensibilidad. Luego, encontraremos una segunda parte en la cual desarrollaremos los conceptos de functores, para después pasar a los conceptos de límite y colimite, en una tercera parte. Veremos el marco teórico primero y luego las aplicaciones a la economía. Utilizaremos una base de datos de elaboración propia a partir de datos recopilados de diversas fuentes que se mencionan luego (y que puede encontrar al final de este documento). Usaremos Python para desarrollar la parte práctica. La idea es dejar claro por qué la teoría de categorías es directamente vinculable a lenguajes de programación como el mencionado, dado su dinamismo, practicidad y relación con modelos de machine learning.

## Objetivo general

El objetivo principal de este trabajo es desarrollar un modelo macroeconómico bimonetario basado en la teoría de categorías, que permita analizar la estabilidad económica, la influencia de las políticas monetarias y las expectativas de devaluación en la economía argentina. Este modelo busca superar las limitaciones de los enfoques tradicionales al introducir una representación dinámica y estructurada de las relaciones entre variables económicas clave.

## Objetivos específicos

Para alcanzar este objetivo general, la investigación se estructura en tres partes, cada una con sus propios objetivos específicos:

Parte 1: Introducción a la Teoría de Categorías y Análisis Básico del Modelo

- Justificar la necesidad de un enfoque categórico en economía, explicando las limitaciones de los modelos algebraicos tradicionales y destacando la flexibilidad de la teoría de categorías (este punto se menciona implícitamente, y se tocará también en próximas partes).
- Establecer la estructura básica del modelo, definiendo la categoría económica con sus objetos, morfismos iniciales y diagramas.
- Realizar un análisis de sensibilidad preliminar, evaluando cómo las principales variables económicas responden a cambios en las expectativas y tasas de interés.
- Comparar la categoría definida con modelos económicos tradicionales, destacando sus ventajas en términos de adaptabilidad y capacidad de representación de sistemas dinámicos.

Parte 2: Incorporación de Functores para Modelar Cambios en la Economía

- Introducir los conceptos de functores de olvido y aprendizaje, explicando cómo estos pueden utilizarse para simplificar o enriquecer la estructura del modelo en función de nueva información o cambios en políticas económicas.
- Refinar el análisis de sensibilidad, utilizando los functores para introducir nuevas políticas económicas en el modelo y evaluar sus efectos sobre la estabilidad macroeconómica.
- Desarrollar un esquema de pronóstico basado en la estructura categórica, aprovechando los functores para modelar la evolución de variables clave seleccionadas.
- Evaluar la consistencia del modelo con datos históricos y escenarios hipotéticos, asegurando que la estructura categórica se mantenga coherente ante distintos contextos económicos.

Parte 3: Cálculo del Equilibrio y Medición de Expectativas de Devaluación

- Introducir los conceptos de límites y colimites de la teoría de categorías, explicando su utilidad en el análisis de equilibrios macroeconómicos.
- Aplicar los límites para calcular el equilibrio económico, definiendo formalmente las condiciones bajo las cuales el sistema alcanza estabilidad en términos de tasas de interés, expectativas y tipo de cambio.

- Utilizar el colimite para modelar las expectativas de devaluación, estableciendo una metodología estructurada para predecir la evolución de estas expectativas con base en datos económicos y comportamiento histórico.
- Evaluar la capacidad del modelo para predecir dinámicas económicas reales, contrastando sus resultados con indicadores empíricos y comparándolos con enfoques tradicionales de medición de expectativas.

# Parte I: Introducción a la Teoría de Categorías y Análisis Básico del Modelo

En esta parte, se define el marco teórico de la teoría de categorías aplicado a la macroeconomía bimonetaria. Se modelan las principales variables económicas como objetos y las relaciones entre ellas como morfismos, estableciendo una estructura matemática que captura las interacciones fundamentales del sistema. Para validar la consistencia del modelo, se analizan diagramas conmutativos y se estudia la estabilidad de las relaciones económicas mediante estructuras de datos categóricas. Se plantea un primer esquema formal del modelo, que servirá de base para las secciones posteriores, así también como la estructura de datos.

## Hipótesis

El modelo de interacción de los mercados monetarios, de inversión y de demanda agregada tradicional llama la atención por no presentar un marco flexible que permita el dinamismo en el mismo. Puntualmente, en el caso de la Argentina, es obvio destacar que vivimos en un país implícitamente bimonetario, donde las expectativas de inflación juegan un rol clave para los argentinos respecto de qué moneda usar para reservar valor o transar bienes, alterando esto el funcionamiento macroeconómico que se estudia tradicionalmente, tanto a nivel de mercados monetarios, como de inversión y demanda agregada. Por tanto, la hipótesis de la que partiremos es:

*"El uso de la teoría de categorías proporciona una representación más flexible y estructurada de la economía bimonetaria argentina en comparación con los modelos algebraicos tradicionales, permitiendo una mejor interpretación de las relaciones entre variables macroeconómicas clave."*. Nuestra motivación es justificar la necesidad de usar teoría de categorías en lugar de modelos estáticos tradicionales. Se validará observando cómo los objetos y morfismos permiten una representación más estructurada y adaptable de las interacciones económicas.

En otras palabras, esta parte de la investigación intentará verificar si, mediante el uso de algunas herramientas algebraicas avanzadas y un poco de intuición, es posible crear un modelo cuyos componentes sean una categoría $\mathcal{E}$, y morfismos entre objetos de dicha categoría. Introduciremos notaciones algebraicas y gráficos más adelante, por ahora, basta con señalar relaciones (morfismos, que serán mejor planteados y desarrollados en el marco teórico) intuitivas, tales como las siguientes:

-El aumento de las expectativas de inflación y/o el aumento de las expectativas de devaluación peso-dólar generan un aumento en la demanda de dólares y una caída en la demanda de pesos.

-Si las expectativas de inflación son inferiores a un determinado coeficiente y la tasa de interés en pesos es considerablemente superior a la tasa de interés en dólares y, además, las expectativas de devaluación son inferiores a otro coeficiente, aumentará la demanda de pesos, generando que haya más demanda de pesos que de dólares.

-Si las expectativas de inflación son superiores a un determinado coeficiente, disminuirán los créditos ofrecidos, que se verá acompañado de una caída en la demanda de pesos.

Como idea general, observe los siguientes gráficos, que son de elaboración propia usando Python y compárelos con los modelos tradicionales.

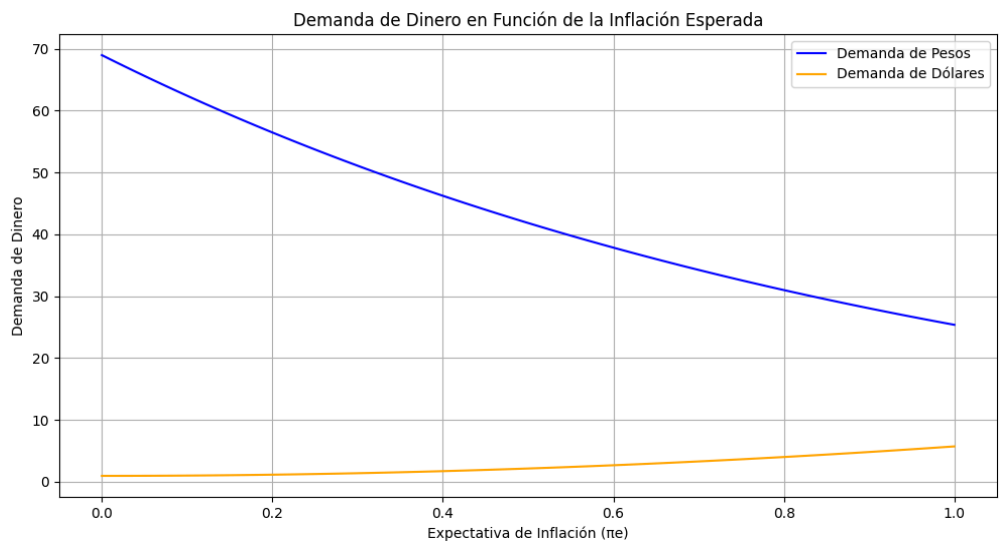

```
import numpy as np
import matplotlib.pyplot as plt

Y = 100
pi_e_values = np.linspace(0, 1, 100)

A = 1
B = 5
k = 2
C = 1

def demanda_pesos(Y, i_P, pi_e):
    return Y / ((1 + i_P) * np.exp(C * pi_e))

def demanda_dolares(pi_e, i_D):
    return (A * (1 + B * pi_e**k)) / (1 + i_D)

i_P = 0.45
i_D = 0.05
```

```python
L_P = demanda_pesos(Y, i_P, pi_e_values)
L_D = demanda_dolares(pi_e_values, i_D)

plt.figure(figsize=(12, 6))
plt.plot(pi_e_values, L_P, label='Demanda de Pesos', color='blue')
plt.plot(pi_e_values, L_D, label='Demanda de Dólares', color='orange')
plt.title('Demanda de Dinero en Función de la Inflación Esperada')
plt.xlabel('Expectativa de Inflación (πe)')
plt.ylabel('Demanda de Dinero')
plt.legend()
plt.grid()
plt.show()

L_P_60 = demanda_pesos(Y, i_P, 0.6)
L_D_60 = demanda_dolares(0.6, i_D)

print(f"Demanda de Pesos a πe=60%: {L_P_60}")
print(f"Demanda de Dólares a πe=60%: {L_D_60}")
```

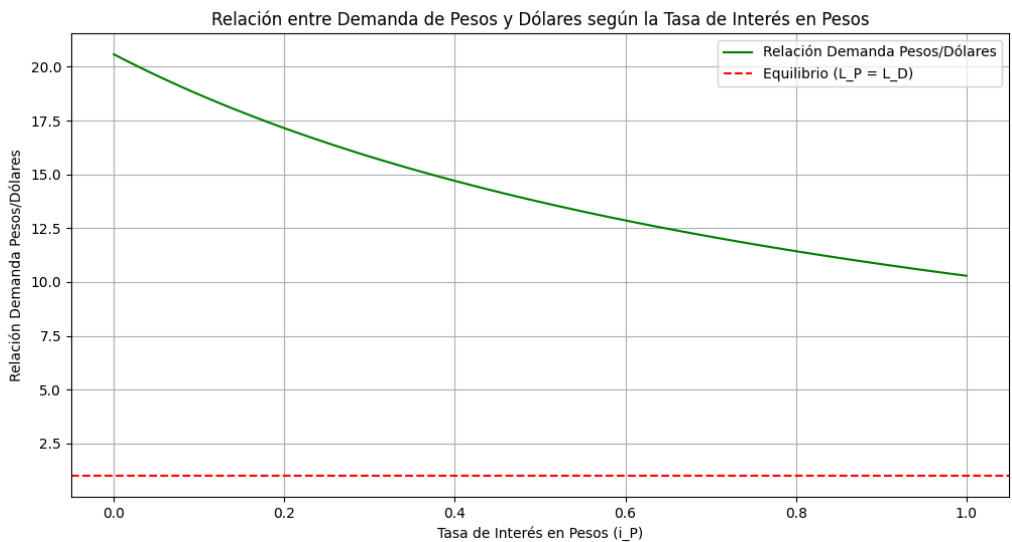

```python
import numpy as np
import matplotlib.pyplot as plt

Y = 100
i_D = 0.05

A = 1
B = 5
k = 2
C = 1

def demanda_pesos(Y, i_P, pi_e):
    return Y / ((1 + i_P) * np.exp(C * pi_e))

def demanda_dolares(pi_e, i_D):
    return (A * (1 + B * pi_e**k)) / (1 + i_D)

i_P_values = np.linspace(0, 1, 100)

relaciones = []
```

```python
for i_P in i_P_values:
    L_P = demanda_pesos(Y, i_P, 0.6)
    L_D = demanda_dolares(0.6, i_D)
    relaciones.append(L_P / L_D)

plt.figure(figsize=(12, 6))
plt.plot(i_P_values, relaciones, label='Relación Demanda Pesos/Dólares', color='green')
plt.title('Relación entre Demanda de Pesos y Dólares según la Tasa de Interés en Pesos')
plt.xlabel('Tasa de Interés en Pesos (i_P)')
plt.ylabel('Relación Demanda Pesos/Dólares')
plt.legend()
plt.grid()
plt.axhline(1, color='red', linestyle='--', label='Equilibrio (L_P = L_D)')
plt.legend()
plt.show()
```

Lo que se puede observar en los gráficos es intuitivo, y está realizado con datos y fórmulas ficticias; no obstante, es similar a lo que podríamos obtener en la realidad, ¿no?

# Marco teórico

Aquí desarrollaremos las bases de un modelo de categorías y señalaremos ciertos conceptos del álgebra para enriquecer los fundamentos sobre los cuales nos moveremos, esto con el fin de mantener un rigor matemático que facilite la futura aplicación a la realidad.

### Introducción a la teoría de categorías

Esta es la herramienta que más utilizaremos para realizar nuestro modelo. Las categorías se pueden interpretar como contextos formados por objetos a los que les ocurren transformaciones (morfismos). Los objetos pueden ser estructuras algebraicas como pueden ser conjuntos, grupos (véase anexo teórico para más detalles), espacios topológicos, vectores, cadenas u otros; mientras que los morfismos son las relaciones que hay en esos objetos, como pueden ser funciones a partir de los conjuntos, isomorfismos en el caso de los grupos, funciones continuas en el caso de los espacios topológicos, etc… En nuestro caso, los objetos pueden ser mercados o economías, a los que representaremos como sea conveniente, usando la estructura algebraica que nos sea más práctica, mayormente conjuntos. Es a estos objetos a los que les ocurren morfismos de forma dinámica y se interrelacionan con los demás objetos dentro de la categoría. Ahora, si un objeto se relaciona con otro objeto, y ese objeto se relaciona con otro, podemos relacionar el primer y el tercer objeto mediante una composición, lo cual es una característica clave de las categorías.
Formalmente, si hay morfismos de la forma

$$f : A \to B \ \land \ g : C \to D$$

entonces debe existir un morfismo compuesto de la forma

$$g \circ f : A \to C$$

y, además, debe haber un elemento neutro que actúa como identidad, es decir,

$$id_A : A \to A$$

Mencionar, además, el concepto de "functor", que es un tipo especial de "transformación" que lleva objetos y morfismos de una categoría a otra; en esencia, una traducción de categorías símiles (lo veremos en la parte correspondiente a functores).
Por último, es de especial importancia destacar los conceptos de límites y colímites, que veremos en la última parte de esta investigación, siendo el primero una forma de mencionar todos los posibles morfismos que pueden suceder (una convergencia) hasta encontrar un equilibrio y el segundo representa el agregado que no fue tenido en cuenta en el límite. En cursos introductorios a la Economía, a la hora de evaluar el impacto de políticas fiscales o monetarias en el modelo keynesiano, realizamos unos gráficos sobre los cuales se realizan transformaciones debidas al impacto de las políticas, pero las mismas no son cosa de un único impacto, sino que se va produciendo una sucesión de impactos. Pues bien, esa convergencia hasta el nuevo equilibrio se denomina "Límite", mientras que la convergencia en la demanda agregada se denomina "Colímite". Este ejemplo es suficiente para poder apreciar el potencial de la teoría de categorías para explicar y calcular equilibrios complejos como los que podemos apreciar en modelos macroeconómicos o en teoría de juegos para calcular el equilibrio de Nash. Se escriben de la siguiente forma, respectivamente:

$$\{f_i\}_{i=1}^{n}$$

$$\{g_j\}_{j=1}^{m}$$

A partir de aquí, introduciremos el aspecto teórico del modelo.

## Objetos, morfismos y diagramas conmutativos

Llamaremos a nuestra categoría $\epsilon$, y contendrá los siguientes objetos:

$L_{ARS} = $ *demanda de pesos*

$L_{USD} = $ *demanda de dólares*

$R_{\frac{ars}{usd}} = $ *demanda relativa de pesos versus dólares*

$M_{\frac{ars}{usd}} = $ *Oferta de dinero en términos de ars − usd*

$\pi_e = $ *expectativas de inflación*

$E = $ *expectativas de devaluación peso − dólar*

$Y = $ *ingreso real*

$i_{ARS} = $ *tasa de interés en pesos*

$i_{USD} = $ *tasa de interés en dólares*

Con los cuales podremos formar morfismos tales como los siguientes:

$f : \pi_e \to L_{ARS}$

$g : i_{USD} \to L_{USD}$

$h : E \to R_{\frac{ars}{usd}}$

Donde f muestra cómo las expectativas de inflación afectan la demanda de pesos, g captura cómo la tasa de interés en dólares afecta la demanda de dólares y h modela cómo las expectativas de devaluación peso-dólar afectan la demanda relativa pesos-dólares. Definiremos el siguiente diagrama conmutativo basado en funciones mapeadas linealmente (polinómicas, en otras palabras), que no es más que un diagrama de funciones, objetos, morfismos o caminos que son de utilidad para modelar la categoría (en términos elementales):

$L_{ARS} = f_1(Y, i_{ARS}, \pi_e)$

$L_{USD} = f_2(i_{USD}, \pi_e, E)$

$R_{\frac{ars}{usd}} = \frac{L_{ARS}}{L_{USD}} = f_3(i_{ARS}, i_{USD}, \pi_e, E, Y)$

$M_{\frac{ars}{usd}} = f_4(C, r_{real}, \pi_e)$

donde C es crédito y $r_{real}$ es la tasa de interés real.

Ahora definiremos functores que representen los cambios en las categorías (por ejemplo, una cambio en las políticas del BCRA):

$$f : \varepsilon \to \varepsilon'$$

A modo de ejemplo, suponga que una nueva política monetaria genera un cambio en la tasa de interés en pesos. Tal cambio podría mapear la demanda de pesos a una nueva función de la siguiente forma, utilizando lo anteriormente mencionado:

$$f(L_{ARS}) \to f'_1\left(Y, i_{ARS} + \Delta i_{ARS}, \pi_e\right)$$

Ya que en este ejemplo se produjo un cambio en la categoría, debemos de alguna manera, mostrar también todos los cambios que se produjeron en la categoría, por tanto realizaremos nuevo diagrama conmutativo, partiendo de la siguiente situación inicial:

$$L_{ARS} \xrightarrow{f_1} R_{\frac{ars}{usd}} \xrightarrow{f_3} M_{\frac{ars}{usd}}$$

$$i_{ARS} \xrightarrow{g} L_{USD} \xrightarrow{h} R_{\frac{ars}{usd}}$$

Dejándonos con los cambios:

$$L'_{ARS} \xrightarrow{f'_1} R'_{\frac{ars}{usd}} \xrightarrow{f'_3} M'_{\frac{ars}{usd}}$$

$$i'_{ARS} \xrightarrow{g'} L'_{USD} \xrightarrow{h'} R'_{\frac{ars}{usd}}$$

## Desarrollo algebraico

Desde el punto de vista topológico, cada función puede pensarse como un mapa continuo que representa el comportamiento de los objetos a cambios en las tasas de interés, expectativas, etc… En el caso de las expectativas de inflación, la situación inicial plantea puntos estables fijos (que luego intentaremos "econometrizar"), significando que para ese punto igualado a Y obtendremos una demanda de pesos estable, reflejando expectativas de inflación estables; no obstante, si aparece un shock externo que modifique la categoría, se producirá una bifurcación en el espacio donde ningún punto estable exista (similar al modelo keynesiano intuitivo), terminando con espirales inflacionarios.

El siguiente paso, ya que asumimos funciones polinómicas, puede ser asumir que las funciones se desarrollen así (a corregir según teoría económica y evidencia empírica):

$$L_{ARS} = \alpha \cdot \pi_e + \beta \cdot Y + \gamma \cdot i_{ARS} + \delta$$

$$L_{USD} = \mu \cdot E + \xi \cdot \pi_e + \nu \cdot i_{USD} + \vartheta$$

donde α, β, ɣ, μ, ξ y ν son coeficientes que capturan la sensibilidad o responsividad del peso o dólar en cada función de demanda respectiva, mientras que δ y ϑ representan aquellos factores que no están modelados (como si fueran demandas bases).
La demanda relativa puede modelarse de esta forma:

$$R_{\frac{ars}{usd}} = \frac{L_{ars}}{L_{usd}} \cdot e^{(i_{ARS} - i_{USD})}$$

Donde el término exponencial representa la atracción relativa del peso frente al dólar en términos de tasas de interés.
Sobre la oferta de dinero, esta puede tomar formas no-lineales, y depende de las variables mencionadas con anteriroidad. Sobra decir que expresarla es mejor en términos empíricos que como una mera ecuación.
Para mejorar el mapeo de las variables, es útil incluir algunas estructuras algebraicas abstractas, como lo es un anillo generado por las variables económicas tal que

$$R = R[\pi_e, Y, i_{ARS}, E, i_{USD}]$$

La utilidad de este anillo es que nos permite plasmar expresiones polinómicas y transformaciones entre variables. Por ejemplo, el anillo tiene elementos de la forma:

$$a \cdot \pi_e + b \cdot Y + c \cdot i_{ARS} + d \cdot E + e \cdot i_{USD} + \ldots$$

donde a, b, c, d y e son coeficientes reales representando elasticidades (sensibilidades).

Otra cosa que podemos hacer es modelar la dinámica de las expectativas (morfismos y relaciones funcionales entre las siguientes funciones) a través de una estructura de grupos para poder apreciar su evolución en el paso del tiempo. Para esto, definiremos el siguiente grupo bajo composición de funciones:

$$G = \langle \pi_e, E \rangle$$

Este grupo nos permite seguir el paso de los impactos sobre las expectativas de inflación y devaluación mientras que estas evolucionan acorde a shocks externos o cambios en políticas del BCRA. También definiremos una operación de grupo ★ (estrella) tal que:

$$\pi_e \star E = \pi_e + E + \eta(\pi_e, E)$$

donde η es un término corrector no-lineal que captura feedback entre las dos expectativas. Este grupo puede ayudar a modelar como las expectativas de inflación y devaluación se alimentan (feed) entre ellas.

Por otro lado, ya que la demanda de dinero y la demanda relativa son combinaciones lineales de las variables centrales, podemos definir un módulo $\mathcal{M}$ sobre el anillo *R*, donde cada función demanda puede ser expresada como un elemento del módulo:

$$L_{ARS} = \alpha \cdot \pi_e + \beta \cdot Y + \gamma \cdot i_{ARS} \in M$$

Usando módulos, podemos apreciar cómo los cambios en las expectativas de inflación, ingreso y tasa de interés en pesos escalarán la demanda de pesos (lineal o no-linealmente) de forma eficiente.

Usando lo que hemos construido hasta ahora, modelaremos la inflación por expectativas en un sistema bimonetario. Primero iniciaremos definiendo una función para las expectativas de inflación, teniendo en cuenta, como se mencionó previamente, que estas son influenciadas por anteriores números, credibilidad del banco central y cambios en las políticas. Por tanto, una relación recursiva puede ser la siguiente:

$$\pi_e(t+1) = \alpha \cdot \pi_e(t) + \beta \cdot i_{real}(t) + \gamma \cdot \Delta E(t) + \varepsilon$$

donde $\Delta E(t)$ representa el cambio en las expectativas de devaluación y $\varepsilon$ es un término de error aleatorio.

Ahora, para la demanda relativa, podemos usar un grupo de acción del grupo expectativo G en el módulo M (un functor, en pocas palabras) tal que:

$$\left(\pi_e, E\right) \cdot L_{ARS} = \alpha \cdot \pi_e \cdot L_{ARS} + \beta \cdot E \cdot L_{ARS}$$

Con esto podremos saber cómo las expectativas de devaluación e inflación transforman el módulo del peso demandado. La respuesta puede ser amplificadora o aplanadora dependiendo del historial de credibilidad del país sobre la inflación.

Esta sencilla construcción algebraica permite, entre otras cosas:

- Analizar series de tiempo dinámicas.
- Realizar análisis de estabilidad.
- Usar homomorfismos de módulo para ver cambios en las condiciones de equilibrio por cambios en políticas, lo cual nosotros representaremos, de modo más simplista, con morfismos en próximas secciones.
- Aplicar límites y colimites para calcular equilibrios y analizar las expectativas de devaluación.

No obstante, esta estructuración algebraica es sólo una de las tantas posibilidades. Quizás este sea el enfoque tradicional, pero, para variar, también usaremos otras construcciones, ya que justamente este punto es donde se puede apreciar la utilidad de la teoría de categorías, su flexibilidad para plasmar y tratar en distintas estructuras algebraicas nuestra información.

# Metodología de la investigación

## Análisis de estabilidad

Sería muy tentador definir el equilibrio como el punto donde la demanda de pesos es igual a la de dólares, o cualquier otro abordamiento similar, pero esto sería un enfoque limitado para países como la Argentina. Es por esta razón que es preferible usar el enfoque de teoría de categorías. En vez de analizar una mera igualdad, analizamos una categoría, viendo como los cambios o morfismos en la misma, como un cambio en las políticas, afecta el resto de funciones y todo el sistema en sí mismo. Es una visión más fructífera, la cual verificaremos luego con datos reales. En términos de categorías, el equilibrio se logra cuando la composición de los morfismos es conmutativa, lo que significa que no hay desequilibrio de flujo neto entre ARS y USD. Las condiciones de equilibrio implicarán establecer un sistema de ecuaciones a partir de los morfismos, centrándose en el tipo de cambio, las expectativas de inflación y la prima de riesgo, como lo pueden ser diagramas conmutativos que muestran cómo diferentes morfismos, como la transformación de la liquidez a través del tipo de cambio, se equilibran con las expectativas de inflación y las tasas de interés, o garantizando que las trayectorias que representan los flujos de liquidez entre las categorías de ARS y USD produzcan los mismos resultados cuando se rastrean a través de diferentes combinaciones de morfismos.

Lo estructurado en la sección anterior es válido a nivel intuitivo, pero no necesariamente significa que vaya a ser muy útil, ya que no tiene en cuenta factores como interacciones cross-market, la prima de riesgo o shocks asimétricos en dólares y en la liquidez de los pesos. Dadas estas limitaciones, podemos llevar el modelo un paso más allá de la siguiente forma, haciendo uso de las categorías que hemos definido, intentando buscar el equilibrio y agregando el factor prima de riesgo:

-Morfismo de expectativas inflacionarias en pesos:

$$\pi_{e,\,ARS} : L_{ARS}(t) \to \pi_{e,\,ARS}(t)$$

$$\pi_{e,\,ARS}(L_{ARS}(t)) = \pi_{e,\,ARS}(t) \cdot L_{ARS}(t)$$

-Morfismo de expectativas inflacionarias en dólares:

$$\pi_{e,\,USD} : L_{USD}(t) \to \pi_{e,\,USD}(t)$$

$$\pi_{e,\,USD}(L_{USD}(t)) = \pi_{e,\,USD}(t) \cdot L_{USD}(t)$$

–Morfismos de la prima de riesgo:

$$\sigma_{\frac{ARS}{USD}} : L_{ARS}(t) \to L_{USD}(t)$$

$$\sigma_{\frac{ARS}{USD}}(L_{ARS}(t)) = L_{ARS}(t) \cdot \frac{1}{1 + \rho(t)}$$

donde $\sigma$ es la función prima de riesgo y donde $\varrho$ es la prima de riesgo en el tiempo t, reflejando el riesgo adicional de tener pesos frente a dólares.

Formalmente, el equilibrio queda de la siguiente manera:

$$L_{ARS}(t) \xrightarrow{R_{\frac{ARS}{USD}}(t)} L_{USD}(t) \land \pi_{e,ARS}(t) \circ L_{ARS}(t) = \pi_{e,USD}(t) \circ L_{USD}(t)$$

Si se desea introducir cambios en las políticas, los morfismos se ajustarán al nuevo contexto, por ejemplo, observe el siguiente diagrama conmutativo, es una forma de representar cómo interactúan los morfismos definidos en las condiciones de equilibrio y tras un cambio en las tasas de interés:

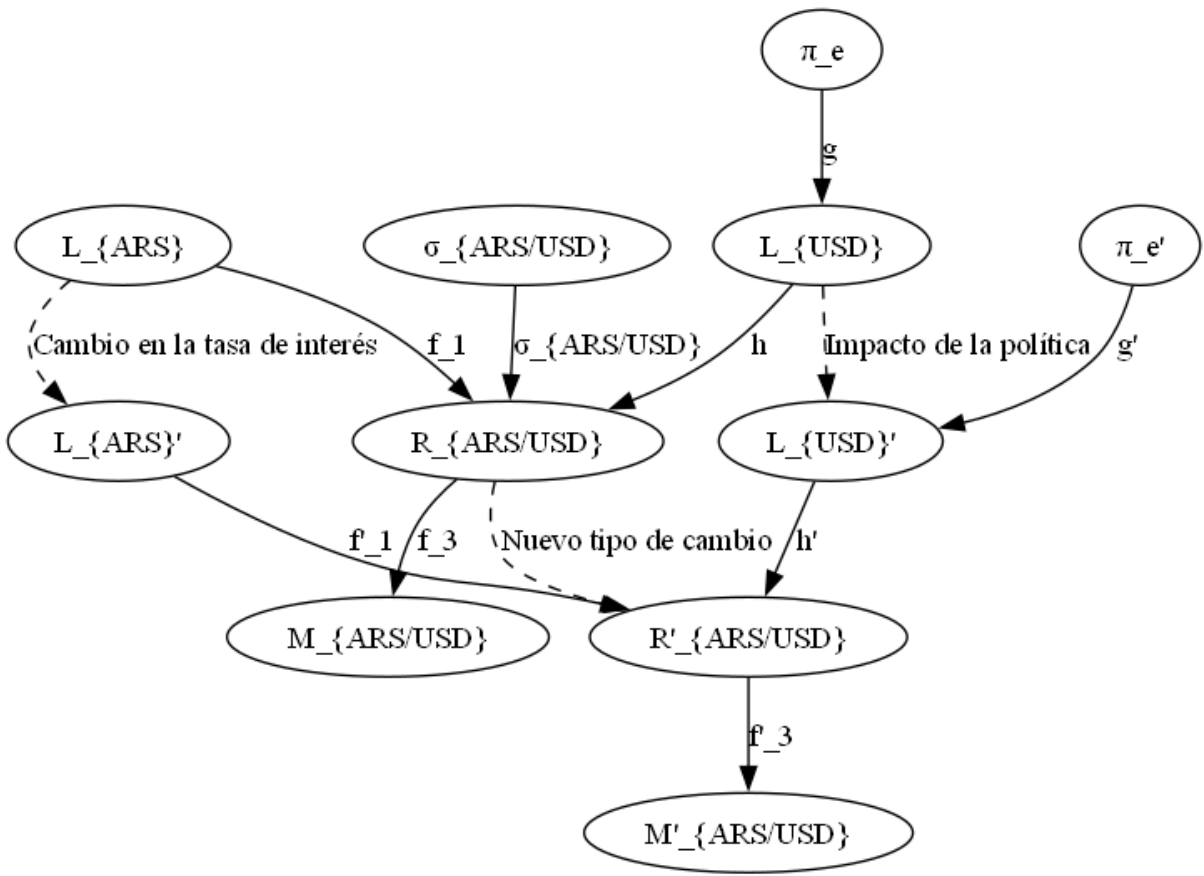

# Estructuras de datos

Iniciaremos con la verificación del funcionamiento del modelo usando datos reales. Primero haremos una lista de los datos utilizados, sus respectivas fuentes y algunas acotaciones, en caso de que los mismos presenten información macroeconómica relevante:

-Tipo de cambio ARS/USD. Usaremos los datos encontrados en Yahoo Finance correspondiente al período 1/1/2018 al 29/12/2023. Los datos fueron extraídos usando Python. Aquí hay un gráfico de su evolución:

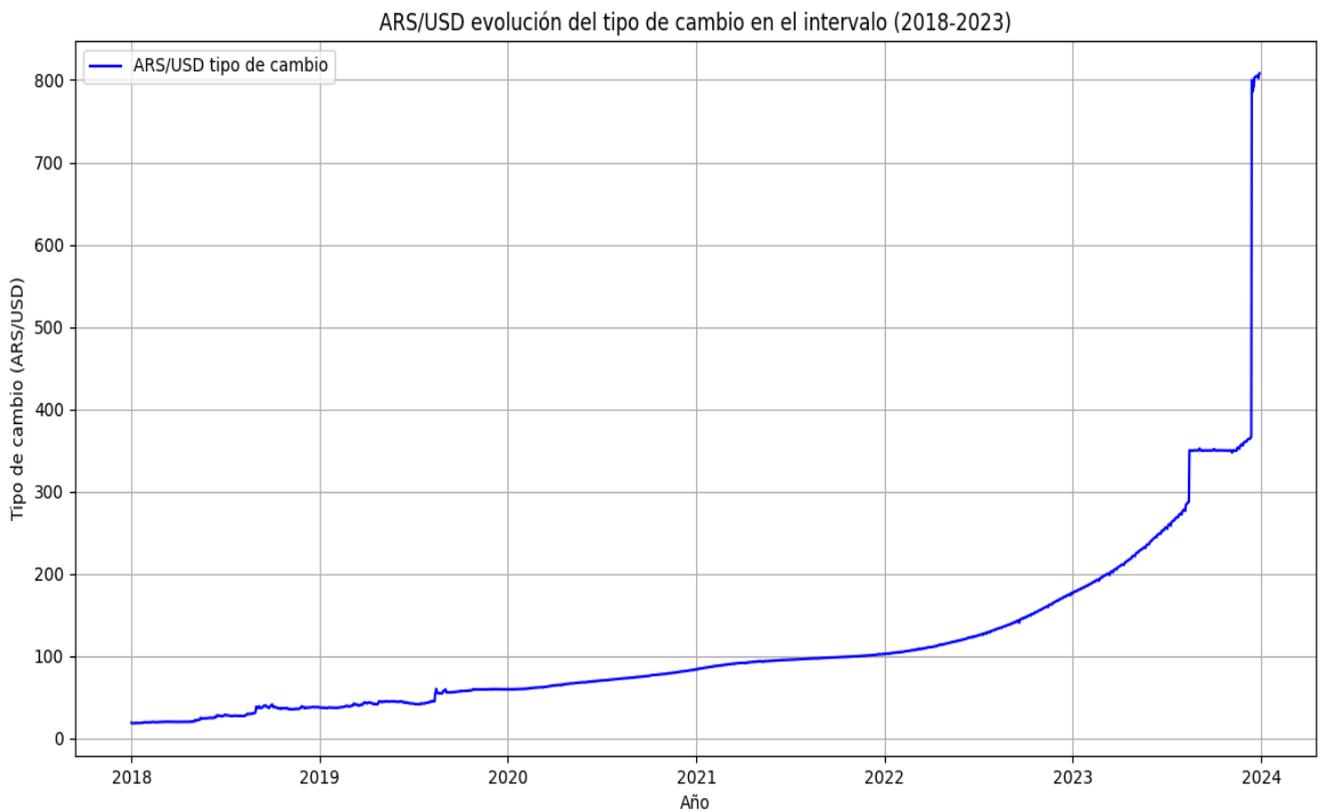

No se adjuntará el código en Python debido a que requiere un archivo con los datos de Yfinance, pero tanto la extracción de datos como el gráfico pueden realizarse fácilmente utilizando inteligencias artificiales como Chat GPT.

-IPC Argentina (2018-2023): Usaremos datos de la FRED Economic Data. Nuevamente, aquí hay un gráfico interesante sobre la evolución del IPC en la Argentina:

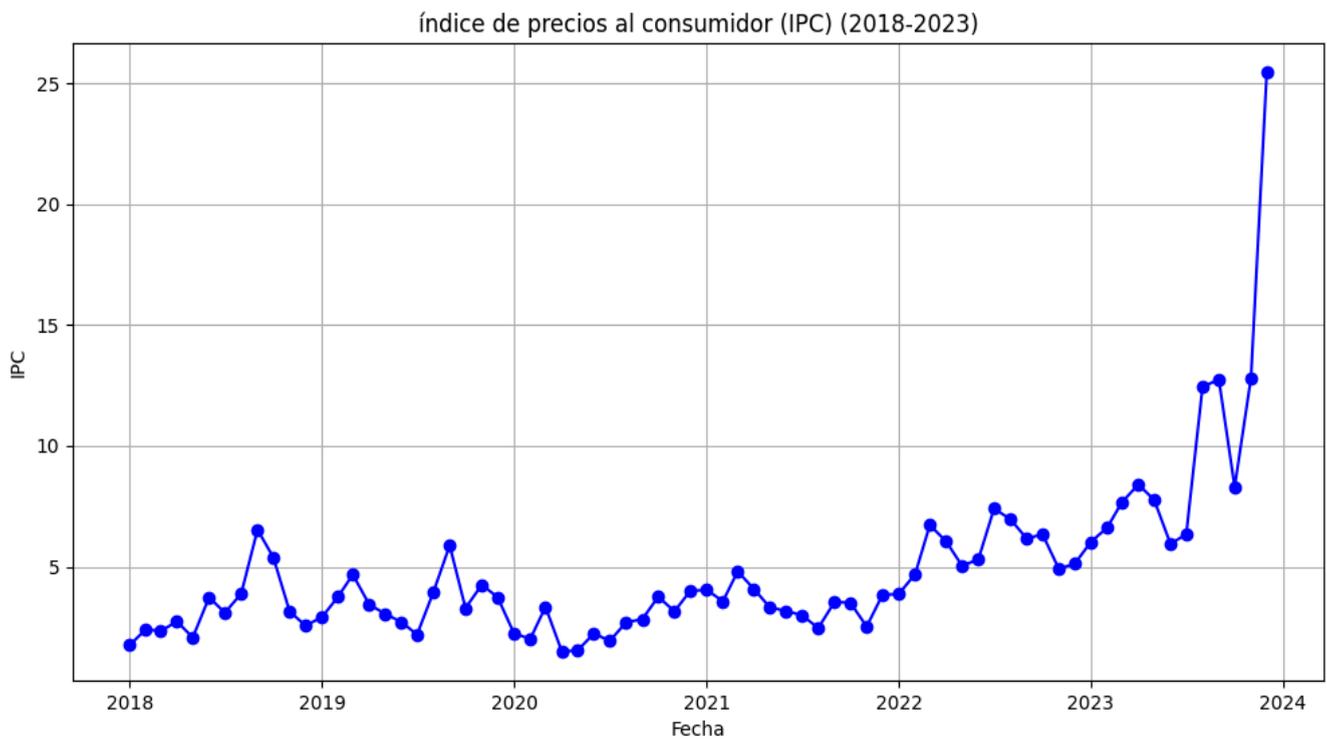

-M2, tasas de interés (long y short) e IPC de Estados Unidos: Datos obtenidos de la FRED Economic Data. Note cómo ha aumentado la M2 desde 2018, al igual que las tasas de interés.

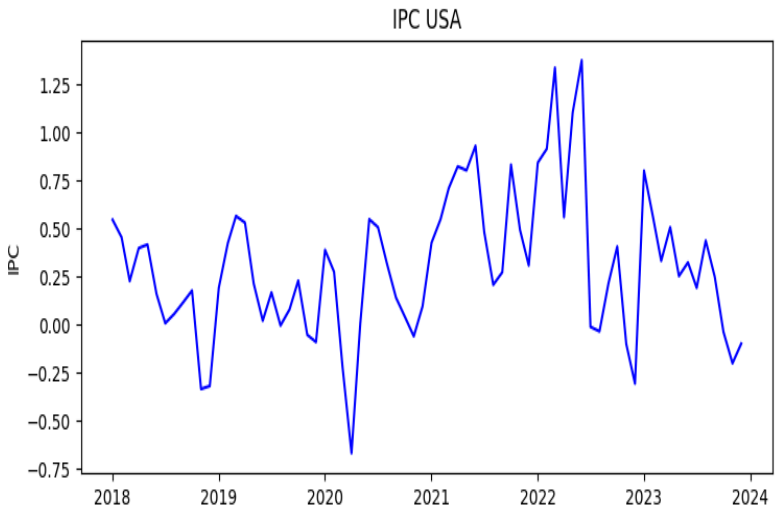
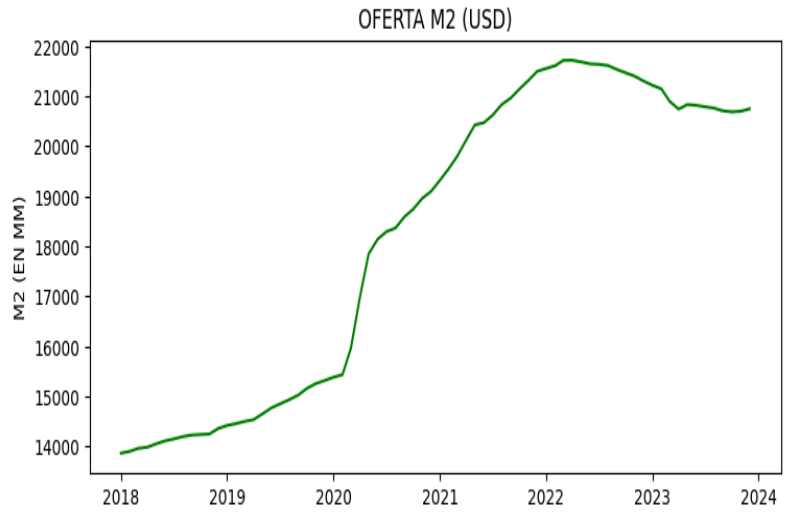
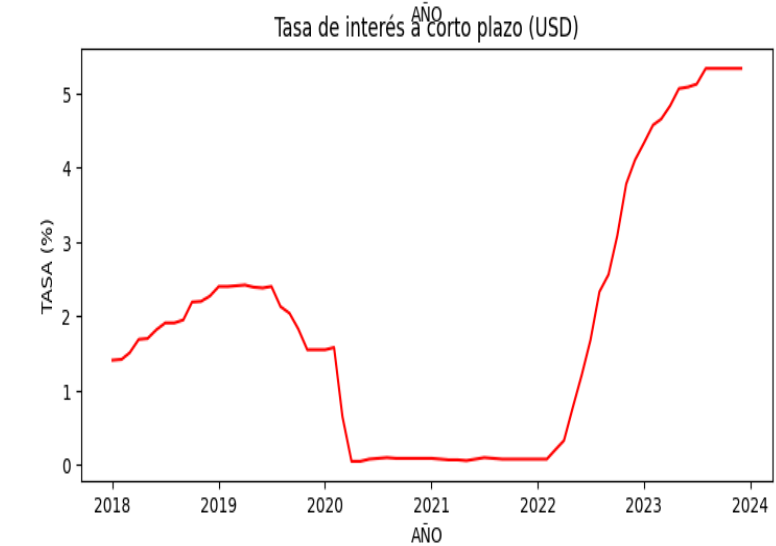
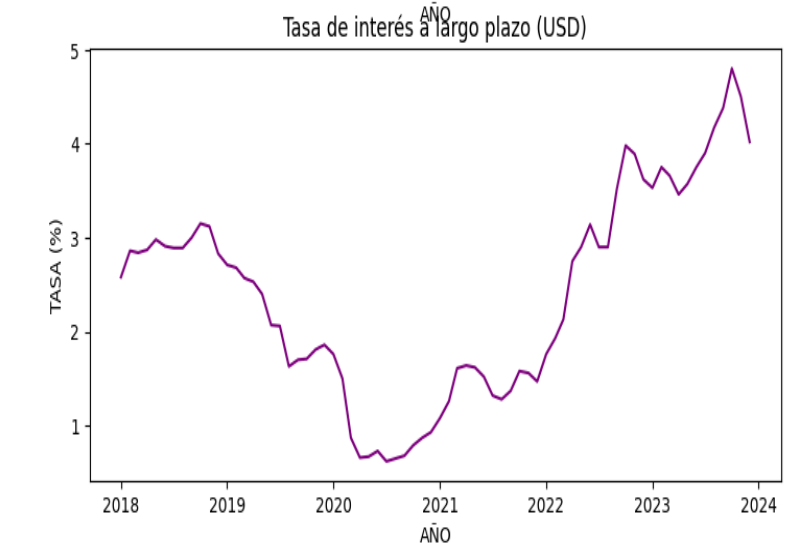

- M2 privado Argentina, promedio móvil de 30 días, variación interanual (en %): Datos obtenidos del BCRA.

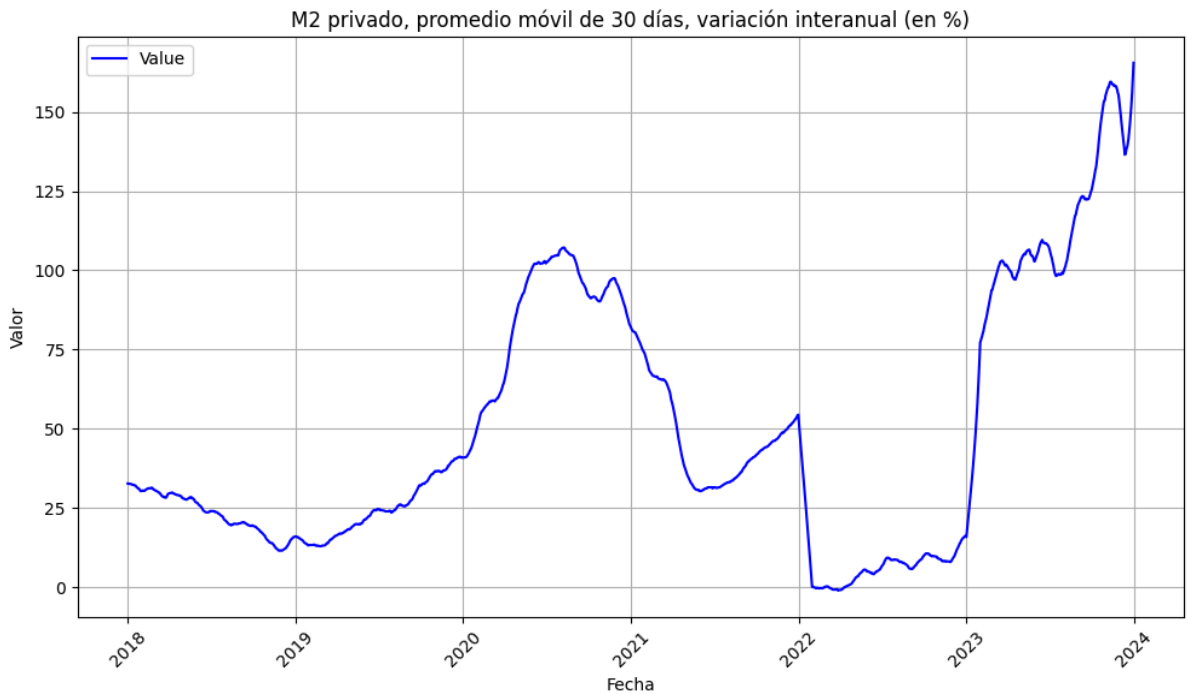

- Tasa de Política Monetaria (en % n.a.) argentina (short interest): Datos obtenidos del BCRA.

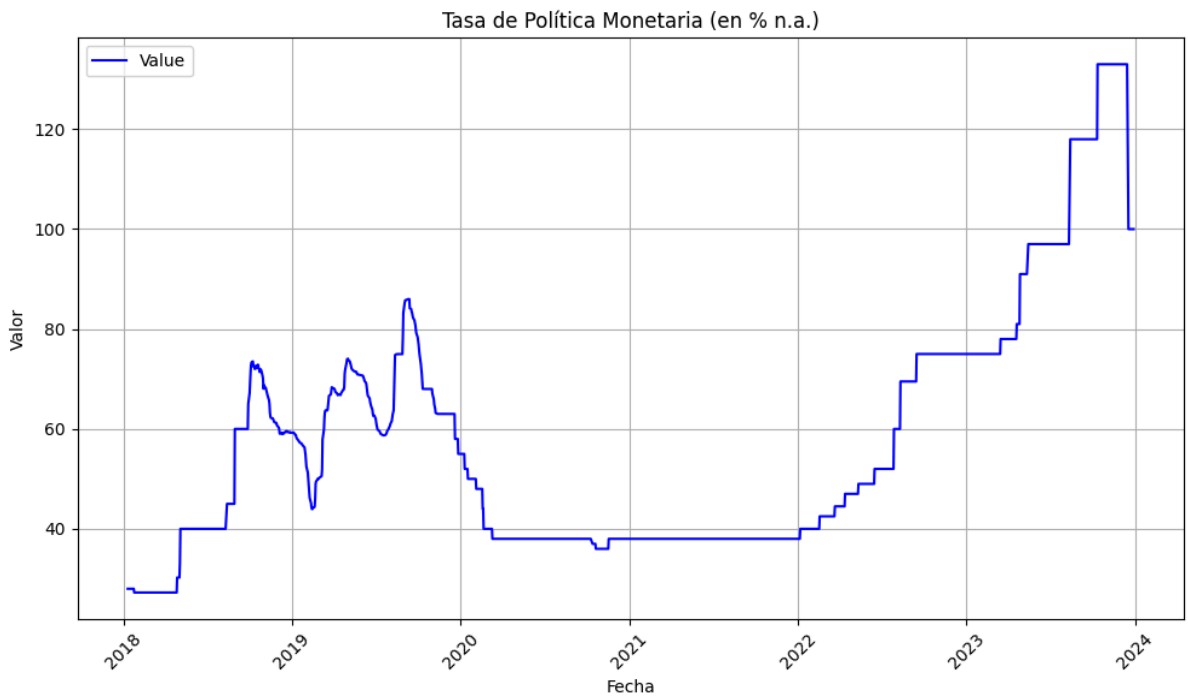

- Tasa de interés préstamos personales argentina (long interest): Datos obtenidos del BCRA.

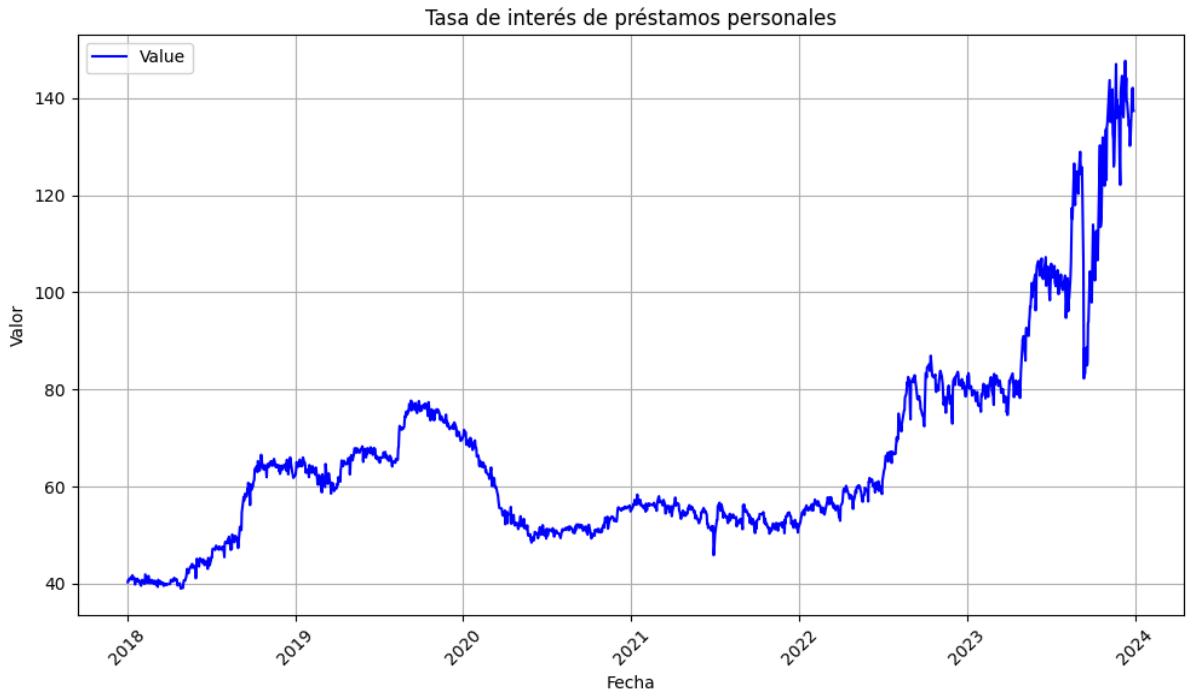

- Expectativas de inflación para Argentina: Data obtenida de los REM del BCRA, usando el valor anual esperado mensualizado (interpolación lineal). No es, claramente, la mejor de las ideas, pero es complicado encontrar mejores fuentes de expectativas. Idealmente, sería bueno conseguir un "sentimiento de mercado" usando métodos de Data Scraping sobre las expectativas diarias de inflación; no obstante, esta aproximación cumple su función y nos abre paso a debates atractivos. Puntualmente, es destacable cómo, si uno estudia los datos oficiales, modeliza una función aproximada, y realiza lo mismo con las expectativas, puede apreciar como los valores de las expectativas comienzan a converger hacia el valor fáctico al cabo de 6 meses (apreciable en 2018). Esta información deja mucho potencial para futuras investigaciones, cuyo grado de relevancia sería extremadamente alto para poder mejorar las mediciones de expectativas si se cuenta con modelos de Data Scraping. En países normales, si se pudiera modelar una función de expectativas teniendo en cuenta estos datos de convergencia usando series de tiempo, se conseguirían mejores aproximaciones, pero la creación de tales funciones trasciende los objetivos de esta investigación. Le recomiendo, además, usando el gráfico, prestar atención a aquellos puntos donde se disparan las expectativas, son de gran relevancia para el análisis fundamental.

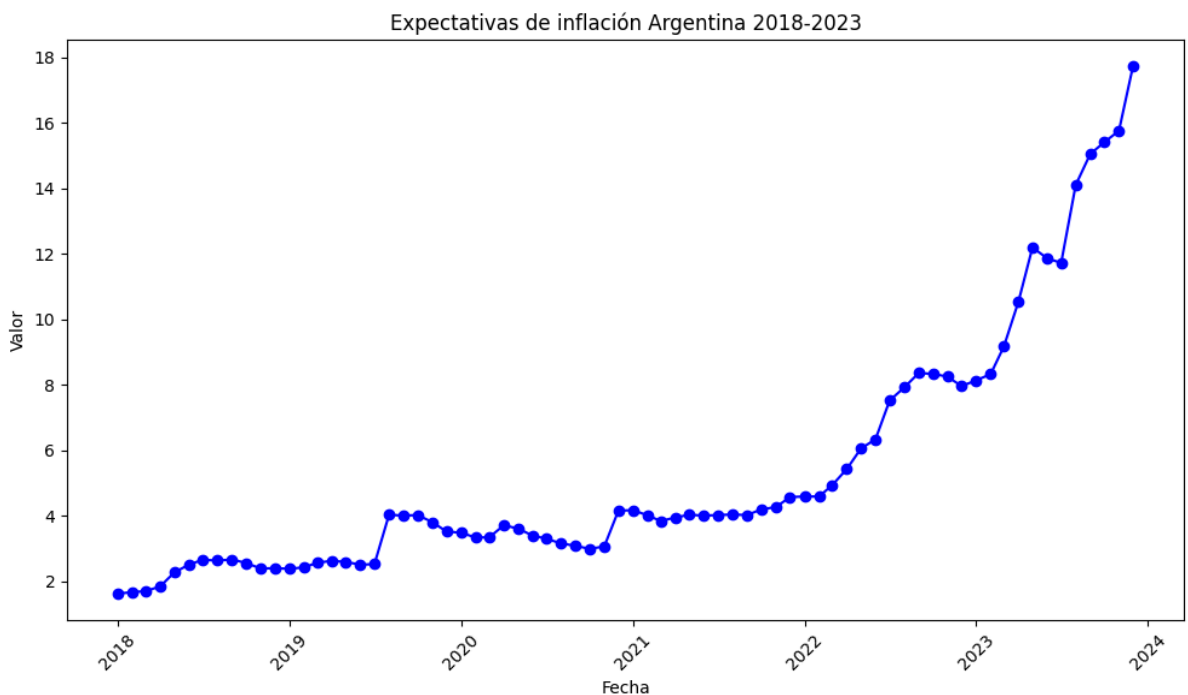

- Para realizar mediciones de la prima de riesgo, utilizaremos el llamado "General government net lending/borrowing for Argentina" (GGNLBAARA188N en el FRED Economic data): Para una descripción de su cálculo, mediciones y objetivos revise https://fred.stlouisfed.org/series/GGNLBAARA188N. También utilizaremos el EMBI+ Arg, aunque, como referencia para elaborar aproximaciones útiles más específicas de la situación del Estado, este indicador es aproximadamente preciso. En definitiva usaremos morfismos en la prima de riesgo para evaluar cómo la categoría reacciona a sus cambios, no necesitamos súper-precisiones, sino aproximaciones. De todos modos, también utilizaremos el índice EMBI+ Arg cuando sea necesario. Antes de pasar al gráfico, es muy importante mencionar que el último valor es una aproximación conservadora, ya que este dato no aparece ni en la FRED ni en el FMI de momento.

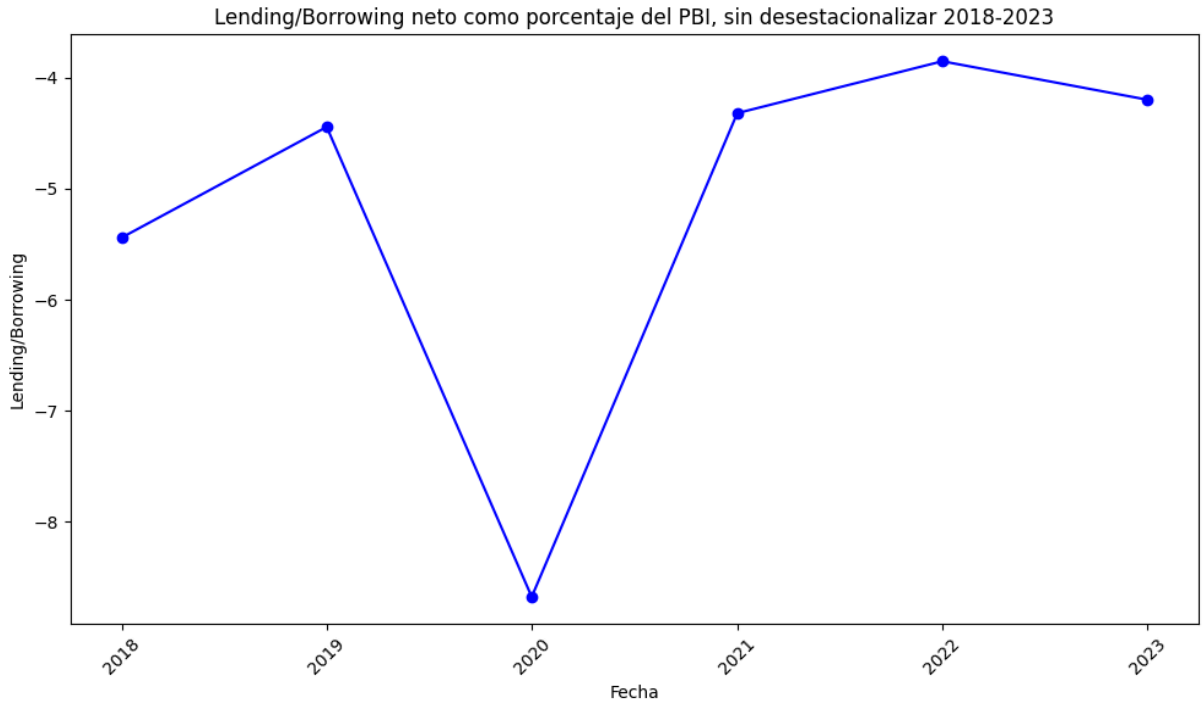

Estos gráficos nos permiten observar relaciones importantes entre sí, nos permiten obtener intuiciones útiles a la hora de entender los morfismos que se producen en la economía por cambios en la liquidez, tasas de interés, tipo de cambio, inflación, expectativas, deuda y préstamos, siendo todas estas variables que tomamos en cuenta para el desarrollo de la categoría, como podrá darse cuenta. Sugiero que revea el marco teórico tras analizar las series de tiempo, de este modo se puede apreciar por qué moldeamos la categoría del modo que lo hicimos. Como podrá darse cuenta, todos los parámetros mencionados, y las propias series de tiempo, están relacionados. Todos forman parte de un sistema dinámico, es por ello que usamos teoría de categorías, la misma nos permite plasmar modelos dinámicos, como iremos descubriendo con el pasar de la investigación.

Utilizaremos las fuentes y estructuras de datos mencionadas para ejecutar pruebas sobre el modelo. Podrá acceder a los Excel con las planillas de datos a través del Drive cuyo vínculo aparecerá al final de esta Parte I.

Para finalizar esta sección, observe los siguientes gráficos y análisis realizados. Usamos el inglés para facilitar el trabajo.

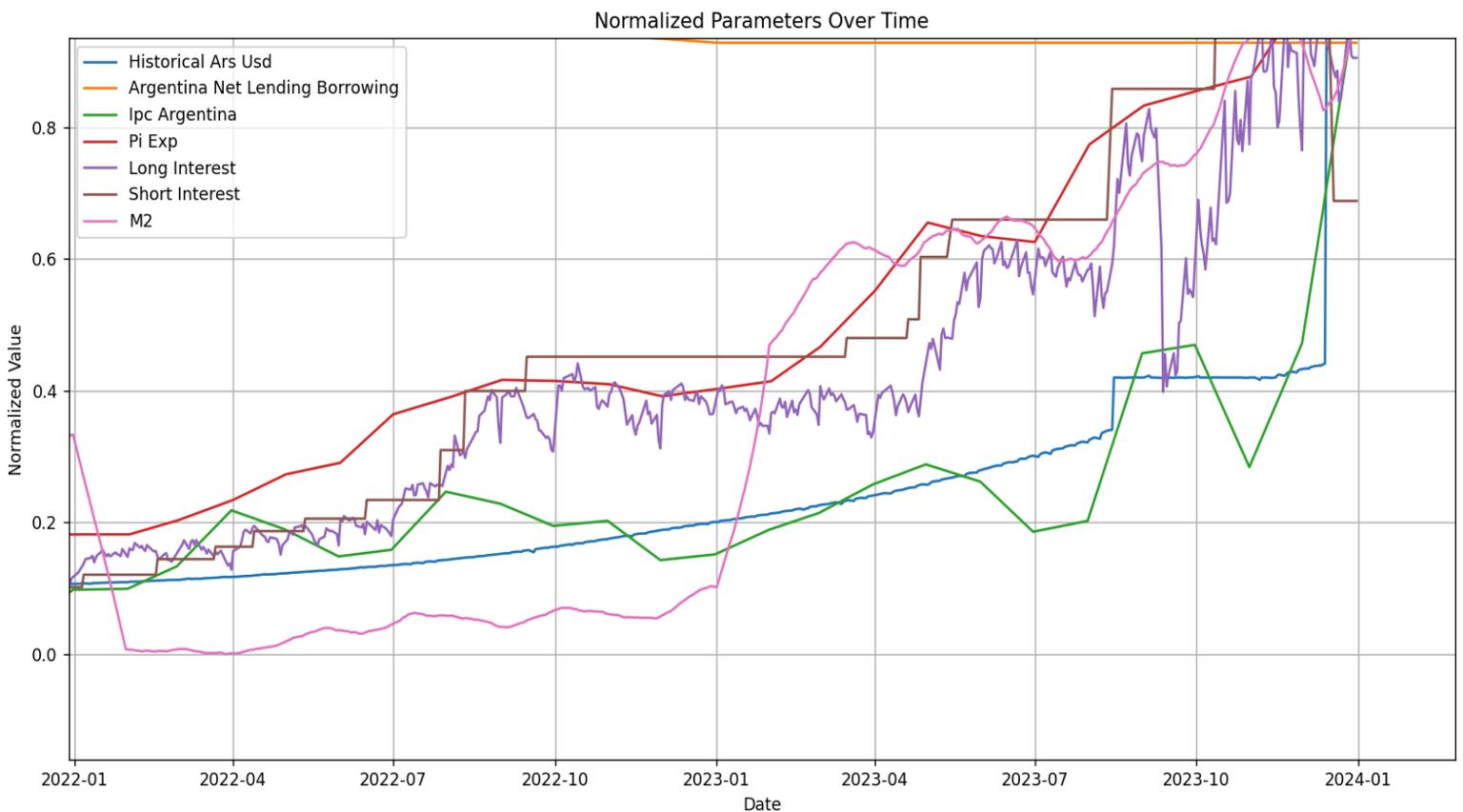

Desde el análisis técnico, se pueden realizar varias observaciones, pero no nos enfocaremos en estos detalles; no obstante, es resaltable como las expectativas de inflación "Pi Exp" parecen comportarse como una resistencia del resto de variables (un techo), siendo muy precisa la relación con el long interest (tasa de interés de préstamos personales), pero sólo en el contexto inflacionario, ya que no se encuentran indicios de que lo mismo suceda en años con inflación menor, recién a partir de octubre del 2020 se hace visible el patrón. Es impresionante cómo ese parámetro, que parece exógeno a las variables macroeconómicas principales, tiene tanta importancia. Por otro lado, la M2 parecería ser una especie de oráculo de lo que va a suceder con las otras variables, siguiendo la teoría económica, el pico de 2023 es un ejemplo, ell valor que toma la M2 en un determinado período, parece representar el que tomarán las variables unos meses o incluso años después.

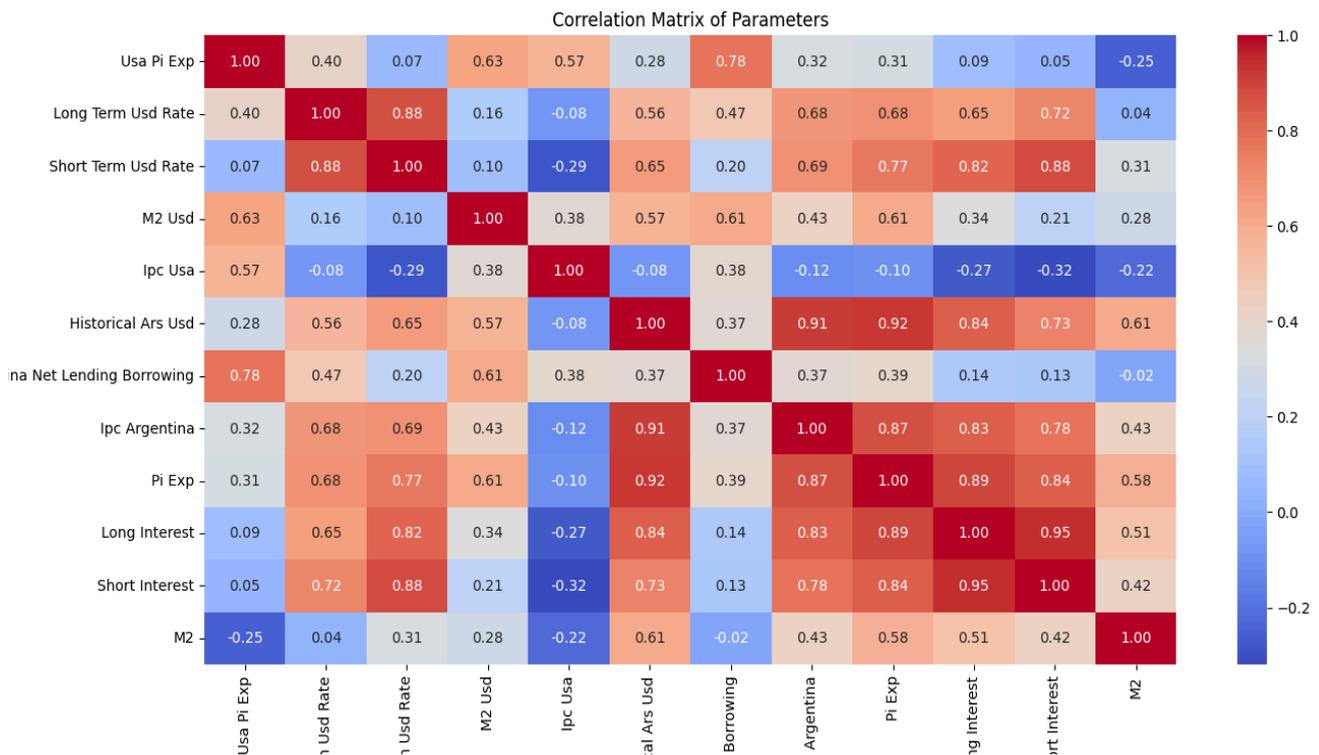

Aquí podemos apreciar una matriz de correlación hecha en Python con unas librerías que permiten hacer uso de diversos algoritmos estadísticos y algebraicos llamada Pandas, MatPlotLib y Seaborn. Si bien las hemos usado para representar toda la información gráfica que se ha visto hasta ahora, parece oportuno mencionarlo ahora para que, en caso de que usted, lector, desee verificar los algoritmos utilizados, o realizar por su cuenta alguna investigación con los datos, sepa que estas librerías pueden facilitarle el camino.

Respecto de lo que podemos observar, note que tenemos dos correlaciones positivas fuertes, la primera siendo las expectativas de inflación en Argentina con el long y el short interest, esto demuestra que **las expectativas de inflación, en efecto, influencian las tasas de interés**; la segunda correlación la vemos entre el IPC de Argentina y las expectativas de inflación, siendo esto obvio, pero aún así significativo para decir que las expectativas no están descarriladas respecto de la inflación real.

También es apreciable una relación entre la M2 de Estados Unidos tanto con las expectativas de inflación de Argentina y con el índice lending-borrowing neto argentino (prestar-devolver), sugiriendo algún tipo de efecto de desbordamiento entre la política monetaria de Estados Unidos en la economía Argentina.

Las expectativas de inflación de Argentina parecen también estar fuertemente vinculadas con el índice anteriormente mencionado, demostrando una correlación entre la política fiscal y las expectativas de inflación.

También vemos una correlación fuerte entre el valor de la relación ARS-USD con la inflación y las expectativas de inflación. Esto refuerza la importancia de la devaluación ante dinámicas inflacionarias como las de Argentina.

Podemos leer información acerca de correlaciones débiles, pero no son de tanta importancia

Para analizar en mayor detalle el impacto de la M2, utilizaremos un modelo de autorregresión vectorial (VAR), el cual es un tipo de modelo de proceso estocástico, útil para este tipo de análisis, véase la página 81 para más detalles. Es sugerible investigar la librería de Python "Statsmodels", la cual cuenta con las herramientas que permiten realizar la prueba de estacionariedad Dickey-Fuller aumentada (ADF test), diferenciar las variables no estacionarias, realizar la prueba de cointegración Johansen, la prueba de causalidad Granger, aplicar el modelo VAR y utilizar la prueba de diagnóstico Ljung-Box. Recuerde estos pasos porque no los volveremos a mencionar más adelante.
He aquí el código para realizar lo mencionado:

```python
import pandas as pd
import numpy as np
from statsmodels.tsa.stattools import adfuller, grangercausalitytests
from statsmodels.tsa.vector_ar.vecm import coint_johansen
from statsmodels.tsa.api import VAR
import matplotlib.pyplot as plt

file_path = 
data = pd.read_excel(file_path)

data.columns = [
    "Date", "US_inflation_exp", "USD_long_rate", "USD_short_rate", "M2_USD",
    "CPI_US", "ARS_USD", "Arg_lending_borrowing", "CPI_Arg",
    "Infl_exp_Arg", "Arg_long_rate", "Arg_short_rate", "M2_Arg"
]

data["Date"] = pd.to_datetime(data["Date"])
data.set_index("Date", inplace=True)

def check_stationarity(series, alpha=0.05):
    result = adfuller(series)
    p_value = result[1]
    return p_value < alpha

stationarity_results = {}
for col in data.columns:
    stationarity_results[col] = check_stationarity(data[col])

print("\nStationarity Results:")
for var, is_stationary in stationarity_results.items():
    print(f"{var}: {'Stationary' if is_stationary else 'Non-stationary'}")

data_diff = data.copy()
for col in data.columns:
    if not stationarity_results[col]:
        data_diff[col] = data[col].diff()

data_diff = data_diff.dropna()
```

```python
def johansen_test(data, alpha=0.05):
    result = coint_johansen(data, det_order=0, k_ar_diff=1)
    traces = result.lr1
    critical_values = result.cvt[:, 1]
    cointegration = traces > critical_values
    return cointegration, result

cointegration, johansen_result = johansen_test(data_diff)

print("\nJohansen Cointegration Test Results:")
for i, is_coint in enumerate(cointegration):
    print(f"Eigenvalue {i+1}: {'Cointegrated' if is_coint else 'Not Cointegrated'}")

if any(cointegration):
    print("\nVariables are cointegrated. Using VECM may be more appropriate.")
else:
    print("\nNo cointegration found. Proceeding with VAR.")

print("\nGranger Causality Test Results:")
max_lag = 5
for col in data_diff.columns:
    if col != "M2_USD":
        test_result = grangercausalitytests(data_diff[["M2_USD", col]], max_lag, verbose=False)
        p_values = [round(test[0]["ssr_ftest"][1], 4) for test in test_result.values()]
        print(f"{col} -> M2_USD Granger Causality p-values: {p_values}")

var_model = VAR(data_diff)
selected_lag = var_model.select_order(maxlags=10)
print("\nSelected Lag Order:", selected_lag.selected_orders)

model_fitted = var_model.fit(selected_lag.selected_orders["aic"])

print("\nVAR Model Summary:")
print(model_fitted.summary())

from statsmodels.stats.diagnostic import acorr_ljungbox

residuals = model_fitted.resid
print("\nLjung-Box Test for Residual Serial Correlation:")
for col in residuals.columns:
    lb_test = acorr_ljungbox(residuals[col], lags=[10], return_df=True)
    print(f"{col}: {lb_test['lb_pvalue'].iloc[0]}")
```

No voy a incluir en este documento los resultados, ya que serían hojas y hojas de información, si incluiré el sumario. Los archivos con la información bruta se encuentran en el Drive.

Summary of Regression Results
==================================
Model:                         VAR
Method:                        OLS

```
--------------------------------------------------------------------
No. of Equations:      12.0000    BIC:              -77.9114
Nobs:                  2188.00    HQIC:             -78.4065
Log likelihood:        49133.0    FPE:              6.67809e-35
AIC:                   -78.6917   Det(Omega_mle):   5.82697e-35
--------------------------------------------------------------------
```

Results for equation US_inflation_exp

```
==========================================================================================
                          coefficient    std. error    t-stat    prob
------------------------------------------------------------------------------------------
const                       -0.000038      0.001359    -0.028    0.978
L1.US_inflation_exp          0.027291      0.021534     1.267    0.205
L1.USD_long_rate            -0.322031      0.562035    -0.573    0.567
L1.USD_short_rate            1.183810      0.707806     1.673    0.094
L1.M2_USD                    0.000317      0.001046     0.303    0.762
L1.CPI_US                    0.052764      0.069273     0.762    0.446
L1.ARS_USD                  -0.000043      0.000084    -0.511    0.609
L1.Arg_lending_borrowing    -1.110980      1.232987    -0.901    0.368
L1.CPI_Arg                  -0.015840      0.056968    -0.278    0.781
L1.Infl_exp_Arg              0.012127      0.212824     0.057    0.955
L1.Arg_long_rate             0.000284      0.000511     0.556    0.579
L1.Arg_short_rate           -0.000497      0.000920    -0.541    0.589
L1.M2_Arg                    0.000968      0.004621     0.209    0.834
L2.US_inflation_exp          0.048160      0.021487     2.241    0.025
L2.USD_long_rate             0.399066      0.560732     0.712    0.477
L2.USD_short_rate           -1.299426      0.709933    -1.830    0.067
L2.M2_USD                   -0.000553      0.001042    -0.531    0.596
L2.CPI_US                   -0.047821      0.069106    -0.692    0.489
L2.ARS_USD                   0.000003      0.000084     0.030    0.976
L2.Arg_lending_borrowing     1.205282      1.233733     0.977    0.329
L2.CPI_Arg                   0.018818      0.057428     0.328    0.743
L2.Infl_exp_Arg             -0.088526      0.213348    -0.415    0.678
L2.Arg_long_rate             0.000106      0.000504     0.211    0.833
L2.Arg_short_rate            0.000059      0.000923     0.064    0.949
L2.M2_Arg                    0.000191      0.004655     0.041    0.967
==========================================================================================
```

Results for equation USD_long_rate

```
==========================================================================================
                          coefficient    std. error    t-stat    prob
------------------------------------------------------------------------------------------
const                       -0.000041      0.000054    -0.747    0.455
L1.US_inflation_exp          0.000109      0.000863     0.127    0.899
L1.USD_long_rate             0.984356      0.022513    43.724    0.000
L1.USD_short_rate           -0.000126      0.028352    -0.004    0.996
```

| | coefficient | std. error | t-stat | prob |
|---|---|---|---|---|
| L1.M2_USD | -0.000009 | 0.000042 | -0.214 | 0.830 |
| L1.CPI_US | 0.002461 | 0.002775 | 0.887 | 0.375 |
| L1.ARS_USD | -0.000002 | 0.000003 | -0.595 | 0.552 |
| L1.Arg_lending_borrowing | -0.003390 | 0.049389 | -0.069 | 0.945 |
| L1.CPI_Arg | 0.018328 | 0.002282 | 8.032 | 0.000 |
| L1.Infl_exp_Arg | 0.001219 | 0.008525 | 0.143 | 0.886 |
| L1.Arg_long_rate | -0.000007 | 0.000020 | -0.329 | 0.742 |
| L1.Arg_short_rate | 0.000002 | 0.000037 | 0.049 | 0.961 |
| L1.M2_Arg | 0.000724 | 0.000185 | 3.910 | 0.000 |
| L2.US_inflation_exp | 0.000145 | 0.000861 | 0.168 | 0.866 |
| L2.USD_long_rate | -0.004944 | 0.022461 | -0.220 | 0.826 |
| L2.USD_short_rate | 0.002708 | 0.028437 | 0.095 | 0.924 |
| L2.M2_USD | 0.000016 | 0.000042 | 0.384 | 0.701 |
| L2.CPI_US | -0.002476 | 0.002768 | -0.895 | 0.371 |
| L2.ARS_USD | -0.000001 | 0.000003 | -0.178 | 0.859 |
| L2.Arg_lending_borrowing | 0.004376 | 0.049419 | 0.089 | 0.929 |
| L2.CPI_Arg | -0.017404 | 0.002300 | -7.566 | 0.000 |
| L2.Infl_exp_Arg | 0.001666 | 0.008546 | 0.195 | 0.845 |
| L2.Arg_long_rate | 0.000017 | 0.000020 | 0.827 | 0.408 |
| L2.Arg_short_rate | 0.000027 | 0.000037 | 0.732 | 0.464 |
| L2.M2_Arg | -0.000857 | 0.000186 | -4.596 | 0.000 |

==================================================================================

Results for equation USD_short_rate

==================================================================================

| | coefficient | std. error | t-stat | prob |
|---|---|---|---|---|
| const | 0.000004 | 0.000043 | 0.097 | 0.923 |
| L1.US_inflation_exp | 0.000387 | 0.000675 | 0.573 | 0.567 |
| L1.USD_long_rate | 0.005742 | 0.017629 | 0.326 | 0.745 |
| L1.USD_short_rate | 0.989767 | 0.022201 | 44.583 | 0.000 |
| L1.M2_USD | -0.000005 | 0.000033 | -0.154 | 0.878 |
| L1.CPI_US | -0.003244 | 0.002173 | -1.493 | 0.135 |
| L1.ARS_USD | -0.000000 | 0.000003 | -0.029 | 0.977 |
| L1.Arg_lending_borrowing | -0.000738 | 0.038673 | -0.019 | 0.985 |
| L1.CPI_Arg | 0.002800 | 0.001787 | 1.567 | 0.117 |
| L1.Infl_exp_Arg | 0.001195 | 0.006675 | 0.179 | 0.858 |
| L1.Arg_long_rate | 0.000031 | 0.000016 | 1.937 | 0.053 |
| L1.Arg_short_rate | -0.000005 | 0.000029 | -0.190 | 0.849 |
| L1.M2_Arg | 0.000294 | 0.000145 | 2.031 | 0.042 |
| L2.US_inflation_exp | -0.001579 | 0.000674 | -2.343 | 0.019 |
| L2.USD_long_rate | 0.004585 | 0.017588 | 0.261 | 0.794 |
| L2.USD_short_rate | -0.002239 | 0.022267 | -0.101 | 0.920 |
| L2.M2_USD | 0.000011 | 0.000033 | 0.328 | 0.743 |
| L2.CPI_US | 0.003194 | 0.002168 | 1.474 | 0.141 |
| L2.ARS_USD | 0.000001 | 0.000003 | 0.308 | 0.758 |

| | coefficient | std. error | t-stat | prob |
|---|---|---|---|---|
| L2.Arg_lending_borrowing | -0.003052 | 0.038697 | -0.079 | 0.937 |
| L2.CPI_Arg | -0.003073 | 0.001801 | -1.706 | 0.088 |
| L2.Infl_exp_Arg | 0.000487 | 0.006692 | 0.073 | 0.942 |
| L2.Arg_long_rate | 0.000004 | 0.000016 | 0.251 | 0.802 |
| L2.Arg_short_rate | -0.000006 | 0.000029 | -0.205 | 0.838 |
| L2.M2_Arg | -0.000333 | 0.000146 | -2.280 | 0.023 |

===========================================================================

Results for equation M2_USD

===========================================================================

| | coefficient | std. error | t-stat | prob |
|---|---|---|---|---|
| const | 0.090713 | 0.028640 | 3.167 | 0.002 |
| L1.US_inflation_exp | 0.193199 | 0.453766 | 0.426 | 0.670 |
| L1.USD_long_rate | -1.041962 | 11.843444 | -0.088 | 0.930 |
| L1.USD_short_rate | -1.867132 | 14.915191 | -0.125 | 0.900 |
| L1.M2_USD | 0.981030 | 0.022040 | 44.511 | 0.000 |
| L1.CPI_US | -1.025851 | 1.459752 | -0.703 | 0.482 |
| L1.ARS_USD | -0.000507 | 0.001773 | -0.286 | 0.775 |
| L1.Arg_lending_borrowing | -0.664876 | 25.982041 | -0.026 | 0.980 |
| L1.CPI_Arg | 1.274447 | 1.200458 | 1.062 | 0.288 |
| L1.Infl_exp_Arg | -0.886714 | 4.484711 | -0.198 | 0.843 |
| L1.Arg_long_rate | -0.000316 | 0.010758 | -0.029 | 0.977 |
| L1.Arg_short_rate | 0.003803 | 0.019392 | 0.196 | 0.845 |
| L1.M2_Arg | 0.265910 | 0.097374 | 2.731 | 0.006 |
| L2.US_inflation_exp | 0.925953 | 0.452792 | 2.045 | 0.041 |
| L2.USD_long_rate | -2.868752 | 11.815987 | -0.243 | 0.808 |
| L2.USD_short_rate | -9.952395 | 14.960010 | -0.665 | 0.506 |
| L2.M2_USD | -0.001313 | 0.021955 | -0.060 | 0.952 |
| L2.CPI_US | 1.024839 | 1.456230 | 0.704 | 0.482 |
| L2.ARS_USD | -0.000330 | 0.001769 | -0.186 | 0.852 |
| L2.Arg_lending_borrowing | 7.122076 | 25.997758 | 0.274 | 0.784 |
| L2.CPI_Arg | -1.379798 | 1.210154 | -1.140 | 0.254 |
| L2.Infl_exp_Arg | 0.299375 | 4.495766 | 0.067 | 0.947 |
| L2.Arg_long_rate | -0.003475 | 0.010619 | -0.327 | 0.743 |
| L2.Arg_short_rate | 0.001280 | 0.019449 | 0.066 | 0.948 |
| L2.M2_Arg | -0.305657 | 0.098099 | -3.116 | 0.002 |

===========================================================================

Results for equation CPI_US

===========================================================================

| | coefficient | std. error | t-stat | prob |
|---|---|---|---|---|
| const | 0.000296 | 0.000116 | 2.545 | 0.011 |

|   | coefficient | std. error | t-stat | prob |
|---|---|---|---|---|
| L1.US_inflation_exp | 0.000020 | 0.001841 | 0.011 | 0.991 |
| L1.USD_long_rate | 0.019129 | 0.048054 | 0.398 | 0.691 |
| L1.USD_short_rate | 0.005171 | 0.060518 | 0.085 | 0.932 |
| L1.M2_USD | -0.000020 | 0.000089 | -0.228 | 0.820 |
| L1.CPI_US | 1.960404 | 0.005923 | 330.987 | 0.000 |
| L1.ARS_USD | 0.000000 | 0.000007 | 0.061 | 0.952 |
| L1.Arg_lending_borrowing | -0.008668 | 0.105422 | -0.082 | 0.934 |
| L1.CPI_Arg | -0.001232 | 0.004871 | -0.253 | 0.800 |
| L1.Infl_exp_Arg | 0.002554 | 0.018197 | 0.140 | 0.888 |
| L1.Arg_long_rate | 0.000088 | 0.000044 | 2.011 | 0.044 |
| L1.Arg_short_rate | 0.000005 | 0.000079 | 0.061 | 0.952 |
| L1.M2_Arg | 0.000360 | 0.000395 | 0.911 | 0.362 |
| L2.US_inflation_exp | -0.000750 | 0.001837 | -0.408 | 0.683 |
| L2.USD_long_rate | -0.001983 | 0.047943 | -0.041 | 0.967 |
| L2.USD_short_rate | 0.003066 | 0.060700 | 0.051 | 0.960 |
| L2.M2_USD | 0.000031 | 0.000089 | 0.352 | 0.725 |
| L2.CPI_US | -0.961616 | 0.005909 | -162.748 | 0.000 |
| L2.ARS_USD | -0.000000 | 0.000007 | -0.067 | 0.946 |
| L2.Arg_lending_borrowing | 0.002036 | 0.105485 | 0.019 | 0.985 |
| L2.CPI_Arg | 0.000297 | 0.004910 | 0.060 | 0.952 |
| L2.Infl_exp_Arg | -0.001093 | 0.018241 | -0.060 | 0.952 |
| L2.Arg_long_rate | -0.000038 | 0.000043 | -0.875 | 0.381 |
| L2.Arg_short_rate | -0.000035 | 0.000079 | -0.448 | 0.654 |
| L2.M2_Arg | -0.000401 | 0.000398 | -1.007 | 0.314 |

==========================================================================================

Results for equation ARS_USD

==========================================================================================

|   | coefficient | std. error | t-stat | prob |
|---|---|---|---|---|
| const | 0.676117 | 0.337958 | 2.001 | 0.045 |
| L1.US_inflation_exp | 1.976783 | 5.354460 | 0.369 | 0.712 |
| L1.USD_long_rate | -35.840094 | 139.753075 | -0.256 | 0.798 |
| L1.USD_short_rate | 20.412855 | 175.999801 | 0.116 | 0.908 |
| L1.M2_USD | -0.113786 | 0.260076 | -0.438 | 0.662 |
| L1.CPI_US | 4.155504 | 17.225132 | 0.241 | 0.809 |
| L1.ARS_USD | -0.065425 | 0.020926 | -3.127 | 0.002 |
| L1.Arg_lending_borrowing | -11.210215 | 306.589039 | -0.037 | 0.971 |
| L1.CPI_Arg | -0.349693 | 14.165446 | -0.025 | 0.980 |
| L1.Infl_exp_Arg | -13.767907 | 52.919750 | -0.260 | 0.795 |
| L1.Arg_long_rate | 0.312852 | 0.126942 | 2.465 | 0.014 |
| L1.Arg_short_rate | 0.190050 | 0.228821 | 0.831 | 0.406 |
| L1.M2_Arg | 12.236265 | 1.149021 | 10.649 | 0.000 |
| L2.US_inflation_exp | -4.297712 | 5.342967 | -0.804 | 0.421 |
| L2.USD_long_rate | 10.677663 | 139.429076 | 0.077 | 0.939 |
| L2.USD_short_rate | -21.766734 | 176.528667 | -0.123 | 0.902 |
| L2.M2_USD | 0.072902 | 0.259074 | 0.281 | 0.778 |
| L2.CPI_US | -4.869176 | 17.183567 | -0.283 | 0.777 |
| L2.ARS_USD | -0.022120 | 0.020879 | -1.059 | 0.289 |
| L2.Arg_lending_borrowing | 22.873032 | 306.774497 | 0.075 | 0.941 |
| L2.CPI_Arg | 22.181588 | 14.279864 | 1.553 | 0.120 |
| L2.Infl_exp_Arg | -3.366526 | 53.050199 | -0.063 | 0.949 |
| L2.Arg_long_rate | -0.539050 | 0.125301 | -4.302 | 0.000 |
| L2.Arg_short_rate | 0.318120 | 0.229501 | 1.386 | 0.166 |
| L2.M2_Arg | -13.148619 | 1.157569 | -11.359 | 0.000 |

=====================================================================================

Results for equation Arg_lending_borrowing
=====================================================================================

| | coefficient | std. error | t-stat | prob |
|---|---|---|---|---|
| const | -0.000004 | 0.000024 | -0.185 | 0.853 |
| L1.US_inflation_exp | 0.000117 | 0.000377 | 0.310 | 0.756 |
| L1.USD_long_rate | 0.000038 | 0.009849 | 0.004 | 0.997 |
| L1.USD_short_rate | -0.000481 | 0.012404 | -0.039 | 0.969 |
| L1.M2_USD | -0.000000 | 0.000018 | -0.005 | 0.996 |
| L1.CPI_US | 0.000165 | 0.001214 | 0.136 | 0.892 |
| L1.ARS_USD | 0.000000 | 0.000001 | 0.020 | 0.984 |
| L1.Arg_lending_borrowing | 0.997116 | 0.021607 | 46.147 | 0.000 |
| L1.CPI_Arg | -0.001925 | 0.000998 | -1.928 | 0.054 |
| L1.Infl_exp_Arg | -0.000077 | 0.003730 | -0.021 | 0.983 |
| L1.Arg_long_rate | 0.000004 | 0.000009 | 0.473 | 0.636 |
| L1.Arg_short_rate | -0.000001 | 0.000016 | -0.072 | 0.942 |
| L1.M2_Arg | 0.000092 | 0.000081 | 1.132 | 0.258 |
| L2.US_inflation_exp | 0.000069 | 0.000377 | 0.182 | 0.855 |
| L2.USD_long_rate | -0.000860 | 0.009826 | -0.087 | 0.930 |
| L2.USD_short_rate | 0.000264 | 0.012441 | 0.021 | 0.983 |
| L2.M2_USD | 0.000001 | 0.000018 | 0.068 | 0.946 |
| L2.CPI_US | -0.000156 | 0.001211 | -0.129 | 0.898 |
| L2.ARS_USD | 0.000000 | 0.000001 | 0.057 | 0.954 |
| L2.Arg_lending_borrowing | -0.001716 | 0.021620 | -0.079 | 0.937 |
| L2.CPI_Arg | 0.001835 | 0.001006 | 1.823 | 0.068 |
| L2.Infl_exp_Arg | 0.000179 | 0.003739 | 0.048 | 0.962 |
| L2.Arg_long_rate | -0.000001 | 0.000009 | -0.129 | 0.897 |
| L2.Arg_short_rate | 0.000000 | 0.000016 | 0.007 | 0.994 |
| L2.M2_Arg | -0.000083 | 0.000082 | -1.013 | 0.311 |

=====================================================================================

Results for equation CPI_Arg
=====================================================================================

|                          | coefficient | std. error | t-stat  | prob  |
|--------------------------|-------------|------------|---------|-------|
| const                    | 0.000550    | 0.000518   | 1.061   | 0.289 |
| L1.US_inflation_exp      | 0.005363    | 0.008206   | 0.654   | 0.513 |
| L1.USD_long_rate         | -0.075492   | 0.214169   | -0.352  | 0.724 |
| L1.USD_short_rate        | -0.007112   | 0.269717   | -0.026  | 0.979 |
| L1.M2_USD                | -0.000080   | 0.000399   | -0.202  | 0.840 |
| L1.CPI_US                | 0.039058    | 0.026397   | 1.480   | 0.139 |
| L1.ARS_USD               | -0.000002   | 0.000032   | -0.071  | 0.943 |
| L1.Arg_lending_borrowing | -0.035067   | 0.469843   | -0.075  | 0.941 |
| L1.CPI_Arg               | 1.004863    | 0.021708   | 46.289  | 0.000 |
| L1.Infl_exp_Arg          | -0.000490   | 0.081099   | -0.006  | 0.995 |
| L1.Arg_long_rate         | -0.000906   | 0.000195   | -4.657  | 0.000 |
| L1.Arg_short_rate        | -0.000249   | 0.000351   | -0.710  | 0.478 |
| L1.M2_Arg                | 0.005566    | 0.001761   | 3.161   | 0.002 |
| L2.US_inflation_exp      | 0.000115    | 0.008188   | 0.014   | 0.989 |
| L2.USD_long_rate         | -0.038926   | 0.213673   | -0.182  | 0.855 |
| L2.USD_short_rate        | -0.021551   | 0.270528   | -0.080  | 0.937 |
| L2.M2_USD                | -0.000006   | 0.000397   | -0.016  | 0.987 |
| L2.CPI_US                | -0.039412   | 0.026334   | -1.497  | 0.134 |
| L2.ARS_USD               | 0.000018    | 0.000032   | 0.556   | 0.578 |
| L2.Arg_lending_borrowing | 0.076000    | 0.470127   | 0.162   | 0.872 |
| L2.CPI_Arg               | -0.021410   | 0.021884   | -0.978  | 0.328 |
| L2.Infl_exp_Arg          | 0.049534    | 0.081299   | 0.609   | 0.542 |
| L2.Arg_long_rate         | -0.000056   | 0.000192   | -0.293  | 0.769 |
| L2.Arg_short_rate        | 0.000723    | 0.000352   | 2.056   | 0.040 |
| L2.M2_Arg                | -0.006331   | 0.001774   | -3.569  | 0.000 |

========================================================================================

Results for equation Infl_exp_Arg
========================================================================================

|                          | coefficient | std. error | t-stat  | prob  |
|--------------------------|-------------|------------|---------|-------|
| const                    | 0.000282    | 0.000144   | 1.960   | 0.050 |
| L1.US_inflation_exp      | 0.000487    | 0.002279   | 0.214   | 0.831 |
| L1.USD_long_rate         | 0.008651    | 0.059475   | 0.145   | 0.884 |
| L1.USD_short_rate        | 0.012568    | 0.074900   | 0.168   | 0.867 |
| L1.M2_USD                | 0.000009    | 0.000111   | 0.078   | 0.938 |
| L1.CPI_US                | -0.001805   | 0.007331   | -0.246  | 0.805 |
| L1.ARS_USD               | 0.000002    | 0.000009   | 0.266   | 0.790 |
| L1.Arg_lending_borrowing | 0.002896    | 0.130475   | 0.022   | 0.982 |
| L1.CPI_Arg               | 0.001820    | 0.006028   | 0.302   | 0.763 |
| L1.Infl_exp_Arg          | 0.985022    | 0.022521   | 43.738  | 0.000 |
| L1.Arg_long_rate         | -0.000112   | 0.000054   | -2.064  | 0.039 |
| L1.Arg_short_rate        | 0.000023    | 0.000097   | 0.234   | 0.815 |
| L1.M2_Arg                | -0.002275   | 0.000489   | -4.653  | 0.000 |

| | coefficient | std. error | t-stat | prob |
|---|---|---|---|---|
| L2.US_inflation_exp | 0.002294 | 0.002274 | 1.009 | 0.313 |
| L2.USD_long_rate | -0.007057 | 0.059337 | -0.119 | 0.905 |
| L2.USD_short_rate | -0.023795 | 0.075125 | -0.317 | 0.751 |
| L2.M2_USD | -0.000044 | 0.000110 | -0.399 | 0.690 |
| L2.CPI_US | 0.001959 | 0.007313 | 0.268 | 0.789 |
| L2.ARS_USD | -0.000000 | 0.000009 | -0.025 | 0.980 |
| L2.Arg_lending_borrowing | 0.004530 | 0.130554 | 0.035 | 0.972 |
| L2.CPI_Arg | -0.003261 | 0.006077 | -0.537 | 0.592 |
| L2.Infl_exp_Arg | -0.011266 | 0.022577 | -0.499 | 0.618 |
| L2.Arg_long_rate | -0.000105 | 0.000053 | -1.960 | 0.050 |
| L2.Arg_short_rate | -0.000006 | 0.000098 | -0.065 | 0.948 |
| L2.M2_Arg | 0.002558 | 0.000493 | 5.194 | 0.000 |

==========================================================================================

Results for equation Arg_long_rate

==========================================================================================

| | coefficient | std. error | t-stat | prob |
|---|---|---|---|---|
| const | 0.084421 | 0.058259 | 1.449 | 0.147 |
| L1.US_inflation_exp | 0.584189 | 0.923022 | 0.633 | 0.527 |
| L1.USD_long_rate | 35.529489 | 24.091175 | 1.475 | 0.140 |
| L1.USD_short_rate | 6.200184 | 30.339526 | 0.204 | 0.838 |
| L1.M2_USD | -0.021556 | 0.044833 | -0.481 | 0.631 |
| L1.CPI_US | 0.265784 | 2.969335 | 0.090 | 0.929 |
| L1.ARS_USD | 0.000435 | 0.003607 | 0.121 | 0.904 |
| L1.Arg_lending_borrowing | 14.971026 | 52.851003 | 0.283 | 0.777 |
| L1.CPI_Arg | 0.706775 | 2.441894 | 0.289 | 0.772 |
| L1.Infl_exp_Arg | 14.831496 | 9.122511 | 1.626 | 0.104 |
| L1.Arg_long_rate | -0.026746 | 0.021883 | -1.222 | 0.222 |
| L1.Arg_short_rate | 0.110703 | 0.039445 | 2.807 | 0.005 |
| L1.M2_Arg | 0.143197 | 0.198073 | 0.723 | 0.470 |
| L2.US_inflation_exp | 0.171485 | 0.921041 | 0.186 | 0.852 |
| L2.USD_long_rate | -35.632078 | 24.035323 | -1.482 | 0.138 |
| L2.USD_short_rate | -6.501345 | 30.430694 | -0.214 | 0.831 |
| L2.M2_USD | 0.012881 | 0.044660 | 0.288 | 0.773 |
| L2.CPI_US | -0.357367 | 2.962170 | -0.121 | 0.904 |
| L2.ARS_USD | -0.001449 | 0.003599 | -0.403 | 0.687 |
| L2.Arg_lending_borrowing | -15.105506 | 52.882973 | -0.286 | 0.775 |
| L2.CPI_Arg | -0.635588 | 2.461618 | -0.258 | 0.796 |
| L2.Infl_exp_Arg | -14.214891 | 9.144998 | -1.554 | 0.120 |
| L2.Arg_long_rate | -0.000508 | 0.021600 | -0.024 | 0.981 |
| L2.Arg_short_rate | 0.099665 | 0.039562 | 2.519 | 0.012 |
| L2.M2_Arg | -0.068003 | 0.199546 | -0.341 | 0.733 |

==========================================================================================

Results for equation Arg_short_rate
===============================================================================

|  | coefficient | std. error | t-stat | prob |
|---|---|---|---|---|
| const | 0.014375 | 0.030568 | 0.470 | 0.638 |
| L1.US_inflation_exp | 0.608408 | 0.484305 | 1.256 | 0.209 |
| L1.USD_long_rate | -1.522169 | 12.640513 | -0.120 | 0.904 |
| L1.USD_short_rate | 2.712079 | 15.918990 | 0.170 | 0.865 |
| L1.M2_USD | 0.002909 | 0.023524 | 0.124 | 0.902 |
| L1.CPI_US | -0.928230 | 1.557994 | -0.596 | 0.551 |
| L1.ARS_USD | -0.000307 | 0.001893 | -0.162 | 0.871 |
| L1.Arg_lending_borrowing | -1.594462 | 27.730644 | -0.057 | 0.954 |
| L1.CPI_Arg | -0.295744 | 1.281249 | -0.231 | 0.817 |
| L1.Infl_exp_Arg | -1.118786 | 4.786534 | -0.234 | 0.815 |
| L1.Arg_long_rate | -0.001386 | 0.011482 | -0.121 | 0.904 |
| L1.Arg_short_rate | 0.264677 | 0.020697 | 12.788 | 0.000 |
| L1.M2_Arg | 0.226561 | 0.103928 | 2.180 | 0.029 |
| L2.US_inflation_exp | -0.467720 | 0.483266 | -0.968 | 0.333 |
| L2.USD_long_rate | 6.598419 | 12.611208 | 0.523 | 0.601 |
| L2.USD_short_rate | 0.464582 | 15.966825 | 0.029 | 0.977 |
| L2.M2_USD | -0.003691 | 0.023433 | -0.158 | 0.875 |
| L2.CPI_US | 0.913603 | 1.554235 | 0.588 | 0.557 |
| L2.ARS_USD | -0.023758 | 0.001888 | -12.581 | 0.000 |
| L2.Arg_lending_borrowing | 1.267327 | 27.747418 | 0.046 | 0.964 |
| L2.CPI_Arg | 0.055614 | 1.291598 | 0.043 | 0.966 |
| L2.Infl_exp_Arg | 3.199584 | 4.798333 | 0.667 | 0.505 |
| L2.Arg_long_rate | -0.006714 | 0.011333 | -0.592 | 0.554 |
| L2.Arg_short_rate | 0.073626 | 0.020758 | 3.547 | 0.000 |
| L2.M2_Arg | -0.207554 | 0.104701 | -1.982 | 0.047 |

===============================================================================

Results for equation M2_Arg
===============================================================================

|  | coefficient | std. error | t-stat | prob |
|---|---|---|---|---|
| const | 0.008405 | 0.006347 | 1.324 | 0.185 |
| L1.US_inflation_exp | 0.000537 | 0.100557 | 0.005 | 0.996 |
| L1.USD_long_rate | 2.744285 | 2.624567 | 1.046 | 0.296 |
| L1.USD_short_rate | -0.913896 | 3.305282 | -0.276 | 0.782 |
| L1.M2_USD | 0.003337 | 0.004884 | 0.683 | 0.494 |
| L1.CPI_US | -0.154891 | 0.323489 | -0.479 | 0.632 |
| L1.ARS_USD | 0.000072 | 0.000393 | 0.183 | 0.855 |
| L1.Arg_lending_borrowing | 0.197658 | 5.757752 | 0.034 | 0.973 |
| L1.CPI_Arg | 0.059409 | 0.266028 | 0.223 | 0.823 |
| L1.Infl_exp_Arg | 0.153876 | 0.993835 | 0.155 | 0.877 |
| L1.Arg_long_rate | 0.000642 | 0.002384 | 0.269 | 0.788 |
| L1.Arg_short_rate | -0.000894 | 0.004297 | -0.208 | 0.835 |
| L1.M2_Arg | 0.850589 | 0.021579 | 39.418 | 0.000 |
| L2.US_inflation_exp | -0.002917 | 0.100341 | -0.029 | 0.977 |
| L2.USD_long_rate | -2.698547 | 2.618482 | -1.031 | 0.303 |
| L2.USD_short_rate | 1.251207 | 3.315214 | 0.377 | 0.706 |
| L2.M2_USD | -0.003445 | 0.004865 | -0.708 | 0.479 |
| L2.CPI_US | 0.140564 | 0.322708 | 0.436 | 0.663 |
| L2.ARS_USD | -0.000230 | 0.000392 | -0.587 | 0.557 |
| L2.Arg_lending_borrowing | -0.272864 | 5.761235 | -0.047 | 0.962 |
| L2.CPI_Arg | 0.120328 | 0.268176 | 0.449 | 0.654 |
| L2.Infl_exp_Arg | -0.371734 | 0.996284 | -0.373 | 0.709 |
| L2.Arg_long_rate | 0.001239 | 0.002353 | 0.527 | 0.599 |
| L2.Arg_short_rate | -0.004985 | 0.004310 | -1.157 | 0.247 |
| L2.M2_Arg | 0.100017 | 0.021739 | 4.601 | 0.000 |

=========================================================================================

Correlation matrix of residuals

| | US_inflation_exp | USD_long_rate | USD_short_rate | M2_USD | CPI_US | ARS_USD | Arg_lending_borrowing | CPI_Arg | Infl_exp_Arg | Arg_long_rate | Arg_short_rate | M2_Arg |
|---|---|---|---|---|---|---|---|---|---|---|---|---|
| US_inflation_exp | 1.000000 | 0.015165 | 0.064099 | -0.013507 | 0.028460 | 0.056527 | 0.010010 | -0.020662 | -0.000013 | -0.002687 | -0.003647 | -0.022410 |
| USD_long_rate | 0.015165 | 1.000000 | 0.142713 | -0.079765 | 0.200065 | -0.027347 | -0.072901 | -0.014264 | -0.290600 | -0.027215 | 0.001975 | -0.011091 |
| USD_short_rate | 0.064099 | 0.142713 | 1.000000 | -0.199085 | 0.243986 | -0.012042 | 0.055266 | 0.002376 | 0.004433 | -0.002940 | -0.006449 | 0.003878 |
| M2_USD | -0.013507 | -0.079765 | -0.199085 | 1.000000 | -0.079060 | -0.018171 | -0.003045 | -0.009889 | 0.133407 | 0.012565 | 0.000427 | 0.031210 |
| CPI_US | 0.028460 | 0.200065 | 0.243986 | -0.079060 | 1.000000 | -0.008058 | -0.060861 | 0.004141 | -0.040599 | -0.019735 | -0.006932 | -0.052472 |
| ARS_USD | 0.056527 | -0.027347 | -0.012042 | -0.018171 | -0.008058 | 1.000000 | -0.004281 | -0.003688 | 0.026924 | -0.076911 | -0.015410 | 0.091873 |
| Arg_lending_borrowing | 0.010010 | -0.072901 | 0.055266 | -0.003045 | -0.060861 | -0.004281 | 1.000000 | -0.001583 | -0.005219 | 0.001122 | -0.000949 | 0.017408 |
| CPI_Arg | -0.020662 | -0.014264 | 0.002376 | -0.009889 | 0.004141 | -0.003688 | -0.001583 | 1.000000 | 0.007786 | 0.157836 | -0.013574 | -0.017708 |
| Infl_exp_Arg | -0.000013 | -0.290600 | 0.004433 | 0.133407 | -0.040599 | 0.026924 | -0.005219 | 0.007786 | 1.000000 | -0.004135 | 0.001108 | 0.011860 |
| Arg_long_rate | -0.002687 | -0.027215 | -0.002940 | 0.012565 | -0.019735 | -0.076911 | 0.001122 | 0.157836 | -0.004135 | 1.000000 | -0.002647 | -0.092931 |
| Arg_short_rate | -0.003647 | 0.001975 | -0.006449 | 0.000427 | -0.006932 | -0.015410 | -0.000949 | -0.013574 | 0.001108 | -0.002647 | 1.000000 | 0.052688 |
| M2_Arg | -0.022410 | -0.011091 | 0.003878 | 0.031210 | -0.052472 | 0.091873 | 0.017408 | -0.017708 | 0.011860 | -0.092931 | 0.052688 | 1.000000 |

Estos datos se pueden seguir elaborando, retocando, realizando más pruebas sobre parámetros específicos, pero no lo veremos. Traduciendo a español los números, puntualmente sobre el IPC de Argentina:

1) La información obtenida del IPC dice que el long interest (Lag 1) es negativo y grande (p<0,001), sugiriendo que un aumento en el long interest conduce a una disminución en el IPC argentino. Desde un punto de vista intuitivo, un aumento del long interest incrementa los costos de pedir préstamos para las familias y las empresas, lo que puede reducir el gasto y la inversión, que trae consigo una disminución en la demanda de bienes y servicios; es decir, menor presión alcista sobre los precios, lo que podría frenar la inflación (reflejada en el IPC). Sería útil ver si esto se sostiene, aunque sea más atenuada la relación, en contextos hiperinflacionarios.
2) La relación con la M2 (Lag 1 y 2) argentina es positiva y significativa (p<0,01), sugiriendo que un incremento en la oferta de dinero está asociado con un aumento en el IPC. Demostrando lo que se puede observar gráficamente.
3) Lo mismo se puede mencionar sobre el short interest (Lag 2).

Sobre las expectativas de inflación de Argentina,

1) Un coeficiente negativo y significativo de las variables L1.Arg_long_rate y L2.Arg_long_rate sugiere que un aumento en las tasas de interés a largo plazo de Argentina en el pasado, tanto en lag inmediato como en lag de dos períodos, está asociado con una disminución en las expectativas de inflación en el presente. En otras palabras, las tasas de interés más altas a largo plazo podrían estar actuando como una señal de que los mercados anticipan que la inflación futura será más controlada.
2) En la M2 de Argentina (Lag 1) se aprecia una relación significativa y negativa, sugiriendo una relación inversa entre la oferta de dinero y las expectativas de inflación. En Lag 2 se observa una relación positiva y significativa sugiriendo un efecto rezago positivo de la oferta de dinero sobre las expectativas (nuevamente, demostrando la hipótesis).

Sobre el long interest de Argentina,

1) La mayoría de predictores p tienen valores altos, sugiriendo una importancia estadística baja.
2) En Lag 1, el long interest de Estados Unidos, aunque no del todo estadísticamente significante, tiene un coeficiente positivo grande, indicando una potencial relación entre las tasas de dicho país y Argentina, mostrando las conexiones entre el dólar y el peso.

En conclusión, a pesar de usar un modelo simple y poco refinado, podemos ver que es coherente con la teoría económica, y demuestra que las ideas y objetivos de esta investigación están bien encaminadas. Utilizando esta información, podemos desarrollar ecuaciones que nos permitan desarrollar al mundo práctico la categoría $\varepsilon$ y sus morfismos.

## Planteamiento teórico del modelo práctico

Ahora la pregunta es: ¿Cómo podemos aplicar los conceptos? Pues bien, hay muchas respuestas. Cuando intentamos resolver problemas muy generales en álgebra, recurrimos al álgebra abstracta, ya que la misma nos permite concebir una idea más allá de simples números o letras, nos permite ver estructuras genéricas, como una categoría; no obstante, a la hora de resolver problemas más específicos, como lo que necesitamos en esta sección, se recurre al álgebra Real. Particularmente, veremos cómo plantear los morfismos, teniendo en cuenta lo mencionado en la sección teórica, pero también teniendo en cuenta lo que vimos empíricamente.

Empecemos por las demandas:

Como hemos mencionado, la demanda depende del ingreso, las tasas de interés y las expectativas de inflación, por lo que, en términos prácticos, tenemos algo como esto:

$$L_{ars} = Y\alpha_1 + i_{ars}\alpha_2 - \pi^{exp}_{arg} a_3$$

$$L_{usd} = Y\beta_1 + i_{usd}\beta_2 - \pi^{exp}_{usa}\beta_3$$

La demanda relativa, más simple, se puede representar como el cociente entre la demanda de pesos y la demanda de dólares.

Para las expectativas de devaluación, utilicé la siguiente fórmula, derivada de la paridad de tasas de interés cubierta, ajustada al EMBI+ Arg:

$$E = \pi^{exp}_{arg} - \pi^{exp}_{usa} + i^{short}_{arg} - i^{short}_{usa} - \frac{EMBI+ ARG}{100}$$

Para trabajar con pronósticos de inflación, de tal modo que podamos ver cómo el modelo se desempeña en relación a esta, podemos usar la siguiente fórmula:

$$\pi = \pi^{exp}\gamma_1 + M_{ars/usd}\gamma_2$$

Sobre el ingreso:

$$Y = M_{ars}\delta_1 + lending/borrowing\,\delta_2$$

Como he mostrado a la hora de recolectar la información, usaremos M = M2 (oferta de dinero de Argentina o Estados Unidos siendo igual a la M2 de dicho país).

Estos son nuestros objetos de la categoría, definidos a través de las mencionadas fórmulas, consistentes con la teoría económica y con lo obtenido empíricamente. Los morfismos ya los hemos mencionado antes, así también como los diagramas conmutativos que muestran el flujo entre morfismos.

## Aplicación del modelo

Finalmente, llegó el momento de ver qué podemos hacer con la categoría. Evaluaremos las ecuaciones usando todos los datos que tenemos, para el segmento enero 2018 - diciembre 2023. Es importante que nos familiaricemos con los siguientes parámetros y sus abreviaciones, ya que los trataremos bastante de aquí en adelante:

| | A | B | C | D | E | F | G | H | I | J | K | L | M | N | O | P |
|---|---|---|---|---|---|---|---|---|---|---|---|---|---|---|---|---|
| 1 | Date | Usa Pi Exp | Term Usd | Term Usd | M2 Usd | Ipc Usa | orical Ars | et Lending | c Argentin | Pi Exp | ong Intere | ort Intere | M2 | dp_argenti | Gdp_usa | E |
| 2 | 2018-01-01 00:00:00 | 1.9 | 2.58 | 1.41 | 13861.8 | 0.544775 | 19.087 | -5.441 | 1.757354 | 1.658333 | 40.31 | 28 | 32.7 | 5.25E+11 | 2.07E+13 | 22.73833 |
| 3 | 2018-01-02 00:00:00 | 1.9 | 2.589032 | 1.410323 | 13863.01 | 0.541829 | 18.591 | -5.43827 | 1.757354 | 1.658333 | 40.31 | 28 | 32.7 | 5.25E+11 | 2.07E+13 | 22.71694 |
| 4 | 2018-01-03 00:00:00 | 1.87 | 2.598065 | 1.410645 | 13864.23 | 0.538884 | 18.3905 | -5.43554 | 1.757354 | 1.658333 | 40.75 | 28 | 32.6 | 5.24E+11 | 2.07E+13 | 22.74173 |
| 5 | 2018-01-04 00:00:00 | 1.92 | 2.607097 | 1.410968 | 13865.44 | 0.535939 | 18.4435 | -5.43281 | 1.757354 | 1.658333 | 40.62 | 28 | 32.6 | 5.24E+11 | 2.07E+13 | 22.6963 |
| 6 | 2018-01-05 00:00:00 | 1.95 | 2.616129 | 1.41129 | 13866.65 | 0.532993 | 18.614 | -5.43008 | 1.757354 | 1.658333 | 40.99 | 28 | 32.6 | 5.24E+11 | 2.07E+13 | 22.65895 |
| 7 | 2018-01-06 00:00:00 | 1.936667 | 2.625161 | 1.411613 | 13867.86 | 0.530048 | 18.60167 | -5.42736 | 1.757354 | 1.658333 | 41.02333 | 28 | 32.6 | 5.24E+11 | 2.07E+13 | 22.67005 |
| 8 | 2018-01-07 00:00:00 | 1.923333 | 2.634194 | 1.411935 | 13869.08 | 0.527103 | 18.58933 | -5.42463 | 1.757354 | 1.658333 | 41.05667 | 28 | 32.6 | 5.24E+11 | 2.07E+13 | 22.68734 |
| 9 | 2018-01-08 00:00:00 | 1.91 | 2.643226 | 1.412258 | 13870.29 | 0.524157 | 18.577 | -5.4219 | 1.757354 | 1.658333 | 41.09 | 28 | 32.5 | 5.23E+11 | 2.07E+13 | 22.71439 |
| 10 | 2018-01-09 00:00:00 | 1.95 | 2.652258 | 1.412581 | 13871.5 | 0.521212 | 19.035 | -5.41917 | 1.757354 | 1.658333 | 41.36 | 28 | 32.4 | 5.23E+11 | 2.07E+13 | 22.68812 |
| 11 | 2018-01-10 00:00:00 | 1.94 | 2.66129 | 1.412903 | 13872.72 | 0.518266 | 18.938 | -5.41644 | 1.757354 | 1.658333 | 40.85 | 28 | 32.4 | 5.23E+11 | 2.07E+13 | 22.70864 |
| 12 | 2018-01-11 00:00:00 | 1.89 | 2.670323 | 1.413226 | 13873.93 | 0.515321 | 18.633 | -5.41371 | 1.757354 | 1.658333 | 40.96 | 28 | 32.4 | 5.23E+11 | 2.07E+13 | 22.76419 |
| 13 | 2018-01-12 00:00:00 | 1.92 | 2.679355 | 1.413548 | 13875.14 | 0.512376 | 18.699 | -5.41098 | 1.757354 | 1.658333 | 41.73 | 28 | 32.3 | 5.22E+11 | 2.07E+13 | 22.70578 |
| 14 | 2018-01-13 00:00:00 | 1.923333 | 2.688387 | 1.413871 | 13876.35 | 0.50943 | 18.66333 | -5.40825 | 1.757354 | 1.658333 | 41.46333 | 28 | 32.3 | 5.22E+11 | 2.07E+13 | 22.68258 |
| 15 | 2018-01-14 00:00:00 | 1.926667 | 2.697419 | 1.414194 | 13877.57 | 0.506485 | 18.62767 | -5.40553 | 1.757354 | 1.658333 | 41.19667 | 28 | 32.2 | 5.22E+11 | 2.07E+13 | 22.66488 |
| 16 | 2018-01-15 00:00:00 | 1.93 | 2.706452 | 1.414516 | 13878.78 | 0.50354 | 18.592 | -5.4028 | 1.757354 | 1.658333 | 40.93 | 28 | 32.2 | 5.22E+11 | 2.07E+13 | 22.64718 |
| 17 | 2018-01-16 00:00:00 | 1.94 | 2.715484 | 1.414839 | 13879.99 | 0.500594 | 18.732 | -5.40007 | 1.757354 | 1.658333 | 41.25 | 28 | 32.1 | 5.22E+11 | 2.07E+13 | 22.62281 |
| 18 | 2018-01-17 00:00:00 | 1.9 | 2.724516 | 1.415161 | 13881.21 | 0.497649 | 18.8805 | -5.39734 | 1.757354 | 1.658333 | 40.49 | 28 | 32.1 | 5.21E+11 | 2.07E+13 | 22.64371 |
| 19 | 2018-01-18 00:00:00 | 1.99 | 2.733548 | 1.415484 | 13882.42 | 0.494704 | 18.8655 | -5.39461 | 1.757354 | 1.658333 | 39.85 | 28 | 32 | 5.21E+11 | 2.07E+13 | 22.5253 |
| 20 | 2018-01-19 00:00:00 | 2 | 2.742581 | 1.415806 | 13883.63 | 0.491758 | 18.875 | -5.39188 | 1.757354 | 1.658333 | 40.94 | 28 | 31.8 | 5.21E+11 | 2.07E+13 | 22.51894 |
| 21 | 2018-01-20 00:00:00 | 2 | 2.751613 | 1.416129 | 13884.85 | 0.488813 | 18.8915 | -5.38915 | 1.757354 | 1.658333 | 40.80667 | 28 | 31.6 | 5.21E+11 | 2.07E+13 | 22.50487 |
| 22 | 2018-01-21 00:00:00 | 2 | 2.760645 | 1.416452 | 13886.06 | 0.485868 | 18.908 | -5.38642 | 1.757354 | 1.658333 | 40.67333 | 28 | 31.5 | 5.21E+11 | 2.07E+13 | 22.48051 |
| 23 | 2018-01-22 00:00:00 | 2 | 2.769677 | 1.416774 | 13887.27 | 0.482922 | 18.9245 | -5.3837 | 1.757354 | 1.658333 | 40.54 | 28 | 31.4 | 5.2E+11 | 2.07E+13 | 22.45912 |
| 24 | 2018-01-23 00:00:00 | 1.92 | 2.77871 | 1.417097 | 13888.48 | 0.479977 | 19.1335 | -5.38097 | 1.757354 | 1.658333 | 41.04 | 28 | 31.2 | 5.2E+11 | 2.07E+13 | 22.55375 |

Es fácil deducir qué es cada cosa, lo que no aparece con "usa" o "usd", son los parámetros para Argentina. También aparece el EMBI+ Arg (aunque no se alcanza a ver en la imagen). Note que fueron normalizados por interpolación lineal, y si bien no aparecen así en el Excel, también tratamos con los parámetros no estacionarios, utilizando diferenciación.

Además, también se incluye en el mismo el archivo "category_forecast_results", que contiene los mismos parámetros, pero acorde a los resultados del modelo. A partir de estos, podemos ver cómo se comportó el modelo en relación a los datos reales. A continuación ilustraciones:

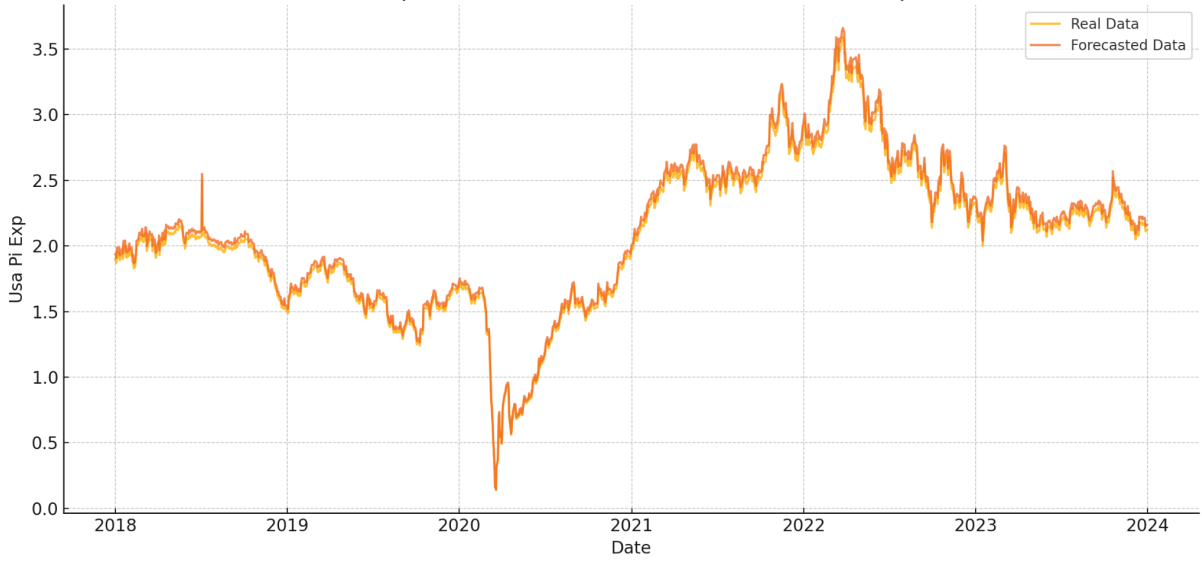
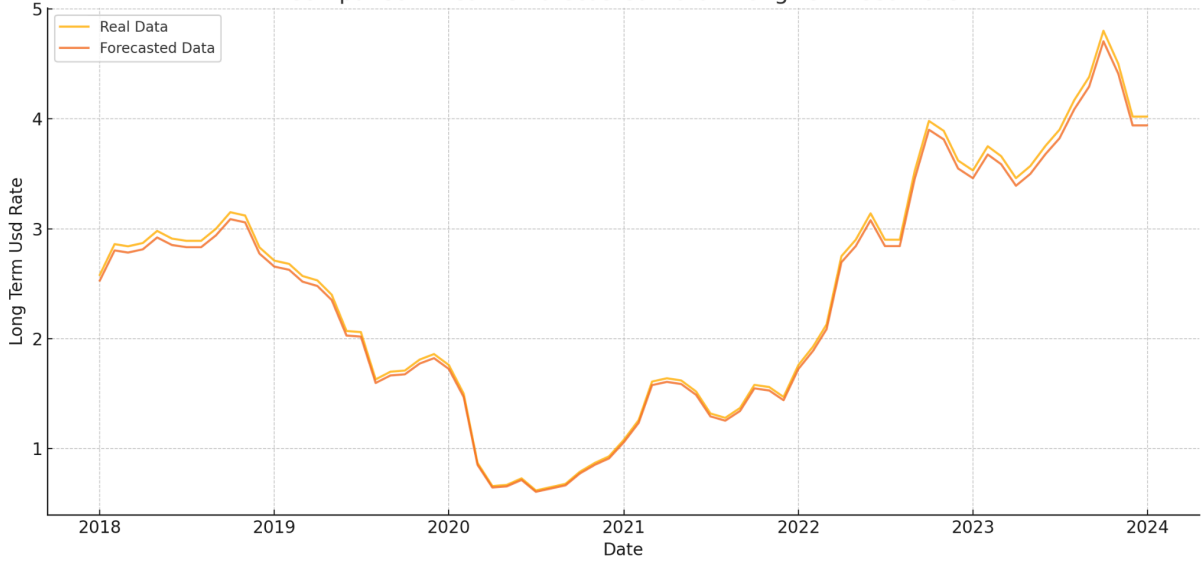

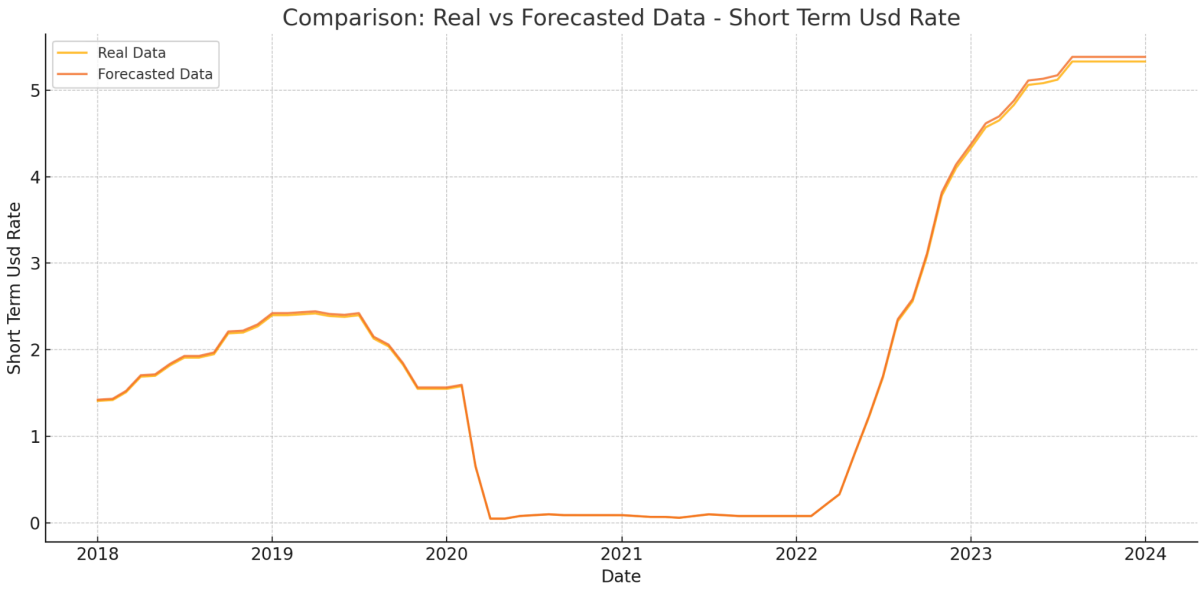

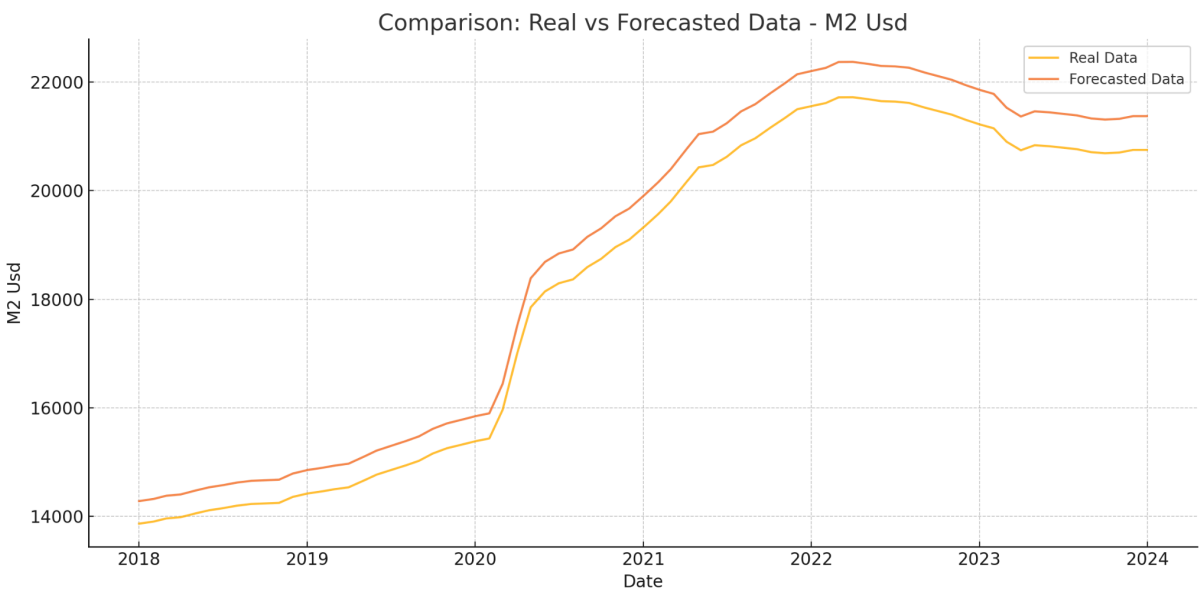

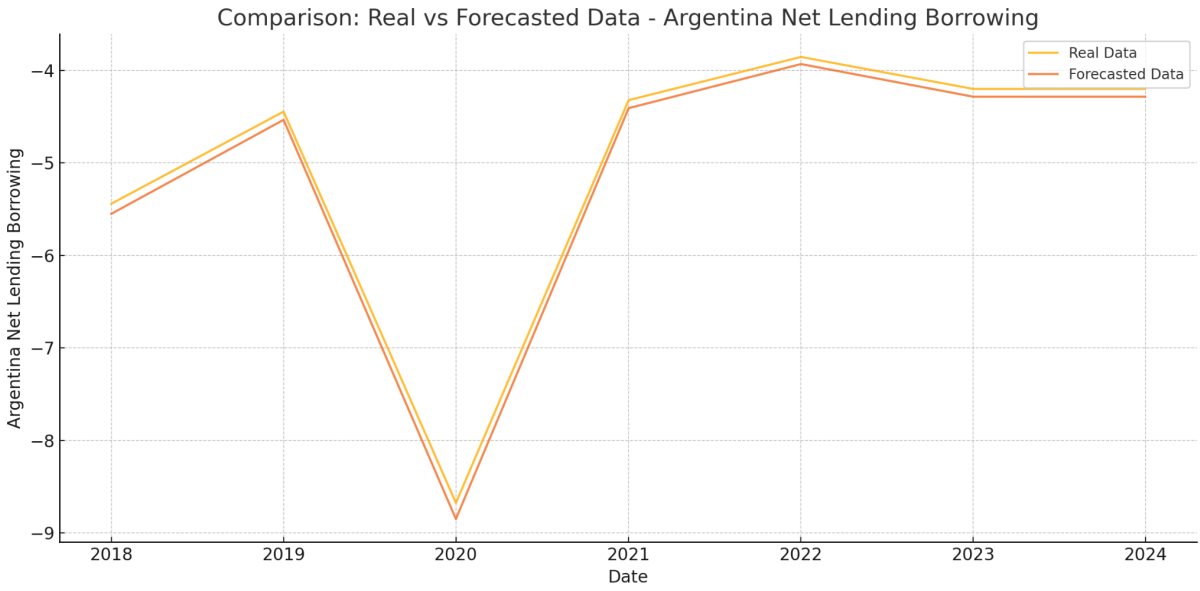

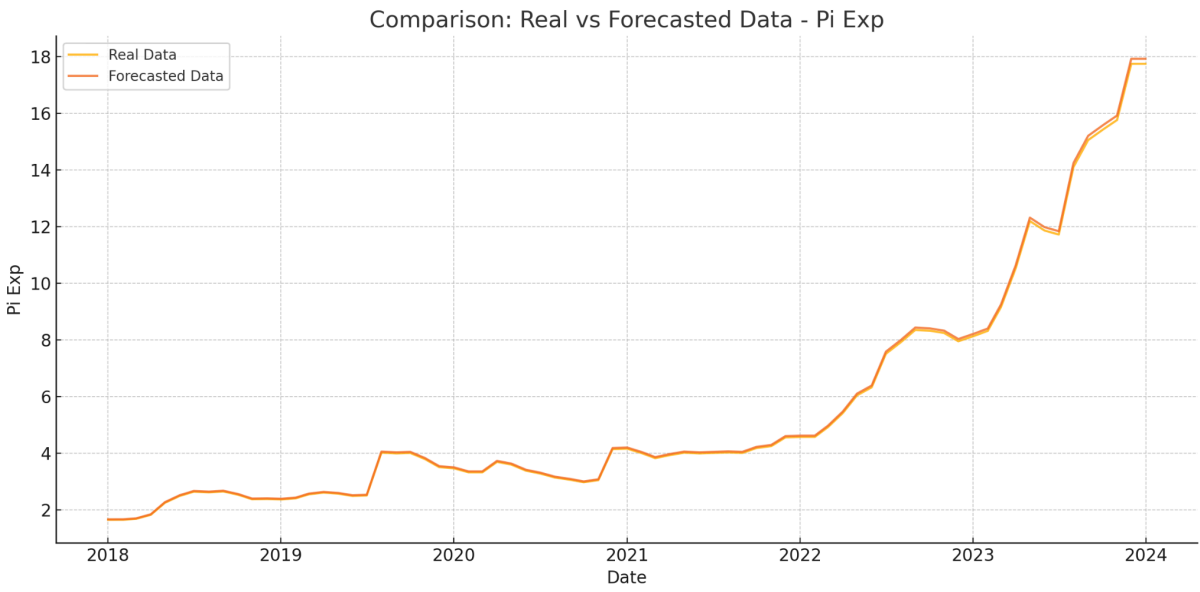

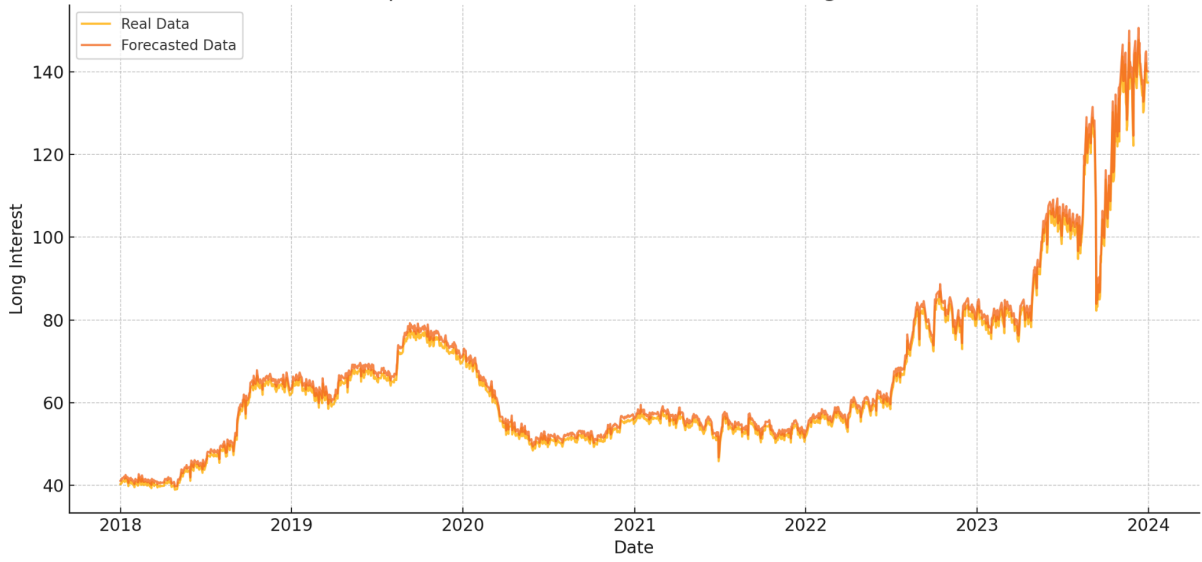

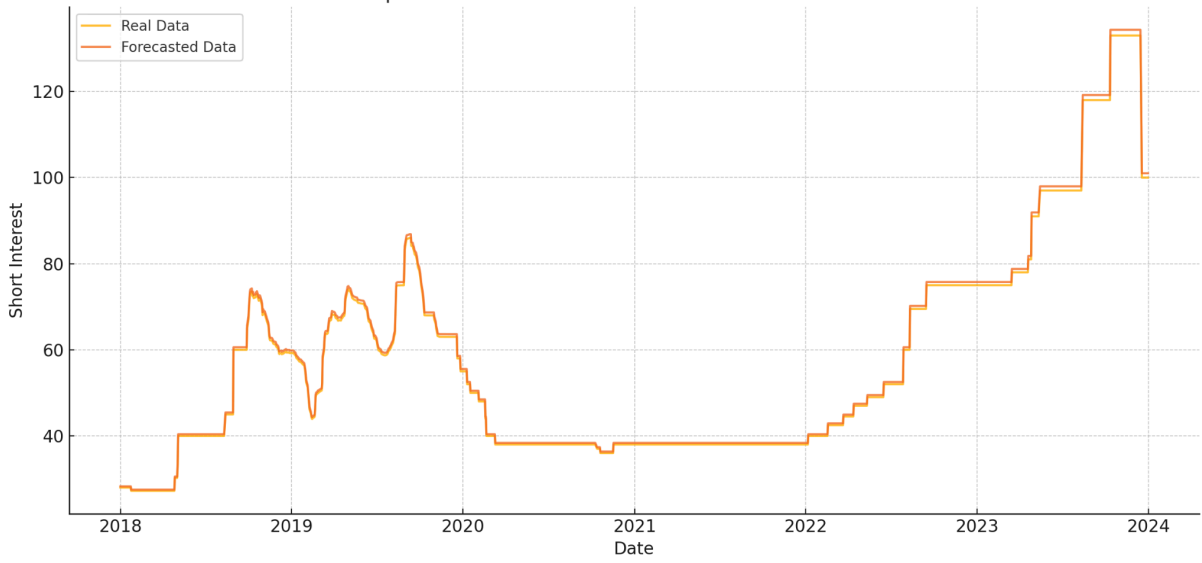

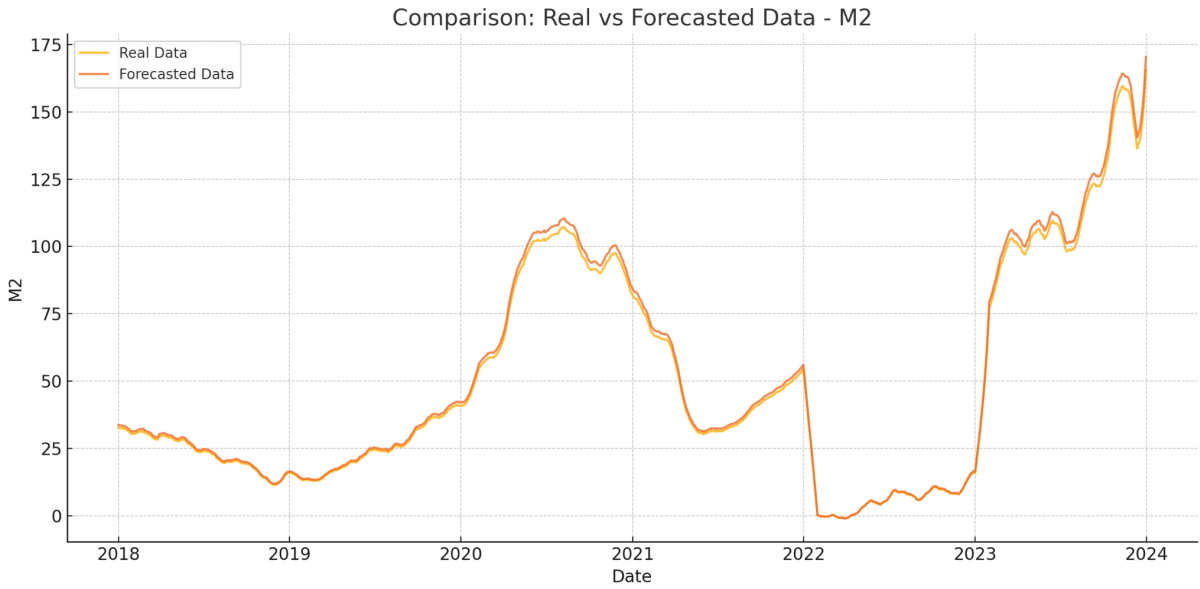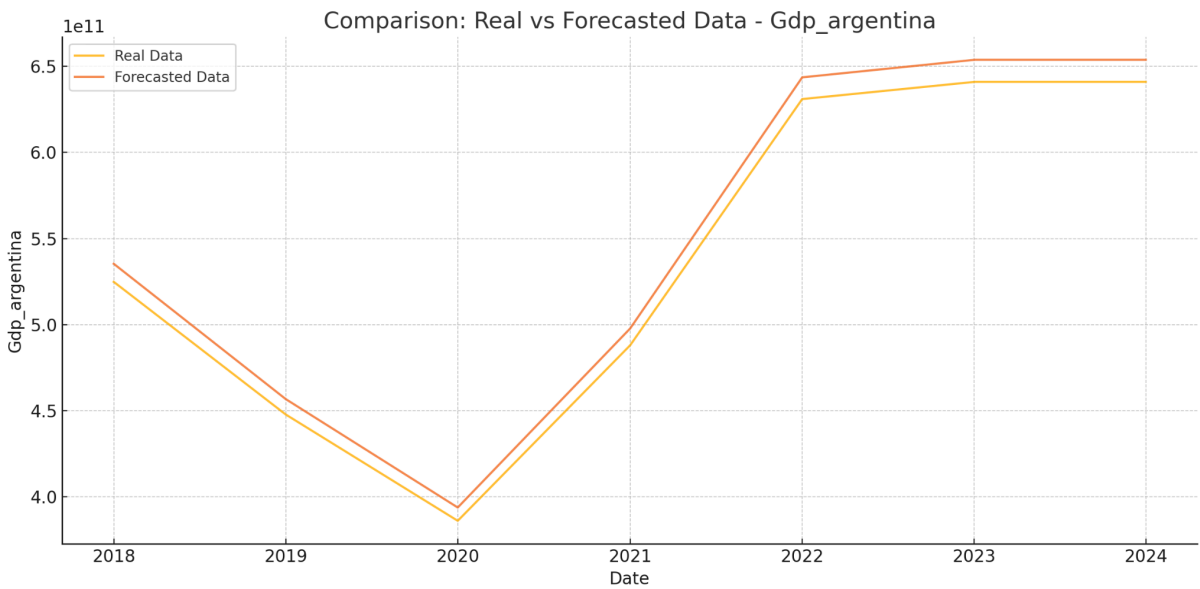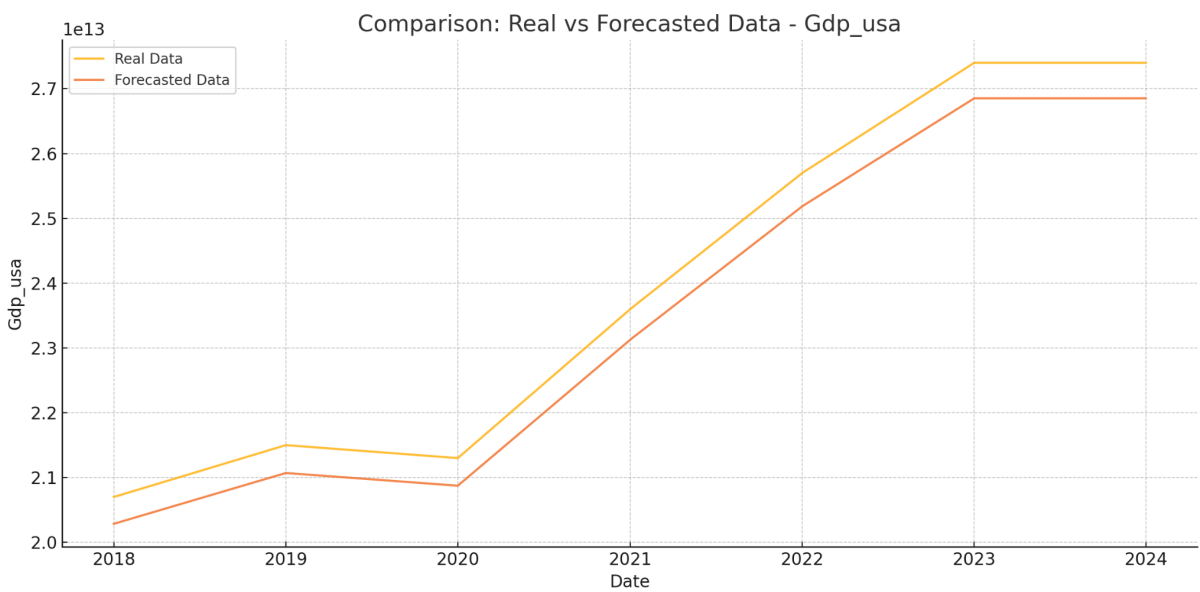

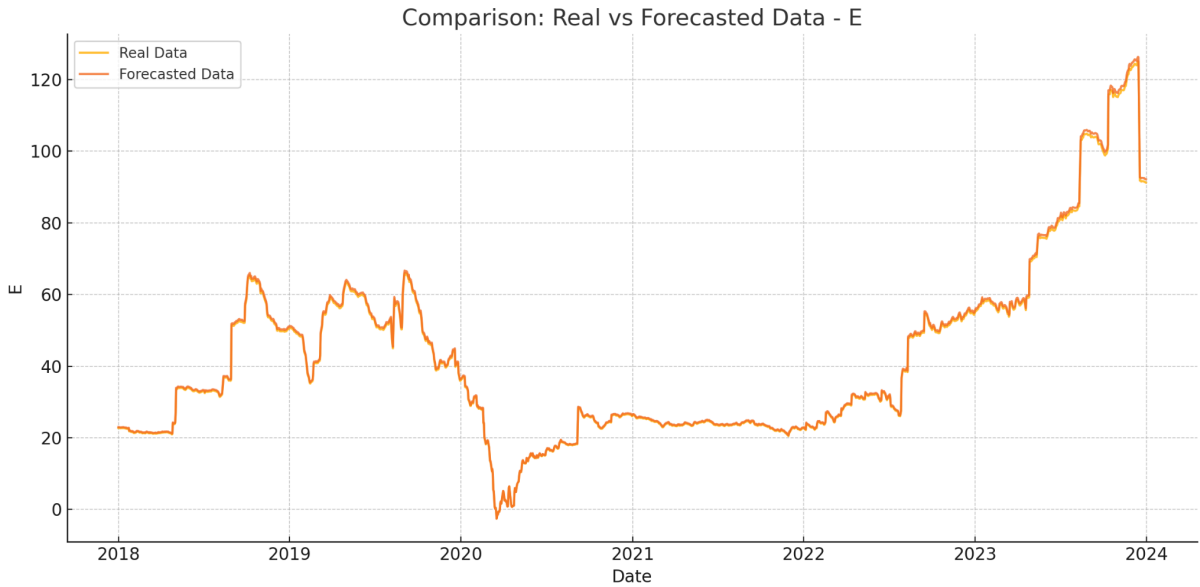

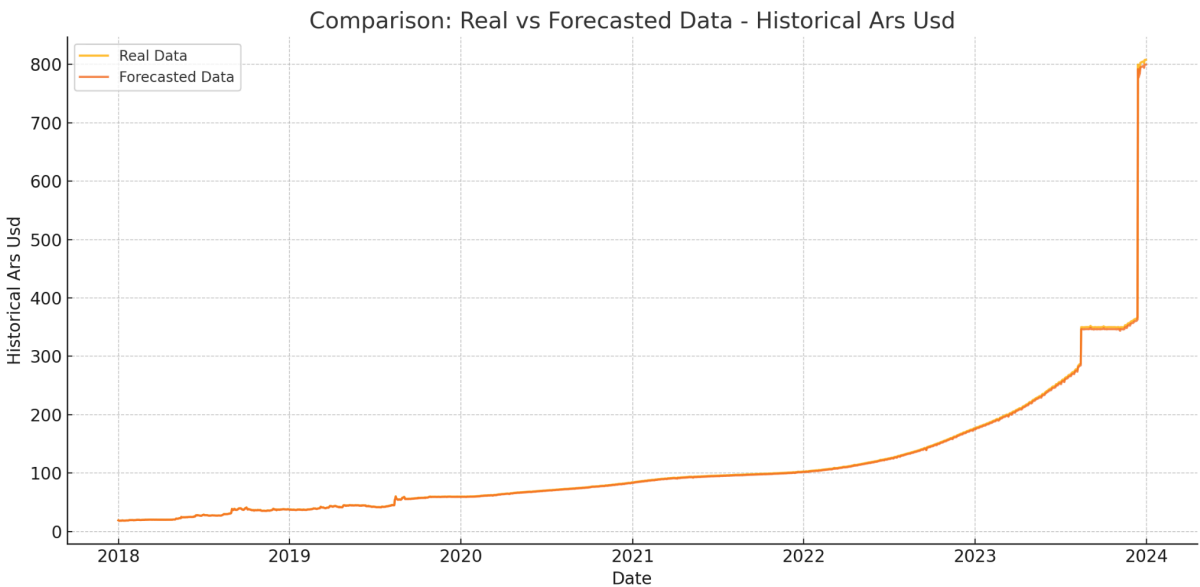

## Conclusiones

En caso de que se lo pregunte, los resultados del modelo, en todos los parámetros, fueron calculados con la teoría que hemos desarrollado hasta ahora. Si despeja en las fórmulas y morfismos, puede obtener las formas de las expectativas de inflación, short, long, etc…
Como se puede apreciar, el modelo funciona, se pueden apreciar fuertes correlaciones entre los datos reales y los estimados por las fórmulas. Se producen algunas discrepancias en periodos de mayor inestabilidad, como en el 2023, en variables como E, M2 y las tasas de interés; no obstante, el modelo sirve para predecir tendencias, sobre todo en períodos de mayor estabilidad económica, aunque también se comporta relativamente bien en períodos de inestabilidad. Como podría esperarse, la demanda relativa predecida por el modelo muestra comportamientos similares a la relación histórica de ars-usd.
Para hacer más preciso el modelo práctico, se podrían optimizar los coeficientes, utilizando técnicas de regresión o machine learning; podríamos incorporar más objetos, como el riesgo político (que tiene fuertes influencias en la economía argentina); podríamos cambiar las fórmulas por algunas no lineales, de tal modo que podamos apreciar mejor aquellos morfismos que no son puramente lineales. En general, complejizar más las fórmulas que usamos como representación del modelo ayudarán a aumentar la precisión; sin embargo, así como está, funciona acorde a la teoría económica.

El uso de mejores y más potentes herramientas computacionales permitirá identificar automáticamente nuevos objetos y morfismos pertenecientes a la categoría que no han sido incluidos, así también como construir redes reales que simulen los diagramas conmutativos, al estilo machine learning.

A través del siguiente link se puede acceder a un esquema simplificado de lo que hemos estado viendo hasta ahora. Está basado en teoría de grafos, donde los nodos representan los objetos y los bordes los morfismos (esto es un ejemplo de cómo utilizar otra estructura algebraica distinta a los tradicionales conjuntos). Si desplaza el mouse sobre los nodos y bordes podrá acceder a cierta información descriptiva:

https://lucho1787.github.io/representaci-n-3d-de-Categor-a-bimonetaria/

Dicho link es una página web, hospedada en un servidor llamado GitHub (que se utiliza en el mundo de la programación), estará disponible para siempre, en caso de que le interese.

Como **hallazgos**: Al modelar la economía bimonetaria argentina mediante teoría de categorías, encontramos que las relaciones económicas pueden representarse eficazmente como morfismos entre objetos, lo que permite estructurar de manera más precisa la dinámica de variables como el tipo de cambio, la inflación y la oferta monetaria.

Además, el análisis de diagramas conmutativos mostró que ciertas interacciones económicas poseen una estructura inherentemente estable, mientras que otras dependen fuertemente de condiciones externas, como las expectativas de los agentes. También se identificó que algunos de estos diagramas pueden no conmutar bajo ciertas políticas monetarias, lo que sugiere que la intervención del banco central puede alterar relaciones fundamentales dentro del sistema.

Este enfoque proporcionó una base teórica sólida para avanzar en la modelización de cambios en la economía a través de functores en la siguiente parte.

## ¿Por qué teoría de categorías?

A diferencia de los modelos tradicionales, las categorías permiten dos cosas, a gran escala: Generalizar y flexibilizar el manejo de objetos o elementos, así también como el universo de relaciones en el que ellos se mueven.

La economía es un sistema dinámico, cambia constantemente, se expande, se contrae, se modifica, lo cual supone que a la hora de intentar entenderla no podamos limitarnos únicamente a un plano unidimensional (como lo sería un sistema lineal), sino que debemos utilizar un sistema enedimensional flexible, donde cada dimensión interactúa y genera morfismos en las otras, y tal cosa puede entenderse como una categoría. Si bien el poder predictivo de las categorías se ve limitado por el nivel de las fórmulas o ecuaciones que utilizamos para modelar los objetos, en el caso de que decidamos usar como base teoría de conjuntos, el poder interpretativo de las mismas es superior al de modelos tradicionales, ya que permite visualizar rápidamente cambios y relaciones.

La teoría de categorías, debido a su extremo nivel de generalidad, tiene el potencial suficiente para explicar en última instancia cómo funciona el universo, de principio a fin, en cada nivel. Es como un machine learning matemático. Mezclar una IA poderosa con teoría de categorías probablemente nos permitirá comprender relaciones profundas no descubiertas.

El potencial real de la teoría de categorías proviene de su perspectiva "atómica" de los objetos (en el sentido anacrónico de que no hay nada más dentro), lo que significa que lo que da sentido a una categoría son sus morfismos, no los objetos. Las categorías son la forma definitiva de generalización, lo que facilita el modelado de sistemas dinámicos.

Por poner un ejemplo: supongamos que estamos discutiendo cómo funciona un ecosistema, podemos definir algunos objetos (cuya cantidad aumentará a medida que los descubramos) y sus relaciones o morfismos. En la sabana, podemos ver cómo se comporta un león alrededor de las cebras y cómo se comportan ellas alrededor de las hojas, y cómo las hojas... En última instancia, podemos modelar la categoría. Quizás usando estadística básica y álgebra se pueda modelar cómo y cuándo algo puede pasar, pero sólo la categoría puede darnos relaciones entre objetos, del primero al último, por composiciones, operadores de aprender y olvidar, functores, etc...

Las categorías son parecidas a la cartografía. No se puede hacer un mapa sólo usando aritmética o álgebra simple, ya que todas estas ramas de las matemáticas dan aproximaciones simples, como el eje de un mapa o la grilla. Las observaciones del mundo real son el input que mapeamos, los objetos y morfismos que se aprecian. El mapa es la categoría. Toda ciencia usa este tipo de observación o intuición, que se puede plasmar usando la teoría de categorías. El álgebra básica explicará cómo funciona un morfismo, pero podemos profundizar más en las relaciones, y eso sólo se puede hacer con teoría de categorías.

Combinar teoría de categorías con inteligencia artificial, el siguiente paso de este trabajo, podría abrir caminos para modelar procesos de aprendizaje, adaptabilidad y transformaciones, especialmente a través de conceptos como los funtores (asignaciones que preservan la estructura) y las mónadas (que modelan el estado y los efectos secundarios). En la IA, las mónadas ya aparecen en lenguajes de programación funcional como Haskell, lo que sugiere que la teoría de categorías podría convertirse en una columna vertebral formal para las arquitecturas de IA. Es fascinante pensar cómo los sistemas de IA modelados con categorías podrían evolucionar para no solo calcular sino también "razonar" sobre sus estructuras, adaptándose dinámicamente a nuevas observaciones. Un enfoque de este tipo podría revolucionar campos como la física teórica, donde ya se está trabajando, biología computacional e incluso las neurociencias.

## Drive con toda la información utilizada

[data utilizada durante el modelo de categorías](#)

# Parte II: Incorporación de Functores para Modelar Cambios en la Economía

En esta parte se introducen los conceptos de functores para modelar cambios estructurales en la economía. Se implementan functores de aprendizaje y olvido para representar cómo los agentes económicos ajustan sus expectativas frente a nuevas políticas y choques externos. La metodología incluye experimentos de sensibilidad para evaluar el impacto de variaciones en tasas de interés y política fiscal sobre la estabilidad macroeconómica, observando cómo los morfismos evolucionan bajo distintos escenarios.

## Hipótesis

*"Los functores de olvido y aprendizaje permiten modelar de manera más precisa el impacto de las políticas económicas sobre la economía bimonetaria argentina, facilitando un análisis dinámico y predictivo de los cambios en estabilidad y sensibilidad de las variables clave."*

La motivación de esta hipótesis es que los modelos tradicionales no permiten una fácil incorporación de nueva información o la simplificación de estructuras complejas sin perder coherencia. Por ello, verificaremos si los functores permiten introducir nuevas políticas y medir cómo afectan a la economía sin necesidad de reconstruir todo el modelo desde cero.

## Marco teórico

A partir de aquí, explicar conceptos del álgebra abstracta se vuelve difícil, y no es fructífero discutir cuestiones teóricas o definiciones, por lo que nos concentraremos en la intuición económica. Para no perder rigor, sugiero la lectura de "Category theory in context", de Emily Riehl, aunque no por esto evitaremos citar definiciones.
Un functor es un mapeo entre categorías. Si tenemos las categorías C y D, el functor F asocia cada objeto X de la categoría C con un objeto F(X) en D, así también como los morfismos, de tal modo que se cumplan la propiedad asociativa y de identidad de morfismos. En otras palabras, un functor mapea los objetos y morfismos de una categoría a otra preservando la estructura, he ahí que se tengan que cumplir las propiedades mencionadas.
Si bien pueden parecer muy abstractos, los functores nos permitirán modelar cambios en políticas a corto y largo plazo. Además, permiten sortear uno de los problemas más grandes en las ciencias, las irregularidades en las leyes matemáticas debido a la naturaleza. ¿Usted cree que podemos entender el comportamiento de redes neuronales, que luego se traducen en una interacción de mercado, sólo a través de meras ecuaciones algebraicas tradicionales? ¿Cree que podemos traducir la conducta humana en un par de símbolos? Claramente, es complicado. Tome como ejemplo la ecuación clásica del salario

$$w = \frac{\partial Q}{\partial L} \cdot P$$

Esta ecuación se mantiene, aproximadamente, en el sector privado, donde los empleados se esfuerzan más para aumentar su sueldo; no obstante, cuando nos movemos al sector público, la relación parece invertirse, la productividad pareciera tomar el rol del salario y viceversa. Por supuesto, esto no siempre es así. Que el morfismo se invirtiera sería muy simple, pero en realidad existen empleados privados "vagos" y empleados públicos "laboriosos". Por tanto, si bien la ecuación puede funcionar, de modo aproximado para algunos contextos, en otros deja mucho que desear.
En esta situación entran los functores, especialmente los siguientes:

## Introducción a los functores de aprendizaje y olvido

Como mencionamos, un functor es un mapeo entre categorías que mantiene la estructura, tanto de objetos como de morfismos. Consideremos una categoría donde los objetos sean sistemas económicos y los morfismos representen las intervenciones políticas. Un functor puede mapear este sistema a otra categoría, como un gráfico de relaciones causales entre variables (tal como el proporcionado previamente).
Para un ejemplo más abstracto, pensemos en cómo representar un conjunto, de teoría de conjuntos, como matriz. También podemos pensar en cómo representar un conjunto como espacio topológico. Este tipo de situaciones se abordan con functores; de hecho, la propia definición de functor es hacer esta transformación. Sobre el primer ejemplo en este párrafo, imaginemos que tenemos una serie de ecuaciones que modelan una empresa de nuestra pertenencia. Si usted es hábil con las matemáticas, quizás prefiera utilizar matrices para resolver problemas de optimización, o quizás prefiera calcular derivadas e integrales en álgebra lineal para obtener distintos enfoques en varias dimensiones de sus rentabilidades o pérdidas. Sobre el segundo ejemplo, imaginemos que tenemos datos sobre la inflación, o funciones, de todos los países del mundo, y queremos hacer un mapa unificado (al estilo mapa climatológico), esto es equivalente a querer traducir un conjunto a una estructura topológica, que se resuelve, cuando existe mucha información, usando teoría de haces, o *sheaf theory*, en esencia, un functor.

Un tipo especial de functor se llama functor de olvido (*forgetful functor*), el cual simplifica estructuras complejas al "olvidar" algunas de sus propiedades. Mapea objetos en una categoría rica (con datos o restricciones adicionales) a una categoría más simple que retiene sólo las características fundamentales. Por ejemplo, imaginemos analizar la economía de Argentina enfocándonos únicamente en las expectativas de inflación y la oferta monetaria, ignorando detalles más específicos como los patrones de consumo o las dinámicas salariales. El functor de olvido elimina estos detalles para simplificar el análisis. Nos permite mapear un sistema monetario (con tasas de cambio, tasas de interés y datos de inflación) a únicamente la oferta monetaria (M2), aislando un aspecto clave, así también como podría usarse para transformar un espacio vectorial de indicadores financieros en un conjunto de escalares, reteniendo solo magnitudes y descartando información estructural.

Un functor de aprendizaje es el inverso de un functor de olvido. Enriquece estructuras al introducir restricciones adicionales, relaciones o capas de datos. Supongamos que inicialmente modelamos la dinámica inflacionaria de Argentina usando variables básicas (oferta monetaria y déficit fiscal). Un functor de aprendizaje podría introducir dependencias históricas, cambios en políticas o shocks externos (por ejemplo, tasas de interés en EE.UU.). Más específicamente, este tipo de functor permite mapear una serie temporal escalar de inflación en un espacio vectorial con rezagos, incorporando efectos de memoria y dependencias, transformar un conjunto de variables macroeconómicas en un modelo de red con aristas dirigidas, codificando causalidad, entre otra infinidad de posibilidades.

Estos functores se utilizan de modo dual. Como analogía, suponga que estamos viendo un mapa planisferio, podemos enfocarnos en Europa, haciendo zoom, perdiendo información del resto del mundo, y luego podemos volver a la visión planisférica, recuperando la información. Incluso podríamos añadir, o quitar, información si la Tierra sufriera un cambio, o si quisiéramos expandirnos a una visión del sistema solar.
Como hemos mencionado, la economía es una ciencia dinámica, se actualiza constantemente. Incluso los mismos sistemas que planteamos, los mercados que modelamos, cambian constantemente. No podemos decir "usaremos esta ecuación en el momento $x$ y esta otra en el momento $y$" o, peor aún, creer que una única ecuación puede sostener todo un modelo de conductas. En su lugar, se sugiere un enfoque de categorías dinámico, que permita actualizar inmediatamente la información y su modelización (a través de la estructura matemática que consideremos menester). Bajo este contexto es que se considera fundamental el uso de los functores mencionados.

Pongamos un ejemplo más pragmático. En finanzas, generalmente deseamos construir una visión grande de los distintos mercados y sus distintos assets, de tal forma que, mediante análisis fundamental o análisis técnico, podamos encontrar la o las mejores alternativas de inversión. En este tipo de situaciones existen varias formas de aplicar teoría de categorías, veamos una de particular interés en la actualidad:
Aquellos dedicados al trading algorítmico, frecuentemente deben desarrollar métodos de análisis de sentimiento de mercado. A nivel de *hedge fund*, se cuenta con los recursos suficientes como para poder utilizar una técnica conocida como *data scraping*, consistente en realizar un código (generalmente en Python) que se encargue de buscar palabras o frases claves en distintos sitios sociales, principalmente internet, y procesarlas, arrojando en base a esto señales. Por ejemplo, asuma que elaboramos un algoritmo de data scraping sencillo, que recopile en foros especializados información sobre assets de nuestro interés en futuros. Como los mercados operan casi todo el tiempo, permanentemente estaremos recibiendo los distintos tipos de señales sobre nuestros assets, lo cual puede entenderse como una categoría, en la cual se aporta permanentemente nueva información a través de functores de olvido o de aprendizaje. El uso de esta estrategia permite automatizar muchos procesos en trading algorítmico. Además, es el principal razonamiento detrás de una inteligencia artificial, ya que los bots cuentan con memoria del pasado, algoritmos para entender el presente y modelos que les permiten predecir el futuro cercano. En esencia, una categoría no es más que una inteligencia artificial, es por ello que es idílica para un sistema dinámico como lo es la economía, donde no se puede operar siempre bajo las mismas

"leyes", ya que varían según el contexto, del mismo modo que en el ejemplo la nueva información genera señales distintas.

# Metodología de la investigación

Nuestra categoría cuenta con algunos objetos y morfismos, aunque es extremadamente acotada respecto de la realidad, es suficiente para plasmar las aplicaciones que iremos viendo con el pasar de las secciones.

El uso de los functores que hemos repasado es relativamente intuitivo. Cuando se agrega nueva información, que modifica algo en nuestra estructura mórfica u objetal usamos functores de aprendizaje de tal modo que se incorpore la nueva información, sin alterar la estructura básica, y, cuando queremos disgregar información, o se produce alguna pérdida de la misma, usamos functores de olvido. Ahora, es natural preguntarse exactamente cómo se hace esto, y qué implicancias tiene sobre las estructuras matemáticas menos abstractas sobre las que estemos trabajando, como pueden ser los datos puntuales (representados como conjuntos de información), funciones, estructuras topológicas, vectores, grupos, etc…
Para poder observar de modo más práctico estos conceptos, utilizaremos la categoría con la que veníamos trabajando hasta ahora, pero con unas pequeñas modificaciones:
Definiremos 3 categorías, de la siguiente forma,

1. Categoría de sistemas monetarios (M): Sus objetos son variables y estructuras como la oferta monetaria, expectativas de inflación, déficit fiscal, tasas de interés y tasas de cambio. Sus morfismos son transformaciones entre estas variables, como cambios en las políticas y reacciones de mercado.

2. Categoría de estructuras de demanda ($\Delta$): Sus objetos son variables representando la demanda de pesos y de dólares en la economía bimonetaria. Los morfismos son cambios en la demanda debido a cambios en políticas monetarias o shocks externos.

3. Categoría de feedback histórico (H): Sus objetos son datos y tendencias históricas, como datos pasados de inflación, devaluación, etc… Los morfismos son mecanismos para la incorporación de tendencias pasadas a expectativas del presente (como efectos de aprendizaje).

Un functor de olvido podría definirse como $U: M \to \Delta$, simplificando el sistema monetario descartando parte de su complejidad estructural y enfocándose en outputs fundamentales, como la demanda de pesos o dólares. Matemáticamente, dado un objeto **M** en M, el functor U lo mapea a un objeto **D** en $\Delta$, olvidando detalles de alto nivel como déficits fiscales o riesgos externos. Esto donde M = {M2, Expectativas de inflación, Déficit fiscal, ARS/USD, Tasas de interés}, D = {Demanda de pesos, Demanda de dólares}. Para un morfismo $f: M1 \to M2$ (transformaciones en M), el functor lo mapea a un morfismo $U(f): D1 \to D2$, preservando la estructura esencial.

En nuestro contexto, considere un cambio en las políticas que modifique la M2 y el déficit fiscal. El functor de olvido mapea esta transformación a su impacto a la demanda de pesos/dólares, ignorando variables intermedias como expectativas inflacionarias.

Un functor de aprendizaje podría definirse como $L: \Delta \to M$, enriqueciendo la estructura de demanda simplificada al reintroducir complejidad, como dependencias históricas o shocks externos. Dado un objeto **D** en $\Delta$, el functor L lo mapea de vuelta a un objeto **M'** en M, incorporando feedback histórico y factores externos, donde D = {Demanda de pesos, demanda de dólares}, **M'** = {M2, expectativas de inflación, ARS/USD, tasas de interés, feedback histórico}. Para un morfismo $g: D1 \to D2$, el functor lo mapea a un morfismo $L(g): M'1 \to M'2$, añadiendo efectos de retroalimentación. El functor de olvido incorpora tendencias de inflación pasada y del riesgo EMBI+ARG al análisis, enriqueciendo las relaciones entre las demandas ARS/USD y las expectativas de inflación.

Como se puede ver, estos functores forman una dualidad, en la que se complementan a sí mismos, a esto lo llamamos par adjunto. La adjunción asegura consistencia entre modelos de demanda simplificados y sistemas monetarios enriquecidos, permitiendo transiciones entre estas perspectivas sin perder coherencia. Las definiciones formales de adjunción no son de nuestro interés, por eso no se mencionará.

Como ejercicio, podríamos verificar que las transformaciones se sostienen con los datos reales; es decir, probar que la categoría es consistente (del mismo modo que si hacemos zoom en un mapa, y luego volvemos a la visión original, o intentamos hacer predicciones sobre cómo se ve el mapa original, deberíamos tener mismos o muy símiles resultados), pero no lo realizaremos debido a la trivialidad teórica que supone. Por ejemplo, podríamos probar si predicciones sobre la demanda de pesos o dólares simplificada se alinean con modelos enriquecidos que contengan la inflación histórica y el riesgo EMBI+ARG histórico.

Sugiero además, como ejercicios o proyectos, alguna de las siguientes cuestiones:

1. Análisis de sensibilidad para políticas intervencionistas: Identificando cuáles variables (como M2, déficits fiscales, etc…) afectan en mayor medida a la inflación y a las tasas de interés. Recordando el modelo que planteamos al principio, al hacer pruebas estadísticas básicas, a través del modelo VAR, se pudieron ver ciertas correlaciones, efectos rezagos, entre otros indicadores útiles. Con el conocimiento del cuál disponemos a esta altura, podemos profundizar más todavía, realizando el siguiente paso a las matriz de correlación, un análisis de sensibilidad, que aquellos con mayor conocimiento en matemáticas pueden potenciar al utilizar Teoría de Haces. En la próxima sección nos enfocaremos en este proyecto.

2. Modelado de feedback cíclico dinámico: Modelar mecanismos de feedback entre políticas domésticas y shocks externos usando el functor de aprendizaje para codificar dependencias históricas y efectos de desbordamiento (como el que se mencionó en la parte II) a los modelos de demanda simplificados. También se pueden iterar transformaciones entre categorías para estimar ciclos de feedback.

> Por ejemplo, se puede modelar cómo un incremento en el déficit fiscal puede conducir las expectativas de inflación, afectando tasas de cambio y, de pasada, influenciando la política fiscal vía fugas de capital.

3. Optimización de políticas: Testear varias combinaciones de políticas para optimizar la estabilidad macroeconómica, usando el functor de olvido para aislar efectos colaterales en la demanda de las políticas. También se puede usar el functor de aprendizaje para reintroducir complejidad (data histórica, desbordamientos de riesgo) y simular efectos de largo plazo. Por ejemplo, optimizar reducciones de déficit fiscal y ajustes de oferta monetaria para estabilizar la demanda de pesos sin gatillar ataques especulativos.

La idea de esta lista es que podamos ver el potencial de la teoría de categorías para modelar sistemas dinámicos. A medida que sigamos introduciendo conceptos y propiedades, como límites y colímites, el alcance de estos proyectos puede seguir creciendo y volverse más simplificado, de tal forma que el economista no tenga que recurrir a análisis visuales, reconocimiento de patrones sobre la información, o a modelos algebraicos de funciones anacrónicos (justamente evitar lo que tuvimos que hacer en la parte II), aunque este nivel ya requerirá de mayor esfuerzo computacional.

## Análisis de sensibilidad para políticas intervencionistas

Un análisis de sensibilidad evalúa cómo pequeños cambios en los inputs, como pueden ser las políticas, afectan los outputs, como los indicadores económicos. En nuestro caso, el functor de olvido aísla relaciones base, por ejemplo entre la oferta monetaria y la demanda de pesos; el functor de aprendizaje reintroduce complejidad para modelar feedback y shocks externos. Esta perspectiva dual ayuda a distinguir cuáles variables tienen el mayor impacto y cómo sus efectos se propagan.
Tradicionalmente, estos procedimientos se realizan con modelos algebraicos primitivos como funciones o vectores. A nivel más avanzado, se utilizan herramientas computacionales con software topológico, que permite mapear mejor las relaciones y sus impactos. En el máximo nivel de complejidad, se puede utilizar teoría de categorías y teoría de haces, junto con software que permita el modelado a través de machine learning.

Este tipo de análisis generalmente se inicia con un análisis general, de tal modo que sea obtenible una visión de las principales correlaciones, y de tal modo que se puedan formular hipótesis generales, luego se pasa al análisis específico. En nuestro caso, ya realizamos el análisis general a través del modelo de autorregresión vectorial, por lo que procederemos con el siguiente paso para determinar cuáles variables afectan en mayor medida a la inflación y a las tasas de cambio.

Para empezar, sugiero rever la matriz de correlaciones y los resultados del modelo VAR, de tal forma que refresque su memoria sobre las relaciones entre los parámetros. Utilizaremos un código en Python para analizar si las relaciones se mantienen en los siguientes contextos:

1. Un año inestable (2023).
2. Ligeros aumento en la M2, ligero aumento en el long interest, ambos casos combinados.
3. Gran aumento en la M2, gran aumento en el long interest, ambos casos.

```python
import pandas as pd
import numpy as np
import matplotlib.pyplot as plt
from statsmodels.tsa.api import VAR

file_path = 
data = pd.read_excel(file_path, parse_dates=['Date'])

data['Year'] = data['Date'].dt.year
data_2023 = data[data['Year'] == 2023]

variables = ['Ipc Argentina', 'M2', 'Long Interest', 'Short Interest', 'Embi+ARG',
'Historical Ars Usd']
data_2023_var = data_2023[variables].dropna()

model = VAR(data_2023_var)
results = model.fit(maxlags=4)

print(results.summary())

baseline = results.fittedvalues['Ipc Argentina']

# Escenario 1:
scenario_1 = data_2023_var.copy()
scenario_1['M2'] *= 1.50
scenario_1_results = model.fit(maxlags=4).fittedvalues['Ipc Argentina']

# Escenario 2:
scenario_2 = data_2023_var.copy()
scenario_2['Long Interest'] += 5.00
scenario_2_results = model.fit(maxlags=4).fittedvalues['Ipc Argentina']

# Escenario 3:
scenario_3 = data_2023_var.copy()
scenario_3['M2'] *= 1.50
scenario_3['Long Interest'] += 5.00
scenario_3_results = model.fit(maxlags=4).fittedvalues['Ipc Argentina']

aligned_dates = data_2023['Date'].iloc[results.k_ar:]  # Align dates with predictions

plt.figure(figsize=(12, 6))
plt.plot(aligned_dates, baseline, label="IPC Argentina base", color='blue')
```

```python
plt.plot(aligned_dates, scenario_1_results, label="Incremento de 50% en M2",
color='orange', linestyle='--')
plt.plot(aligned_dates, scenario_2_results, label="incremento de 5% en Long Interest",
color='green', linestyle='--')
plt.plot(aligned_dates, scenario_3_results, label="Escenario combinado", color='red',
linestyle='--')
plt.title("Análisis de sensibilidad refinado: IPC Argentina bajo diferentes escenarios")
plt.xlabel("Fecha")
plt.ylabel("IPC Argentina")
plt.legend()
plt.grid(True)
plt.tight_layout()
plt.show()

residuals = results.resid['Ipc Argentina']

plt.figure(figsize=(12, 6))
plt.plot(aligned_dates, residuals, label="Residuos", color='red')
plt.axhline(0, color='black', linestyle='--')
plt.title("Residuos de IPC Argentina (Modelo VAR 2023 con refinamientos)")
plt.xlabel("Fecha")
plt.ylabel("Residuos")
plt.legend()
plt.grid(True)
plt.tight_layout()
plt.show()
```

Este código es el utilizado para el tercer contexto, puede deducirse cómo se verá el primero y el segundo.

En los tres casos, las relaciones se sostienen perfectamente. Aquí está la representación gráfica del tercer caso, también incluye la visualización de los residuos:

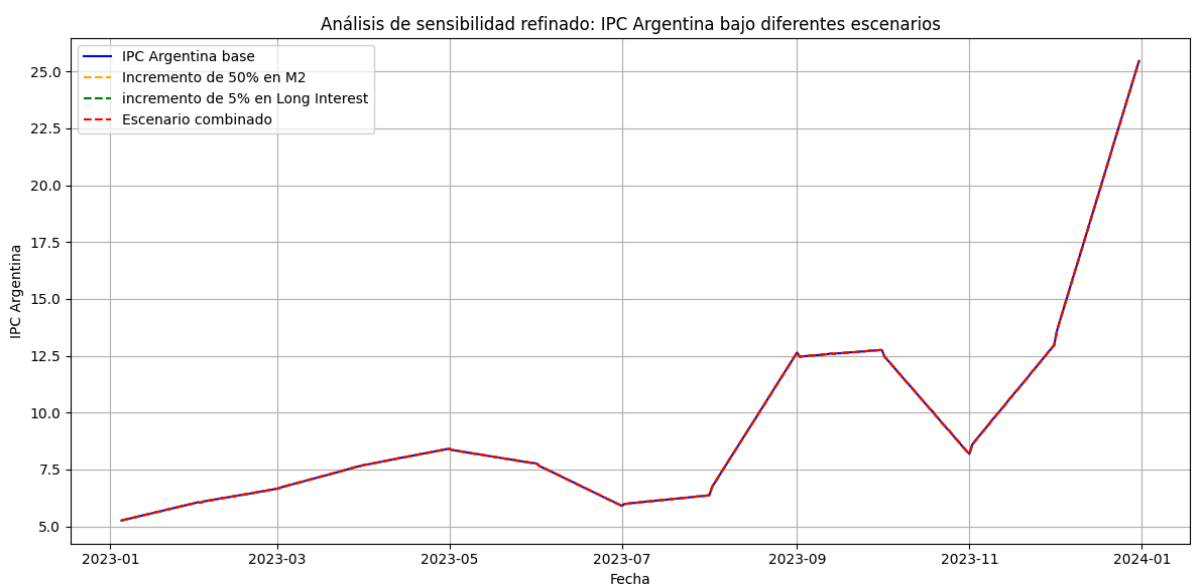

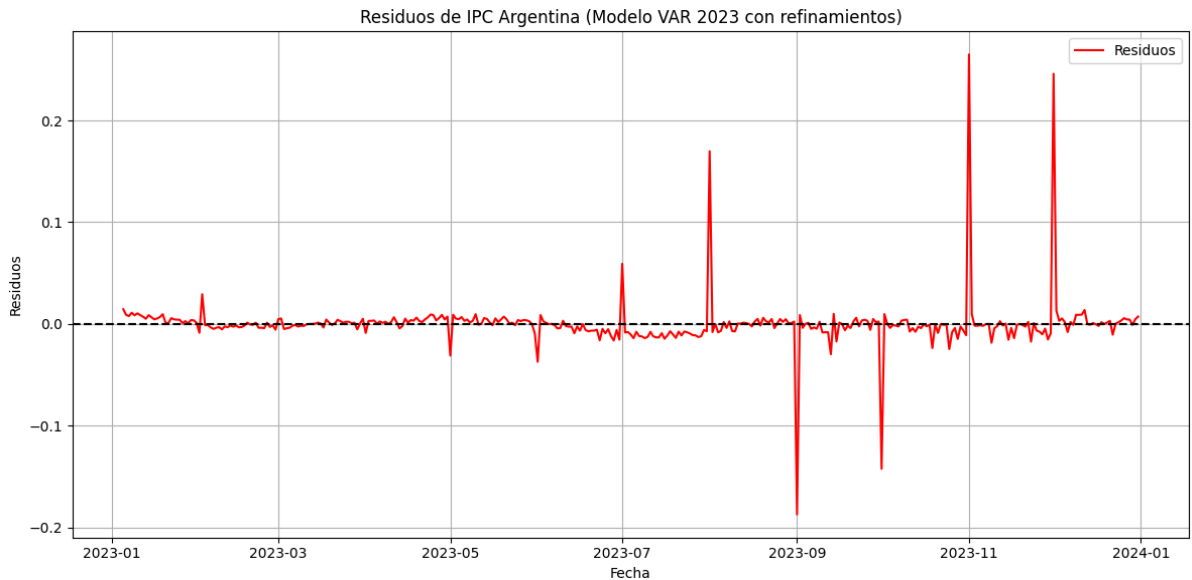

Como puede apreciarse, las relaciones se sostienen, acorde a lo proporcionado por el modelo VAR. Además, los residuos son bastante pequeños. Quizás utilizar otro modelos sea útil, pero, dado el nivel de certeza, tenemos la suficiente evidencia para concluir que las relaciones se mantienen, incluso en contextos de alta inestabilidad.

El siguiente paso es ver en detalle cuál de las variables es la que más afecta al IPC, para ello realizaremos un *feature importance analysis* (análisis de importancia de variables), el cual consiste en un método para determinar esto mismo en un modelo de machine learning. Particularmente, utilizaremos *impulse response functions*. Para detalles del funcionamiento de esta, véase página 82.

A continuación el código utilizado, nos enfocaremos en la M2 y en el long interest:

```python
import pandas as pd
import matplotlib.pyplot as plt
from statsmodels.tsa.api import VAR

file_path = 
data = pd.read_excel(file_path, parse_dates=['Date'])

data['Year'] = data['Date'].dt.year
data_2023 = data[data['Year'] == 2023]

variables = ['Ipc Argentina', 'M2', 'Long Interest', 'Short Interest', 'Embi+ARG',
'Historical Ars Usd']
data_2023_var = data_2023[variables].dropna()

model = VAR(data_2023_var)
results = model.fit(maxlags=4)

forecast_horizon = 10
fevd = results.fevd(forecast_horizon)
```

```python
print(fevd.summary())

fevd.plot(figsize=(12, 6))
plt.title("Forecast Error Variance Decomposition for Ipc Argentina")
plt.xlabel("Periods")
plt.ylabel("Proportion of Variance")
plt.legend(loc="upper right")
plt.grid(True)
plt.tight_layout()
plt.show()

irf = results.irf(forecast_horizon)

irf.plot(orth=False, impulse='M2', response='Ipc Argentina', figsize=(12, 6))
plt.title("Impulse Response of Ipc Argentina to M2 Shock")
plt.tight_layout()
plt.show()

irf.plot(orth=False, impulse='Long Interest', response='Ipc Argentina', figsize=(12, 6))
plt.title("Impulse Response of Ipc Argentina to Long Interest Shock")
plt.tight_layout()
plt.show()

irf.plot(orth=False, impulse='Short Interest', response='Ipc Argentina', figsize=(12, 6))
plt.title("Impulse Response of Ipc Argentina to Short Interest Shock")
plt.tight_layout()
plt.show()
```

El mismo producirá una lista de valores de la siguiente forma:

FEVD for Ipc Argentina

|   | Ipc Argentina | M2 | Long Interest | Short Interest | Embi+ARG | Historical Ars Usd |
|---|---|---|---|---|---|---|
| 0 | 1.000000 | 0.000000 | 0.000000 | 0.000000 | 0.000000 | 0.000000 |
| 1 | 0.994631 | 0.001434 | 0.003780 | 0.000001 | 0.000153 | 0.000002 |
| 2 | 0.991597 | 0.002000 | 0.005581 | 0.000002 | 0.000778 | 0.000043 |
| 3 | 0.989606 | 0.002073 | 0.006817 | 0.000017 | 0.001284 | 0.000204 |
| 4 | 0.988542 | 0.001943 | 0.007235 | 0.000022 | 0.001778 | 0.000480 |
| 5 | 0.987652 | 0.001737 | 0.007339 | 0.000020 | 0.002405 | 0.000847 |
| 6 | 0.986735 | 0.001507 | 0.007192 | 0.000014 | 0.003219 | 0.001333 |
| 7 | 0.985634 | 0.001293 | 0.006894 | 0.000011 | 0.004224 | 0.001945 |
| 8 | 0.984280 | 0.001105 | 0.006497 | 0.000016 | 0.005422 | 0.002681 |

|   | Ipc Argentina | M2 | Long Interest | Short Interest | Embi+ARG | Historical Ars Usd |
|---|---|---|---|---|---|---|
| 9 | 0.982648 | 0.000943 | 0.006040 | 0.000029 | 0.006811 | 0.003529 |

FEVD for M2

|   | Ipc Argentina | M2 | Long Interest | Short Interest | Embi+ARG | Historical Ars Usd |
|---|---|---|---|---|---|---|
| 0 | 0.000382 | 0.999618 | 0.000000 | 0.000000 | 0.000000 | 0.000000 |
| 1 | 0.000105 | 0.997747 | 0.000322 | 0.000057 | 0.000197 | 0.001572 |
| 2 | 0.000056 | 0.995464 | 0.000703 | 0.000822 | 0.000170 | 0.002785 |
| 3 | 0.000089 | 0.993415 | 0.002234 | 0.001633 | 0.000164 | 0.002464 |
| 4 | 0.000140 | 0.990696 | 0.003914 | 0.002445 | 0.000275 | 0.002531 |
| 5 | 0.000182 | 0.987436 | 0.005517 | 0.003186 | 0.000491 | 0.003189 |
| 6 | 0.000219 | 0.983843 | 0.006820 | 0.003881 | 0.000778 | 0.004459 |
| 7 | 0.000261 | 0.979733 | 0.007899 | 0.004501 | 0.001129 | 0.006477 |
| 8 | 0.000315 | 0.975037 | 0.008802 | 0.005058 | 0.001538 | 0.009250 |
| 9 | 0.000385 | 0.969747 | 0.009577 | 0.005554 | 0.001996 | 0.012742 |

FEVD for Long Interest

|   | Ipc Argentina | M2 | Long Interest | Short Interest | Embi+ARG | Historical Ars Usd |
|---|---|---|---|---|---|---|
| 0 | 0.044461 | 0.011327 | 0.944212 | 0.000000 | 0.000000 | 0.000000 |
| 1 | 0.045640 | 0.008283 | 0.943090 | 0.002148 | 0.000193 | 0.000646 |
| 2 | 0.048241 | 0.007578 | 0.933962 | 0.007844 | 0.000615 | 0.001760 |
| 3 | 0.055109 | 0.006344 | 0.907694 | 0.027813 | 0.000695 | 0.002344 |
| 4 | 0.064040 | 0.005652 | 0.874213 | 0.050232 | 0.001759 | 0.004104 |
| 5 | 0.073564 | 0.005706 | 0.839052 | 0.073359 | 0.002920 | 0.005399 |
| 6 | 0.083462 | 0.006176 | 0.807974 | 0.092058 | 0.003662 | 0.006668 |
| 7 | 0.092948 | 0.007184 | 0.778509 | 0.108556 | 0.004224 | 0.008579 |
| 8 | 0.101521 | 0.009045 | 0.750028 | 0.123140 | 0.004678 | 0.011589 |

| | Ipc Argentina | M2 | Long Interest | Short Interest | Embi+ARG | Historical Ars Usd |
|---|---|---|---|---|---|---|
| 9 | 0.108870 | 0.011853 | 0.721507 | 0.136917 | 0.005038 | 0.015815 |

FEVD for Short Interest

| | Ipc Argentina | M2 | Long Interest | Short Interest | Embi+ARG | Historical Ars Usd |
|---|---|---|---|---|---|---|
| 0 | 0.000024 | 0.005015 | 0.000083 | 0.994878 | 0.000000 | 0.000000 |
| 1 | 0.000032 | 0.011938 | 0.000133 | 0.987732 | 0.000019 | 0.000146 |
| 2 | 0.000130 | 0.008225 | 0.000232 | 0.959606 | 0.000112 | 0.031694 |
| 3 | 0.000451 | 0.005947 | 0.000276 | 0.895886 | 0.000168 | 0.097271 |
| 4 | 0.000679 | 0.006744 | 0.000411 | 0.848217 | 0.000123 | 0.143826 |
| 5 | 0.000751 | 0.006244 | 0.000603 | 0.820401 | 0.000108 | 0.171894 |
| 6 | 0.000732 | 0.005259 | 0.001473 | 0.807921 | 0.000103 | 0.184512 |
| 7 | 0.000686 | 0.004905 | 0.002951 | 0.802911 | 0.000111 | 0.188435 |
| 8 | 0.000627 | 0.005604 | 0.004922 | 0.800860 | 0.000156 | 0.187831 |
| 9 | 0.000565 | 0.007242 | 0.007048 | 0.799808 | 0.000261 | 0.185076 |

FEVD for Embi+ARG

| | Ipc Argentina | M2 | Long Interest | Short Interest | Embi+ARG | Historical Ars Usd |
|---|---|---|---|---|---|---|
| 0 | 0.000000 | 0.000175 | 0.000059 | 0.010350 | 0.989416 | 0.000000 |
| 1 | 0.001178 | 0.001346 | 0.000060 | 0.020549 | 0.975158 | 0.001710 |
| 2 | 0.004902 | 0.001705 | 0.001319 | 0.025213 | 0.963885 | 0.002976 |
| 3 | 0.008096 | 0.001653 | 0.001744 | 0.025633 | 0.959080 | 0.003794 |
| 4 | 0.011036 | 0.001419 | 0.002838 | 0.024552 | 0.956357 | 0.003798 |
| 5 | 0.014310 | 0.001211 | 0.005760 | 0.023198 | 0.951904 | 0.003617 |
| 6 | 0.018026 | 0.001049 | 0.010944 | 0.021974 | 0.944607 | 0.003400 |
| 7 | 0.022065 | 0.001004 | 0.017816 | 0.020750 | 0.935213 | 0.003152 |
| 8 | 0.026382 | 0.001090 | 0.025821 | 0.019491 | 0.924340 | 0.002876 |

| | Ipc Argentina | M2 | Long Interest | Short Interest | Embi+ARG | Historical Ars Usd |
|---|---|---|---|---|---|---|
| 9 | 0.030953 | 0.001270 | 0.034431 | 0.018219 | 0.912496 | 0.002630 |

FEVD for Historical Ars Usd

| | Ipc Argentina | M2 | Long Interest | Short Interest | Embi+ARG | Historical Ars Usd |
|---|---|---|---|---|---|---|
| 0 | 0.000000 | 0.018518 | 0.002301 | 0.002603 | 0.000282 | 0.976297 |
| 1 | 0.000000 | 0.112631 | 0.003554 | 0.003631 | 0.001223 | 0.878961 |
| 2 | 0.000369 | 0.130896 | 0.002702 | 0.003709 | 0.002865 | 0.859459 |
| 3 | 0.001360 | 0.124103 | 0.003047 | 0.003129 | 0.003128 | 0.865233 |
| 4 | 0.002800 | 0.120406 | 0.005451 | 0.002763 | 0.003051 | 0.865529 |
| 5 | 0.004757 | 0.114858 | 0.008929 | 0.002523 | 0.003124 | 0.865810 |
| 6 | 0.007491 | 0.109250 | 0.013069 | 0.002354 | 0.003407 | 0.864429 |
| 7 | 0.011391 | 0.104208 | 0.017477 | 0.002228 | 0.003860 | 0.860836 |
| 8 | 0.016688 | 0.099642 | 0.021697 | 0.002135 | 0.004440 | 0.855399 |
| 9 | 0.023615 | 0.095554 | 0.025628 | 0.002074 | 0.005162 | 0.847966 |

Y tres gráficos. Es de especial interés el siguiente:

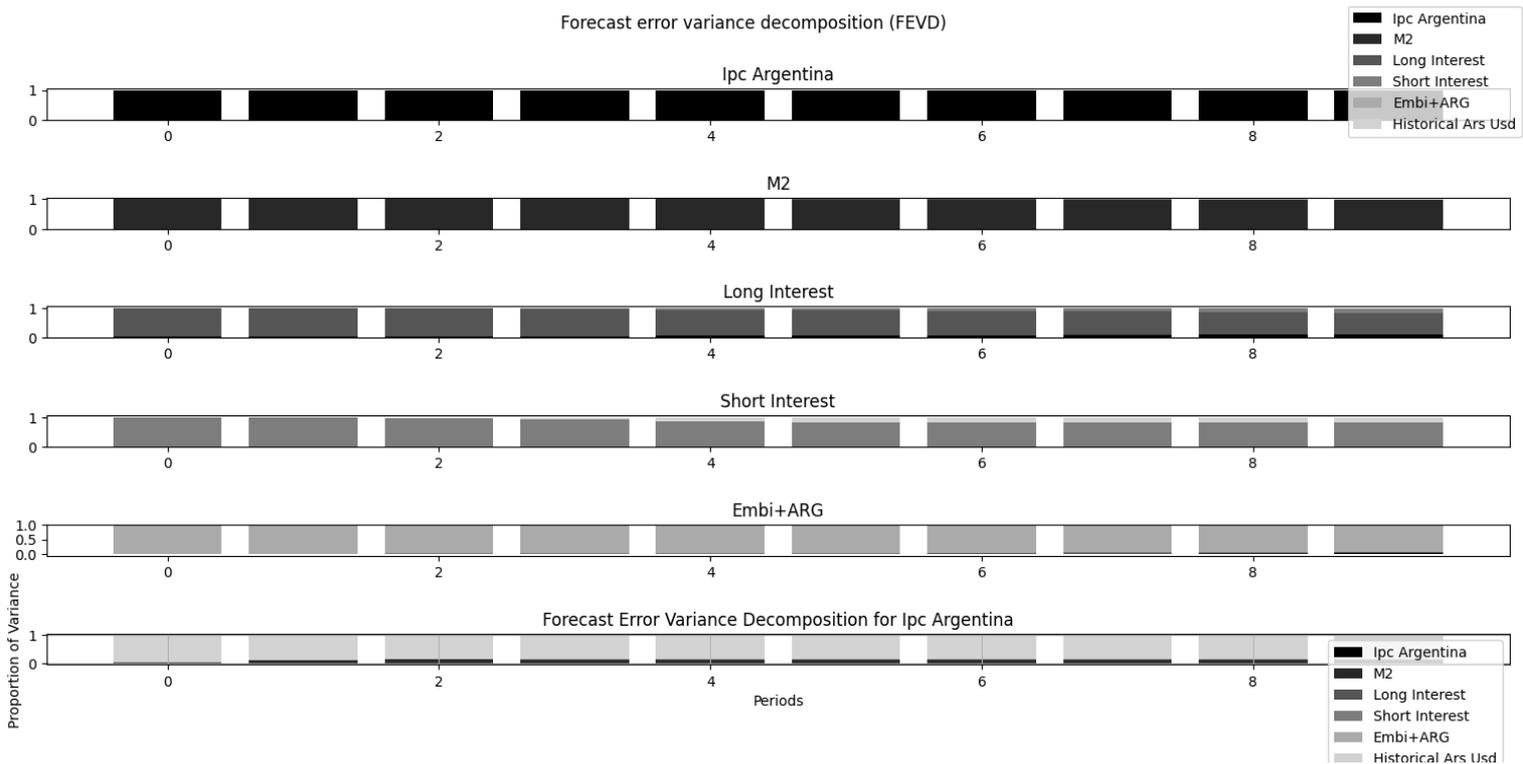

Las interpretaciones de la descomposición de varianzas obtenida son las siguientes:

1. Se puede apreciar una dominancia de la variable IPC Argentina, significando que los valores pasados de la variables dominan la descomposición, indicando un alto nivel de persistencia. La mayor parte de la variación de IPC Argentina es explicada por sí misma, especialmente en períodos tempranos.

2. La contribución de M2 a IPC Argentina es pequeña pero crece significativamente sobre el tiempo, sugiriendo que cambios en la oferta monetaria tienen un impacto rezagado (lag), coincidiendo con lo observado en nuestro modelo VAR, y limitado sobre las expectativas de inflación en el modelo, lo cual se asemeja bastante a la intuición económica fundamentalista.

3. La contribución del Long Interest al IPC Argentina es menor pero consistente. Nuevamente, alineándose con la intuición económica. Las tasas de interés a largo plazo influyen en la inflación a través de la demanda agregada, aunque menos directamente

4. Las variables externas (EMBI+ ARG, Historical ARS/USD) contribuyen mínimamente a IPC Argentina, indicando que factores externos como el riesgo y las tasas de cambio pueden no influir directamente en la inflación en el modelo cortoplacista.

5. Short Interest tiene un impacto casi nulo en IPC Argentina en este modelo.

En la práctica, parece que las expectativas de inflación en Argentina son primariamente conducidas por su propio historial, enfatizando la necesidad de políticas que tengan por objetivo tendencias inflacionarias persistentes. El impacto mínimo de variables vinculadas con políticas monetarias sugiere que intervenciones monetarias tradicionales pueden tener efectividad limitada si no se tocan los problemas estructurales.

En conclusión, según el modelo FEVD (forecast error variance decomposition), la variables más útil para manejar la inflación parece ser el propio historial de inflación. La mayor proporción de varianza en IPC Argentina se explica por sí misma (cerca del 99%). La inflación persistente en Argentina sugiere que es clave enfocar los esfuerzos en las expectativas de inflación atrincheradas y el incremento persistente de los precios, quizás a través de políticas cuyos objetivos sean los factores que impulsan la inflación subyacente, y la gestión de expectativas podrían ser más efectivas que las intervenciones a corto plazo. En definitiva, implementar reformas fiscales y manejar las expectativas, así también como proporcionar políticas antiinflacionarias con **credibilidad** puede reducir las expectativas crecientes.

Adicionalmente, la oferta monetaria tiene un impacto limitado pero que crece con el tiempo, sugiriendo, obviamente, que reducir la tasa de crecimiento en esta puede ayudar a manejar la inflación, aunque esto siendo parte de un plan a largo plazo, dado el efecto rezago. Algo similar puede decirse de las tasas de interés a largo plazo.

Para perfeccionar el modelo, podríamos emplear efectos no lineales o términos de interacción, de tal forma que se revelen dinámicas no a la vista para el modelo simple que utilizamos. Además, se pueden explorar variables adicionales como déficits fiscales o precios en commodities para mayores análisis, aunque no realizaremos estos análisis.

Hasta aquí suele llegar un modelado tradicional (de forma simplificada); no obstante, nosotros aprovecharemos los conceptos de functores en teoría de categorías para analizar sistemáticamente y reconciliar las interacciones entre relaciones domésticas simplificadas e interacciones globales enriquecidas.

El análisis de sensibilidad que conducimos mostró los impactos individuales y colectivos de variables como M2, Long Interest y factores externos en la inflación (IPC Argentina); sin embargo, este análisis nos dotó de datos estáticos. Utilizar functores nos permite descomponer complejidad, permitiéndonos simplificar relaciones a sus factores domésticos fundamentales, lo cual es crítico para el análisis de políticas, enfocándonos primariamente en lo que Argentina puede controlar directamente; y también nos permite reintroducir contexto, expandiendo el modelo al añadir influencias externas, mostrando cómo la economía doméstica de Argentina interactúa con factores globales, mejorando el entendimiento de efectos desbordamiento como el que habíamos mencionado en la Parte II. Es esta adjunción de functores la que permitirá construir un puente en el espacio entre dinámicas locales y globales.

El análisis también demostró que IPC Argentina es primariamente influenciada por sus propios valores pasados y factores monetarios domésticos, pero la efectividad de políticas depende de restricciones externas, como el riesgo y las dinámicas en las tasas de cambio. Es aquí cuando el uso de functores se justifica, ya que el mismo nos muestra, a través del functor U, qué puede controlar Argentina y a qué debe reaccionar, a través del functor L. Esta dualidad es crítica para diseñar políticas coordinadas, como se podrá intuir económicamente. Es esta intuición la que deseamos matematizar.

Definiremos en Python al functor U (modelo base), de olvido, estando este compuesto por M2, Long Interest e IPC Argentina, y definiremos al functor L (modelo enriquecido), de aprendizaje, tal que esté compuesto por EMBI+ ARG y las tasas de Estados Unidos.

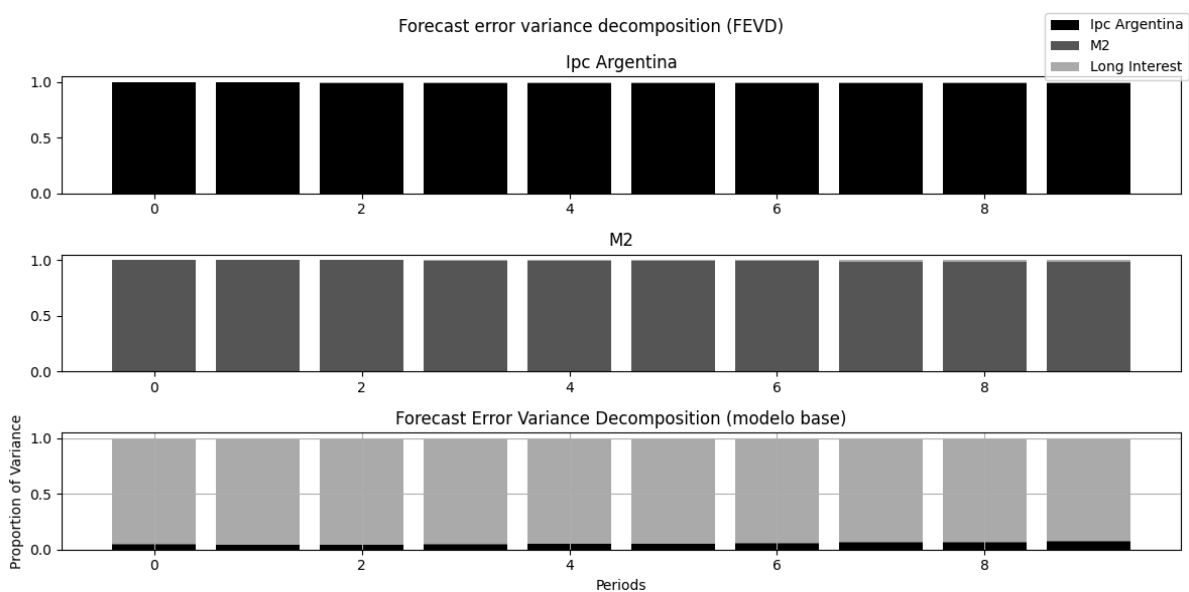

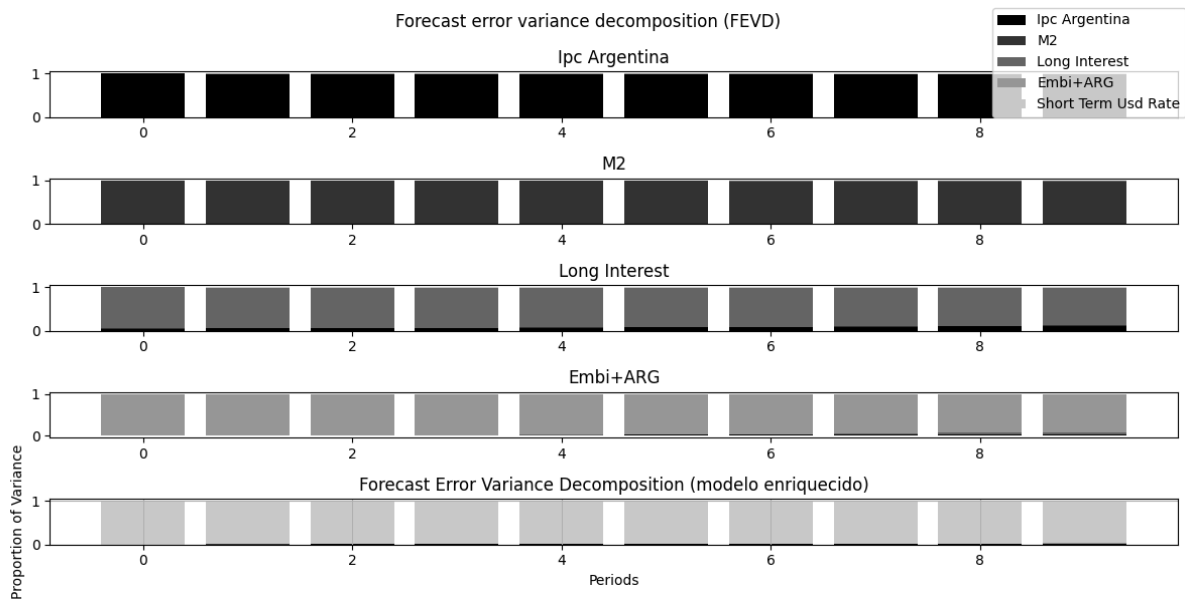

Este sistema muestra los mismos resultados de antes, pero usando el enfoque de functores, siendo más práctico a la hora de diseñar políticas. El modelo enriquecido permite entender la influencia de factores externos, como el riesgo, y el modelo base sirve como fundamento para distinguir entre influencias locales de influencias externas.

Quizás el mayor uso de esta idea sea el testear escenarios para ver cuáles políticas responden mejor, primero de forma aislada en el modelo base, y luego ante la presencia de shocks externos. En Teoría de categorías, estos conceptos se conocen como "interacciones locales-globales" y "feedback y adaptación".

Antes de culminar esta sección, veremos otra aplicación similar de los functores, medir el impacto localizado de diferentes políticas, por ejemplo, podemos ver cómo el functor local o el functor enriquecido responden a shocks en los mercados locales (como podría ser por una devaluación del peso repentina).

El siguiente approach tiene un poder predictivo considerable, es sorprendente lo bien que se manejó en base a los datos con los que hemos estado trabajando.

```python
import pandas as pd
import numpy as np
import matplotlib.pyplot as plt
from xgboost import XGBRegressor
from sklearn.model_selection import train_test_split
from sklearn.metrics import mean_squared_error, mean_absolute_error

file_path = 
data = pd.read_excel(file_path)
data['Date'] = pd.to_datetime(data['Date'])

domestic_vars = ['Ipc Argentina', 'M2', 'Historical Ars Usd']
enriched_vars = ['Ipc Argentina', 'M2', 'Historical Ars Usd', 'Short Term Usd Rate', 'Long Term Usd Rate']

def create_lagged_features(data, variables, lags=3):
```

```python
    lagged_data = data.copy()
    for var in variables:
        for lag in range(1, lags + 1):
            lagged_data[f'{var}_lag{lag}'] = lagged_data[var].shift(lag)
    return lagged_data.dropna()

domestic_data = create_lagged_features(data, domestic_vars)
enriched_data = create_lagged_features(data, enriched_vars)

def train_xgboost_model(data, features, target):
    X = data[features]
    y = data[target]
    X_train, X_test, y_train, y_test = train_test_split(X, y, test_size=0.2, random_state=42)
    model = XGBRegressor(n_estimators=100, random_state=42)
    model.fit(X_train, y_train)

    y_pred = model.predict(X_test)
    mse = mean_squared_error(y_test, y_pred)
    mae = mean_absolute_error(y_test, y_pred)
    print(f"MSE: {mse}, MAE: {mae}")

    return model

domestic_features = [col for col in domestic_data.columns if 'lag' in col]
enriched_features = [col for col in enriched_data.columns if 'lag' in col]
target = 'Ipc Argentina'

domestic_model = train_xgboost_model(domestic_data, domestic_features, target)
enriched_model = train_xgboost_model(enriched_data, enriched_features, target)

def introduce_shock(data, variable, shock_value):
    shocked_data = data.copy()
    shocked_data[variable] *= shock_value
    return shocked_data

shock_variable_domestic = 'Ipc Argentina'
shock_variable_enriched = 'Short Term Usd Rate'
shock_value = 1.5

shocked_domestic_data = introduce_shock(domestic_data, shock_variable_domestic, shock_value)
shocked_enriched_data = introduce_shock(enriched_data, shock_variable_enriched, shock_value)

def forecast_xgboost(model, recent_data, forecast_steps=12):
    forecasts = []
    current_data = recent_data.copy()
    for _ in range(forecast_steps):
        pred = model.predict(current_data.reshape(1, -1))
        forecasts.append(pred[0])
        current_data = np.roll(current_data, -1)
        current_data[-1] = pred[0]
    return forecasts

recent_domestic = shocked_domestic_data.iloc[-1][domestic_features].values
```

```python
recent_enriched = shocked_enriched_data.iloc[-1][enriched_features].values

forecast_horizon = 12
shocked_domestic_forecast = forecast_xgboost(domestic_model, recent_domestic,
forecast_horizon)
shocked_enriched_forecast = forecast_xgboost(enriched_model, recent_enriched,
forecast_horizon)

forecast_index = pd.date_range(start=data['Date'].iloc[-1], periods=forecast_horizon +
1, freq='M')[1:]

plt.figure(figsize=(12, 6))
plt.plot(forecast_index, shocked_domestic_forecast, label="Pronóstico doméstico post
shock", color='blue')
plt.plot(forecast_index, shocked_enriched_forecast, label="Pronóstico enriquecido post
shock", color='red')
plt.title("Pronóstico de inflación: Doméstico vs enriquecido post shock")
plt.xlabel("Fecha")
plt.ylabel("Inflación (Ipc Argentina)")
plt.legend()
plt.grid(True)
plt.tight_layout()
plt.show()
```

En el código utilizamos una librería que soporta *machine learning* (XGBoost), este puntualmente es una herramienta poderosa que identifica patrones en los datos para realizar predicciones precisas.

El modelo toma datos económicos históricos (por ejemplo, inflación, oferta monetaria, tipos de cambio) y aprende cómo estas variables interactúan a lo largo del tiempo. Usa este aprendizaje para predecir qué ocurrirá en el futuro bajo diferentes escenarios (como "políticas de choque"). Es como analizar cómo responde el PIB a cambios en las tasas de interés a lo largo de décadas, pero el modelo realiza este análisis para todas las variables disponibles simultáneamente y puede simular diferentes políticas. No nos enfocaremos en entender puntualmente cómo funciona el modelo, ya que posee una dificultad considerable; no obstante, si le interesa profundizar, sugiero leer el detalle de la librería. Sólo se mencionará lo siguiente respecto del código: Entrenamos dos modelos, el modelo doméstico, enfocado únicamente en variables nacionales (por ejemplo, inflación, oferta monetaria, tipo de cambio) y el modelo enriquecido, que Incluyó factores globales (por ejemplo, EMBI+ para riesgo país, tasas de interés en USD). Luego aplicamos políticas de choque (controles monetarios y fiscales estrictos) al conjunto de datos. Usamos el modelo para predecir la inflación bajo estas políticas para 2024 y años posteriores. Los resultados muestran que los modelos enriquecidos son más sensibles y adaptativos a los cambios de política, ya que incluyen factores globales que a menudo se pasan por alto en los modelos domésticos tradicionales, aunque lo realmente importante fue la precisión del modelo, observe:

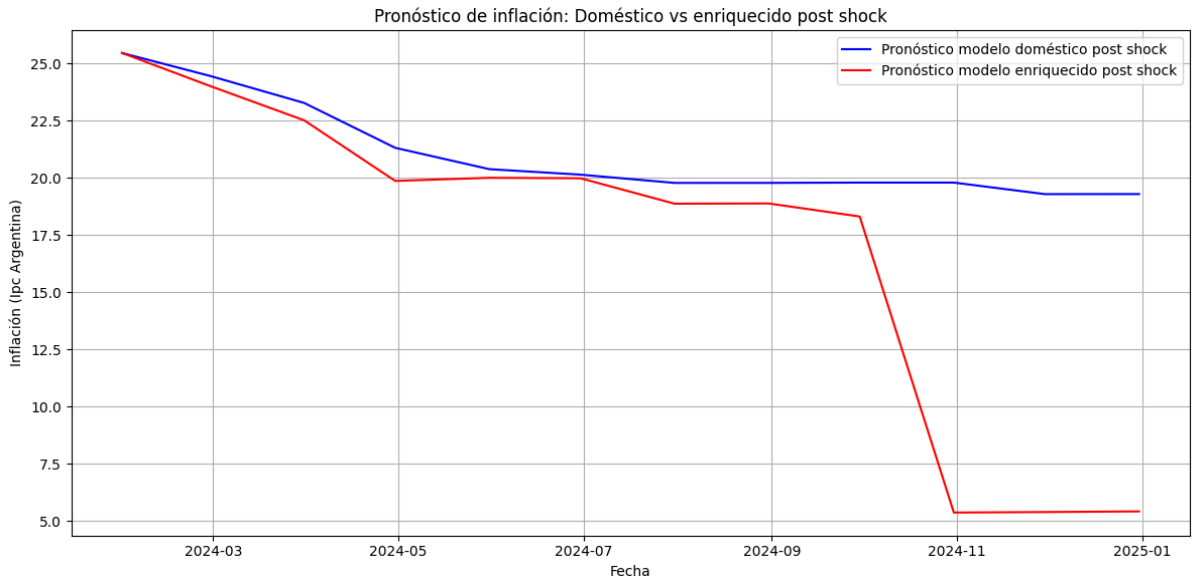

El modelo enriquecido estimó una inflación de aproximadamente 5% para fines de 2024 (mientras que la inflación real según el INDEC en diciembre de 2024 fue del 2,7%) aplicando las siguientes políticas:

- Reducción del 20% en M2 (simulando una disminución en la base monetaria).
- Reducción del 50% en Argentina Net Lending Borrowing (simulando una reducción en el déficit fiscal).

Esto demuestra por qué es útil el enfoque de functores:

1) Nos permite realizar rápidamente simulaciones como la vista, dada la flexibilidad de la teoría de categorías.
2) Nos permite observar varios puntos de vista (doméstico, global o cualquier otro que se le ocurra).
3) Nos permite aplicar políticas sobre varios campos a la vez y observar cómo interactúan las variables en un sistema dinámico.

Los economistas a menudo usan modelos teóricos o regresiones simples para predecir cómo las políticas pueden impactar la economía; sin embargo, estos métodos suelen asumir que las relaciones son fijas (por ejemplo, multiplicadores constantes en modelos IS-LM) y tienen dificultades para manejar interacciones complejas y no lineales (como bucles de retroalimentación entre inflación y tipos de cambio); los modelos de machine learning, en contraste, como el que usamos destacan porque se adaptan a relaciones complejas, detectando efectos no lineales y dinámicos en los datos que los modelos tradicionales podrían pasar por alto, además permiten simular escenarios. XGBoost puntualmente nos permite simular escenarios hipotéticos, como cómo respondería la inflación a políticas de choque basadas en patrones históricos.

Si tuviéramos todos los datos disponibles respecto de las medidas tomadas durante el mandato de Milei, así también como una mayor cantidad de objetos y morfismos, podríamos haber conseguido una estimación absurdamente precisa, demostrando, nuevamente, que la economía no puede verse en el álgebra tradicional, ya que no vivimos en un mundo lineal, sino en uno dinámico, donde la información tiende al infinito.

Como puede ver, el machine learning y la teoría de categorías, así también como otras teorías avanzadas de las matemáticas, se llevan bien dado al match en nivel flexibilidad y dinamismo. Realizar este tipo de análisis con modelos algebraicos y estadísticos tradicionales tomaría eones.

Por otro lado, sería bastante útil realizar un modelo como el presente que se actualice todo el tiempo, que incluya más objetos, functores, morfismos y datos actualizados de los mismos, así también como datos de las políticas que el gobierno tomó, toma y planea tomar, de tal modo que se puedan obtener predicciones más fiables y de forma inmediata, aunque realizar algo semejante trasciende los objetivos de esta investigación, además de que un artilugio así tendría un nivel de poder alucinante, a tal punto que permitiría realizar *inside trading* sobre varios mercados, como el de bonos, commodities y acciones en empresas locales, dependiendo qué tan hábil sea con las relaciones fundamentales y técnicas su usuario.

# Parte III: Cálculo del Equilibrio y Medición de Expectativas de Devaluación

En esta última parte, se aplican límites para definir el equilibrio macroeconómico dentro del marco categórico. Se emplean colimites en conjunto con análisis de series temporales y modelos de proyección como el VAR para estimar expectativas de devaluación. Se estudia la influencia de los agentes en la formación de expectativas y se estudia la capacidad predictiva del modelo categórico.

## Hipótesis

*"El uso de límites y colimites en la teoría de categorías proporciona una metodología más precisa para calcular equilibrios macroeconómicos y para modelar las expectativas de devaluación en la economía argentina."*

Esta hipótesis nace de que los modelos econométricos tradicionales tienen dificultades para capturar la dinámica de expectativas de devaluación y cómo evolucionan en el tiempo, así también como para modelar un equilibrio desde una visión holística. Verificaremos si los límites y colimites pueden estructurar de manera más efectiva el cálculo del equilibrio y si permiten medir la evolución de expectativas de devaluación de manera más coherente con datos históricos y tendencias.

## Marco teórico

En economía, solemos lidiar con sistemas donde varios agentes, mercados o políticas interactúan dinámicamente. Estas interacciones tienden a estados de equilibrio o integración, tales como un precio de mercado o una política nacional unificada. Los conceptos de límite y colimite en teoría de categorías proveen un marco matemáticamente robusto para analizar estos fenómenos.
Esta sección introduce estas herramientas de forma rigurosa pero intuitiva, preparándonos para su aplicación formal a problemas económicos.

Para empezar, en su base fundamental, los límites y colimites son construcciones matemáticas que resumen cómo objetos interactúan dentro de un sistema estructurado como categoría. En una categoría, los objetos se vinculan mediante morfismos, representando relaciones o transformaciones entre estos objetos.

Un límite puede ser pensado como un camino para "sintetizar acuerdo" entre una red de objetos y morfismos. Identifica un objeto singular que sirve como el más compatible o "universal" resumen del sistema. Es un objeto (o solución) que captura el más completo modo de satisfacer las relaciones en un diagrama (recordemos que las categorías pueden representarse como diagramas de flechas, donde los morfismos son las flechas).

Por definición, dado un diagrama D en una categoría C, el límite de D es un objeto L en C, junto con morfismos a L desde todos los objetos en D, tal que cualquier otro objeto con morfismos similares se factorice de forma única a través de L. En términos de categorías,

$$lim\, D = (L, \{\pi_i : L \to D_i\}_{i \in I}$$

Donde L satisface la propiedad universal de ser único hasta el nivel de isomorfismo (hasta ser una igualdad reversible, en términos elementales). Llamamos a $\pi_i : L \to D$ proyección para cada objeto $D_i$ en D.

Intuitivamente, esto podría representar encontrar el estado más estable de un sistema, como el precio donde la oferta iguala a la demanda.

Ahora, le pregunto a usted, lector, ¿puede encontrar el equilibrio de cualquier mercado dado a partir de un conjunto de parámetros como los que hemos estado tratando hasta ahora (M2, tasas de intereses, de cambio, PBI, EMBI+ARG, etc…) y sus respectivas series de tiempo? Quizás pueda elaborar estimaciones del mercado de bonos en base a la situación macroeconómica del país, y, usando teoría económica, puede llegar a estimaciones sobre otros mercados o assets con tendencias similares o opuestas; no obstante, llevar a cabo este proceso sería engorroso (debería de construir un mapa de relaciones fundamentales de gran tamaño para aquellos mercados cuyos destinos no están muy claros) y produciría estimaciones imprecisas. Por otro lado, imaginemos un algoritmo de machine learning basado en teoría de categorías. Tal algoritmo podría dar detalles sobre los precios de equilibrio a una velocidad impresionante y sería bastante preciso, esto ya que la teoría de categorías permite, como ya vimos, relacionar áreas aparentemente distantes mediante functores (recuerde la definición formal) y establecer traducciones entre las mismas, así también como puentes. Por más que tengamos una categoría con un número *n* de objetos y morfismos, la estructura diagramática es tal que podemos ver cómo un cambio en el primer objeto repercutirá en el último objeto, además podemos separar la categoría y moldearla de forma que se simplifique el análisis. El uso de la ley de composición de morfismos permite conocer inmediatamente cómo el primer objeto de la categoría se relaciona con el último. Generalmente estamos acostumbrados a ver la composición únicamente en el álgebra básica de conjuntos, por ejemplo:

dadas $f: A \to B$ y $g: B \to C$, se define a la composición como $g \circ f: A \to C$. Es decir, una función tal que toma un elemento de A y lo transforma en uno de B, usando f, para luego transformarlo en uno de C, usando G . La composición nos permite pasar directamente de $A$ a $C$ sin tener que detenernos explícitamente en $B$; es decir, encapsula en un solo paso la transformación completa entre el primer y el último conjunto.

En muchas estructuras matemáticas (como conjuntos, vectores, grupos de transformación, etc…) existe la composición, y, particularmente, siempre debe existir cuando trabajamos con categorías. Por esto mismo es que un sistema basado en su funcionamiento puede acceder en milisegundos a información extremadamente compleja, como lo es la situación planteada en la página anterior.

Continuando, también existe el colimite, el cual trabaja similarmente, pero "agregando divergencia", más que "sintetizar acuerdo". Construye un objeto singular que combina o unifica el diagrama entero. Se puede pensar en ellos como una herramienta que menciona qué pasa cuando fusionamos la categoría, la unificación del todo. Nos resume cómo los objetos en el sistema se combinan, unifican o divergen. La definición formal se puede plantear de un modo similar al de límite:

$$\text{colim}\, D \;=\; (c, \{l_i : D_i \to c\}_{i \in I})$$

Aunque aquí se trata de inyecciones, no proyecciones, para cada objeto $D_i$ en D.

La diferencia entre el límite y el colimite es que el límite trae o tracciona todos los objetos en D a un objeto en L, que puede representar todo el sistema con respecto a las relaciones en D. L conecta cada objeto en el diagrama D, y satisface la propiedad de que cualquier otro objeto tratando de "hacer lo mismo" se debe factorizar a través de L; mientras que el colimite empuja todos los objetos en *D* a un objeto *C* que integra o combina todo en el sistema mientras que respeta las relaciones en *D*, y satisface la propiedad de que cualquier otro objeto tratando de "combinarse del mismo modo" se debe factorizar a través de C. En conclusión, piense en el límite como encontrar el espacio común o acuerdo entre todos los objetos; y piense en el colimite como construir un resumen unificado del sistema.

Para hacer más clara la distinción, la siguiente analogía puede ser útil: El límite se puede interpretar como hallar el precio de equilibrio de un mercado, que es el punto donde todas las curvas de oferta y demanda se cruzan en equilibrio. El colimite se puede interpretar como sumar todos los PBI regionales para obtener el PBI nacional, es un agregado de diferentes componentes. Como podrá darse cuenta, no es lo mismo buscar un equilibrio, un factor común, que una suma o agregado.

## Metodología de la investigación

### Límites

Ahora cabe preguntarse cómo se calculan estos equilibrios o agregados, que es lo que estudiaremos en esta sección.

Los conceptos de límite y colimite son más fáciles de visualizar que los functores, por lo que es más sencillo proceder a las aplicaciones. Para empezar, si consideramos morfismos que representan las relaciones entre las expectativas de inflación, la demanda de dinero y las tasas de interés, el límite de este diagrama será el estado de equilibrio donde estas fuerzas se compensen mutuamente, lo cual podría traducirse en una situación donde la propensión marginal a demandar pesos sea igual a la propensión marginal a demandar dólares.

Como hemos visto, el límite consolida múltiples morfismos en un único morfismo "resumen" central que proporciona una visión completa de las relaciones económicas, por ejemplo, si las expectativas de inflación afectan tanto la demanda de pesos como el tipo de cambio, un límite representaría el punto donde estas influencias se estabilizan, permitiendo una visión general de la estabilidad monetaria. Por otro lado, el colimite captura la "salida" o divergencia del modelo económico, por ejemplo, cuando el peso experimenta hiperinflación, la salida de su valor relativo al dólar podría capturarse como un colimite, que representa el efecto agregado de diversos factores económicos (como las expectativas de devaluación). Además, los colimites pueden modelar situaciones donde múltiples acciones económicas (morfismos) contribuyen a un solo resultado, por ejemplo, múltiples políticas fiscales y monetarias, representadas como diferentes morfismos, podrían converger en un único resultado como la estabilización de la tasa de inflación o un cambio significativo en la preferencia de moneda.

Buscando un mayor nivel de pragmatismo, lo que haremos será, por un lado, usar los límites para identificar puntos de estabilidad en el sistema monetario, como cuando las expectativas de inflación dejan de causar grandes desviaciones en la demanda de moneda, ya que esto ayudará a identificar las condiciones económicas que conducen al equilibrio; por otro lado, aplicaremos colimites para entender cómo los choques agregados, como un aumento repentino del déficit fiscal o choques externos, como subidas de tasas de interés en EE.UU., se propagan a través del sistema, dado que el uso de colimites puede ayudar a analizar "puntos de salida" o comportamientos extremos, como la inflación descontrolada o las crisis de devaluación.

A continuación, empezaremos con los límites, luego con los colimites. Observe el siguiente código:

```python
import pandas as pd
import numpy as np
import matplotlib.pyplot as plt
from scipy.optimize import minimize

file_path = 
data = pd.read_excel(file_path)
data['Date'] = pd.to_datetime(data['Date'])

parameters = ['Usa Pi Exp', 'Long Term Usd Rate', 'Short Term Usd Rate', 'M2 Usd', 'Ipc Usa',
              'Historical Ars Usd', 'Argentina Net Lending Borrowing', 'Ipc Argentina',
              'Pi Exp', 'Long Interest', 'Short Interest', 'M2',
              'Gdp_argentina', 'Gdp_usa', 'E', 'Embi+ARG']

data = data[['Date'] + parameters].dropna()

def equilibrium_condition(params, gdp_usa, gdp_argentina, embi, historical_ars_usd, long_term_usd_rate):
    e = params[0]
    penalty = 0.0

    gdp_ratio = gdp_usa / gdp_argentina
    morphism_1 = (e - gdp_ratio) ** 2

    morphism_2 = (e - embi * historical_ars_usd) ** 2

    morphism_3 = (e - long_term_usd_rate) ** 2
```

```python
        penalty += morphism_1 + morphism_2 + morphism_3
        return penalty

results = []
for _, row in data.iterrows():
    gdp_usa = row['Gdp_usa']
    gdp_argentina = row['Gdp_argentina']
    embi = row['Embi+ARG']
    historical_ars_usd = row['Historical Ars Usd']
    long_term_usd_rate = row['Long Term Usd Rate']

    res = minimize(
        equilibrium_condition,
        x0=[historical_ars_usd],
        args=(gdp_usa, gdp_argentina, embi, historical_ars_usd, long_term_usd_rate),
        method='Nelder-Mead'
    )
    results.append(res.x[0])

data['equilibrio tipo de cambio'] = results

plt.figure(figsize=(12, 6))
plt.plot(data['Date'], data['Historical Ars Usd'], label='tipo de cambio observado', color='blue')
plt.plot(data['Date'], data['equilibrio tipo de cambio'], label='equilibrio tipo de cambio', color='red')
plt.title('tipo de cambio observado vs equilibrio')
plt.xlabel('fecha')
plt.ylabel('Tipo de cambio (ARS/USD)')
plt.legend()
plt.grid(True)
plt.tight_layout()
plt.show()
```

Este código nos permite estimar el tipo de cambio de equilibrio para el peso, relativo al dólar. Compara el equilibrio a el tipo de cambio observado sobre el tiempo, resaltando cuán lejos la data se desvía del equilibrio dado a distorsiones económicas como inflación, déficit fiscal o shocks externos (justamente los problemas de Argentina, he ahí que nos interese esto). Note que usamos morfismos dentro del código, e interpretamos los parámetros como objetos. El primer morfismo asegura que el tipo de cambio refleje el tamaño relativo de las economías de Estados Unidos y Argentina; el segundo morfismo conecta el tipo de cambio al riesgo de Argentina (EMBI) y la tasa observada; el tercer morfismo asegura que el tipo cambio refleje tasas a largo plazo de Estados Unidos (long interest), como proxy de estabilidad global. Cada morfismo calcula una penalidad o castigo (penalty), que es una diferencia cuadrada (squared difference) tal que mida cuán lejos la tasa predecida está de satisfacer la relación específica. El objetivo es minimizar la penalidad total para encontrar el equilibrio en el tipo de cambio que mejor balancee estas relaciones, usando algoritmos de optimización numéricos, como lo es el Nelder-Mead.
Es un modo teóricamente y prácticamente elegante de plantear nuestro objetivo, encontrar el equilibrio de la categoría; no obstante, la Argentina no responde a estos planteamientos, observe:

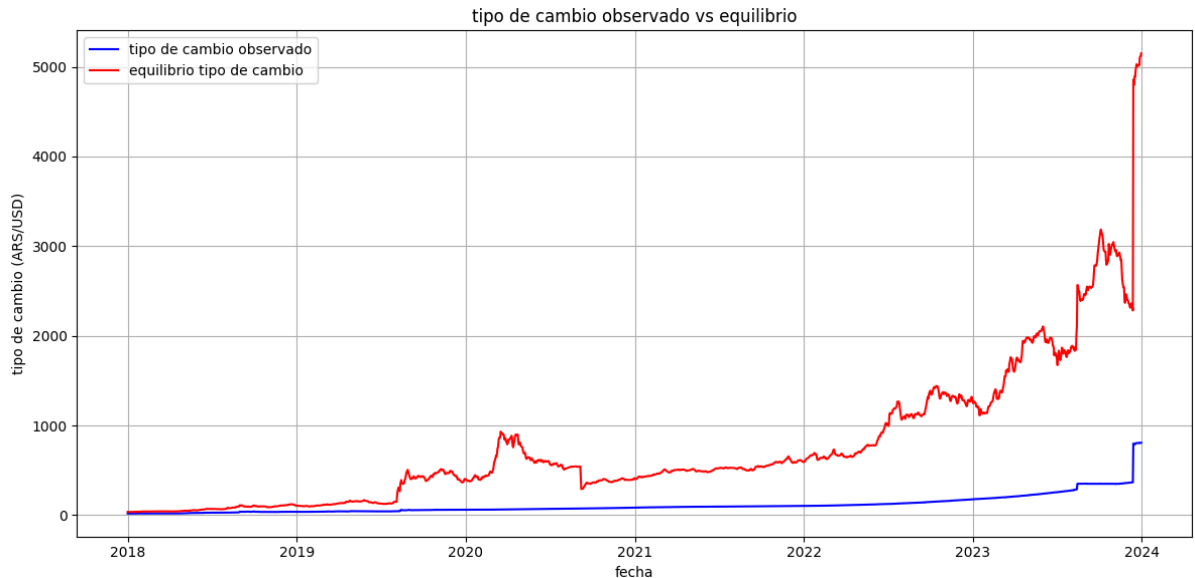

Como podemos apreciar, hay diferencias considerables entre el equilibrio y la tasa observada, aunque esto no es un error, tiene perfecto sentido, ya que el desequilibrio es un indicador crítico de problemas estructurales en una economía. En el caso de Argentina, el gráfico resalta el severo desalineamiento entre las tasas observadas y de equilibrio, lo cual puede ser un indicador de:
1. Persistentes desbalances macroeconómicos: Déficits crónicos, exceso de oferta monetaria y la alta inflación están muy probablemente conduciendo la tasa observada lejos del equilibrio teórico.
2. Distorsiones en las políticas: Controles artificiales en las tasas de cambio o políticas referidas al déficit o la moneda insostenibles pueden provocar esta divergencia.
3. Expectativas de mercado: La línea roja puede estar reflejando lo que el mercado esperaría en un escenario balanceado, mientras que la azul podría reflejar la realidad moldeada por intervenciones e ineficiencias de mercado.

Como podrá darse cuenta, estos tres puntos resumen a la Argentina, y es por eso que este análisis es valioso, refleja lo que hay que "arreglar" para tener un país más "estable" si se quiere. Es esto mismo lo que los modelos de equilibrio están diseñados para identificar y manejar. Puede pensarse en el equilibrio no como mercados siempre balanceados, sino como una construcción teórica para medir cuán lejos una economía se está desviando del balance.
En resúmen, el análisis de equilibrio sirve para:

1. Cuantificar la diferencia entre los valores observados y las condiciones de equilibrio.
2. Usar la diferencia como métrica para evaluar la severidad de los desbalances económicos.
3. Identificar dónde hacer ajustes para solucionarlos mediante políticas.

Calcular la diferencia (*gap*), así también como hacer un algoritmo de machine learning para ver qué pasaría si se realizan ajustes como los propuestos en la parte II es algo que ya realizamos, y arrojaría resultados triviales si se realizara de vuelta.

Para esta altura, es comprensible cómo la teoría de categorías provee al economista de un marco teórico dinámico, con el potencial de abarcar todos los objetos y morfismos económicos si se dispone del poder computacional suficiente. Esta pequeña categoría que hemos construido ha sido suficiente para apreciar problemas estructurales en Argentina, y, de ser más grande, podríamos encontrar ineficiencias no sólo macroeconómicas sino también microeconómicas y mesoeconómicas. También note que no es un sustituto para el álgebra o la estadística tradicional, sino una mejora en rigurosidad conceptual

## Colimites

Pasemos a ver los colimites: Los mismos pueden representar los efectos agregados de varias variables y transformaciones. En términos intuitivos, un colimite agrega múltiples objetos (variables) en un único objeto que captura su comportamiento combinado, también describen propiedades emergentes que resultan de múltiples fuerzas, como las expectativas de inflación, tasas de interés y oferta monetaria trabajando conjuntamente para influenciar las expectativas de devaluación.

Podemos usar colimites para varias cosas prácticas, entre ellas:

1. Expectativas agregadas: Si diferentes agentes (mercados, gobierno, inversores) tienen diferentes expectativas de inflación, un colimite puede representar la expectativa general de mercado.
2. Captura flujos monetarios: Combinando diferentes fuentes de liquidez (M2, short interest, net lending) a una única medición de liquidez efectiva.
3. Crea un equilibrio sintético: Tomando el colimite de varios morfismos que plasmen políticas monetarias, podemos computar una postura monetaria efectiva.
4. Expresar una expectativa de tipo de cambio óptimo: En vez de tomar una sola fórmula, el colimite puede ser una tasa emergente computada desde todos los datos disponibles.

A pesar de todos estos usos, nosotros nos enfocaremos en calcular el agregado, que incorpora liquidez, expectativas y actividad económica en un solo indicador. Para no perder rigor, más allá de que no vaya a cambiar los resultados (pero si el enfoque), es útil pensar en que utilizaremos functores, particularmente un functor F que mapea desde la categoría hasta los reales; es decir, mapea variables económicas a series de tiempo expresadas en reales. El colimite de este functor es lo que intentaremos calcular a continuación:

```
import pandas as pd
import numpy as np
from sklearn.decomposition import PCA
import matplotlib.pyplot as plt
from statsmodels.tsa.stattools import grangercausalitytests
from statsmodels.tsa.api import VAR
```

```python
file_path = 
df = pd.read_csv(file_path)

df["Date"] = pd.to_datetime(df["Date"])

economic_variables = [
    "M2", "Pi Exp", "Long Interest", "Short Interest",
    "Historical Ars Usd", "Argentina Net Lending Borrowing",
    "Gdp_argentina", "Gdp_usa"
]

external_factors = ["Embi+ARG"]
pca = PCA(n_components=3)
pca_components = pca.fit_transform(df[economic_variables])

df["Colimit_Aggregate_PCA"] = np.sum(pca_components * pca.explained_variance_ratio_, axis=1)
rolling_corr = df[economic_variables].rolling(window=180, min_periods=1).corr(df["E"])

dynamic_weights = rolling_corr.mean().abs()
dynamic_weights /= dynamic_weights.sum()

df["Colimit_Aggregate_Dynamic"] = np.sum(df[economic_variables] * dynamic_weights, axis=1)

df["Colimit_Aggregate_Scaled"] = (
    (df["Colimit_Aggregate_PCA"] - df["Colimit_Aggregate_PCA"].min()) /
    (df["Colimit_Aggregate_PCA"].max() - df["Colimit_Aggregate_PCA"].min())
) * (df["E"].max() - df["E"].min()) + df["E"].min()

df["Colimit_Aggregate_Smooth"] = df["Colimit_Aggregate_Scaled"].rolling(window=30, min_periods=1).mean()

df_gc = df[["Colimit_Aggregate_Smooth", "E"]].dropna()
granger_test = grangercausalitytests(df_gc, maxlag=5, verbose=True)

all_columns = ["Colimit_Aggregate_Smooth", "E"] + external_factors
df_var = df[all_columns].dropna()

model = VAR(df_var)
results = model.fit(maxlags=5, ic='aic')

forecast_steps = 10
forecast = results.forecast(df_var.values[-5:], steps=forecast_steps)

forecast_df = pd.DataFrame(forecast, columns=df_var.columns)
forecast_df["Date"] = pd.date_range(start=df["Date"].iloc[-1], periods=forecast_steps, freq="D")

plt.figure(figsize=(12, 6))
plt.plot(df["Date"], df["Colimit_Aggregate_Smooth"], label="colimite agregado", color="blue")
plt.plot(df["Date"], df["E"], label="E", color="red", linestyle="dashed")
plt.plot(forecast_df["Date"], forecast_df["E"], label="Pronóstico E (var con Embi))", color="green", linestyle="dotted")
plt.legend()
```

```
plt.title("VAR Model Forecast de expectativas de devaluación (E) con Embi+Arg")
plt.xlabel("Tiempo")
plt.ylabel("Valor")
plt.xticks(rotation=45)
plt.grid()
plt.show()
```

El código realiza un análisis de componentes principales (PCA) para combinar información y pronosticar las expectativas de devaluación. El proceso de combinación utiliza pesos dinámicos derivados de correlaciones móviles entre las diferentes variables. Emplea pruebas de causalidad de Granger para examinar las relaciones causales entre variables y construye un modelo VAR, incorporando el índice Embi+Arg como un factor de riesgo externo, capturando potencialmente la influencia de las condiciones económicas globales en las expectativas de devaluación. El gráfico obtenido es el siguiente:

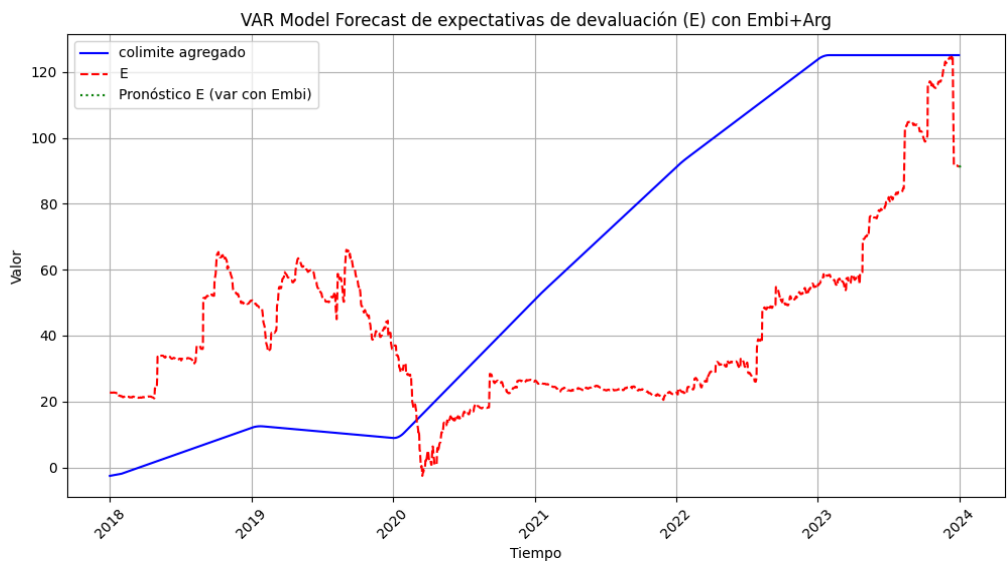

Entonces, construimos un colimite usando PCA y pesos dinámicos para asegurarnos que represente una condición de agregado económico.

El colimite es predictivo en Lag 1 (confirmado por la prueba de causalidad); es decir, es predictivo en el corto plazo, aunque se comporta demasiado linealmente, sugiriendo que captura la tendencia a largo plazo pero no la volatilidad a corto plazo de las expectativas.

En conclusión, el colimite existe como una forma de agregado económico, pero no explica completamente la volatilidad en la devaluación. Para refinar esto, es buena idea extender el poder predictivo del colimite usando el modelo VAR, ya que esto, sumado a incorporar el EMBI+ARG, se mejora considerablemente la pronosticabilidad, aunque carecen de las fluctuaciones afiladas observadas en los datos reales, por lo que sigue siendo mejorable. Por ejemplo, incorporar transformaciones no lineales como el análisis de componentes principales de núcleo, así también como agregar más objetos a la categoría, principalmente datos de mercado puede ser útil.

El colimite determina una postura sintética macro, más que un tipo de cambio absoluto proyectable, lo cual es información útil para saber dónde están los límites de lo que podemos representar y cómo. Los colimites pueden necesitar de más transformaciones functoriales para modelar comportamientos a corto plazo.

Para finalizar esta sección, es menester destacar que los colimites reflejan una oportunidad para capturar una visión holística del comportamiento económico, ya que permite combinar sintéticamente las variables en vez de tratarlas individualmente, distinto a lo que se suele hacer de forma tradicional (indicadores individuales). Además, el colimite permite seguirle el paso a la macroeconomía general, lo cual puede ser de utilidad para diseñar políticas, por ejemplo, dado los resultados de las pruebas de Granger, el colimite podría dar señales de advertencia sobre inestabilidad cambiaria y riesgo de devaluación, de particular interés para funcionarios e inversores, que se puede apreciar en el caso de Argentina. Me atrevería a decir que podríamos estar hablando de un nuevo tipo de indicador macroeconómico, calculado como el colimite sobre la cantidad de variables que se deseen y representaría un "índice de expectativas de devaluación", reflejando el riesgo de devaluación. El mismo generalizaría métodos tradicionales de pronóstico sobre tasas de cambio al agregar múltiples variables a un colimite singular. Nos provee de un modo para cuantificar el riesgo de devaluación, sin tener que depender de indicadores separados como la inflación o tasas de interés. Podría tener mejores resultados que los modelos actuales de predicción a corto plazo, confirmado por las pruebas de Granger. Este indicador sería de especial interés para traders y aquellos enfocados en el desarrollo de políticas (policymakers).

## Conclusiones

El análisis de límites y colimites permitió formalizar un concepto de equilibrio económico estructural, mostrando que la convergencia hacia ciertos estados depende de la forma en que los agentes ajustan sus expectativas.
Se identificó que, en la economía bimonetaria argentina, el equilibrio no es único ni estático, sino que puede oscilar entre múltiples configuraciones dependiendo de la confianza del mercado y las políticas implementadas.
Los modelos de series temporales aplicados a la estructura categórica indicaron que las expectativas de devaluación siguen patrones altamente no lineales, donde pequeñas variaciones en variables clave pueden amplificarse rápidamente en todo el sistema.
Además, los resultados mostraron que los métodos tradicionales no capturan adecuadamente estos efectos, mientras que el enfoque categórico permite visualizar la evolución de expectativas de una manera más estructurada.

# **Bibliografía**

## Artículos y libros

## Sitios web (fuentes de datos)

# **Anexo teórico**

## Técnicas usadas pero no desarrolladas

### Modelo VAR

El Modelo de Vectores Autorregresivos (VAR) es una herramienta econométrica utilizada para modelar la dinámica de sistemas de variables económicas interdependientes. Se trata de una extensión del modelo AR (Autorregresivo) aplicado a múltiples series temporales. Es una herramienta econométrica fundamental para el análisis de relaciones dinámicas entre múltiples series temporales económicas. Al extender el modelo autorregresivo (AR) a un sistema multivariado, el VAR permite la incorporación de variables interdependientes, donde cada variable se expresa como una función lineal de sus propios valores pasados y los valores pasados de todas las demás variables del sistema. Esta estructura captura la interrelación de las variables económicas, donde los valores actuales se ven influenciados tanto por su propia historia como por la historia de otras variables relevantes. Los coeficientes estimados cuantifican la magnitud y dirección de estas interdependencias, proporcionando información sobre la propagación de shocks o cambios a través del sistema a lo largo del tiempo.

La capacidad del modelo VAR para capturar dinámicas complejas e interacciones sin imponer restricciones teóricas a priori lo convierte en una herramienta flexible para el análisis exploratorio y la previsión. Además, mediante análisis de impulso-respuesta y descomposición de la varianza, el VAR permite rastrear los efectos de shocks específicos en variables individuales y evaluar las contribuciones relativas de diferentes shocks a la volatilidad general del sistema, mejorando la comprensión de las interrelaciones y la dinámica subyacente en el sistema económico. Esto sería a nivel algebraico:

$$Y_t = c + A_1 Y_{t-1} + A_2 Y_{t-2} + \ldots + A_p Y_{t-p} + \varepsilon_t$$
$$\varepsilon_t \sim \mathcal{N}(0, \Sigma)$$

En nuestra investigación, el VAR permite modelar cómo las variables macroeconómicas, como la tasa de interés, el tipo de cambio y la inflación, interactúan entre sí a lo largo del tiempo. En particular, puede ser útil para analizar cómo las expectativas de devaluación afectan a las tasas de interés y la demanda de pesos y dólares.
Este modelo requiere series estacionarias; si las variables tienen tendencia, debe aplicarse diferenciación. Además, puede ser sensible al número de rezagos seleccionados, y no impone restricciones estructurales, lo que puede dificultar la interpretación causal.

## Función de Impulso-Respuesta (IRF)

Las Funciones de Impulso-Respuesta permiten analizar cómo un shock en una variable afecta a las demás dentro de un modelo VAR. Se definen como:

$$IRF(h) = \frac{\partial Y_{t+h}}{\partial \varepsilon_t}$$

Donde mide el efecto de un shock en la variable $j$ sobre la variable $i$ después de $h$ períodos. Se obtiene mediante la descomposición de Cholesky o descomposición estructural.

El IRF es clave para evaluar el impacto de un shock en las expectativas de devaluación sobre variables como la demanda de pesos o la inflación.

## Descomposición de Varianza

La descomposición de varianza permite determinar qué porcentaje de la variabilidad de una variable en un modelo VAR se explica por shocks en otras variables. Formalmente, se basa en:

$$VD(h) = \frac{\text{Var}(Y_{t+h}|\varepsilon_t)}{\text{Var}(Y_{t+h})}$$

Puede ayudar a entender si la variabilidad en el tipo de cambio está determinada en mayor medida por expectativas de devaluación, política monetaria o shocks externos.

# Conceptos del álgebra abstracta

## Grupos

Un grupo es una estructura algebraica que consiste en un conjunto $G$ junto con una operación binaria $\cdot$ que satisface ciertas propiedades. Formalmente, un grupo es un par $(G, \cdot)$ donde:

1. Clausura: Para todo a,b ∈ G, el producto a · b ∈ G.
2. Asociatividad: Para todo a,b,c ∈ G, se cumple que (a · b) · c = a · (b · c)

3. Elemento identidad: Existe un elemento e ∈ G tal que para todo a ∈ G, se cumple que e · a = a · e.
4. Elemento inverso: Para todo a ∈ G, existe un a tal que ese a en operación con su inverso de como resultado e, de forma conmutativa.

Si además la operación es conmutativa (a · b = b · a para todo a,b ∈ G), el grupo es abeliano.

Como aplicaciones, el conjunto de tasas de interés con la operación de suma puede modelarse como un grupo aditivo. Además, en teoría monetaria, el grupo multiplicativo de factores de devaluación puede representar dinámicas de inflación acumulada.

## Anillos

Un anillo (R,+, · ) es una estructura algebraica con dos operaciones binarias:

1. (R,+) es un grupo abeliano.
2. (R, · ) es una operación asociativa y distributiva respecto a la suma.
3. Puede o no tener elemento identidad para la multiplicación.

Si la multiplicación es conmutativa, el anillo se llama conmutativo.

En economía, el conjunto de matrices económicas que describen interacciones entre sectores puede verse como un anillo bajo suma y multiplicación de matrices.

## Posibles estructuras algebraicas para categorías

En esta investigación, salvo en contadas ocasiones, utilizamos teoría de conjuntos para desarrollar los objetos y morfismos de la categoría. Esto significa que nuestra estructura consta de objetos, representados por conjuntos, y morfismos, representados por funciones (con dominio e imagen específicos); no obstante, es posible utilizar otras estructuras algebraicas, que pueden ser más o menos útiles dependiendo del contexto. A continuación una lista de algunas posibles estructuras, incluyendo sus objetos y morfismos, así también como el nombre por el que se las suele llamar normalmente:

- *Set*: Tiene conjuntos como sus objetos y funciones, con dominio y codominio especificados, así como sus morfismos.

- *Top*: Tiene espacios topológicos como sus objetos y funciones continuas como sus morfismos.

- *Set y Top\** tienen conjuntos o espacios con un punto base especificado como objetos y funciones que preservan el punto base (funciones continuas) como morfismos.

- *Group*: Tiene grupos como objetos y homomorfismos de grupos como morfismos. Este ejemplo extiende el término general "morfismos" a los datos de una categoría abstracta. Las categorías Ring de anillos unitarios y asociativos y sus homomorfismos de anillos, y Field de cuerpos y homomorfismos de cuerpos, se definen de manera similar.

- Para un anillo unitario pero no necesariamente conmutativo $R$, *Mod R* es la categoría de $R$-módulos izquierdos y homomorfismos de $R$-módulos. Esta categoría se denota por *Vect $k$* cuando el anillo es un cuerpo $k$ y se abrevia como *Ab* en el caso de Mod $Z$, ya que un $Z$-módulo es precisamente un grupo abeliano.

- *Graph*: Tiene grafos como objetos y morfismos de grafos (funciones que asignan vértices a vértices y aristas a aristas, preservando relaciones de incidencia) como morfismos. En la variante *DirGraph*, los objetos son grafos dirigidos, cuyas aristas ahora se representan como flechas, y los morfismos son morfismos de grafos dirigidos, que deben preservar fuentes y destinos.

- *Man*: Tiene variedades (manifolds) suaves, es decir, infinitamente diferenciables, como objetos y aplicaciones suaves como morfismos.

- *Meas*: Tiene espacios medibles como objetos y funciones medibles como morfismos.

- *Poset*: Tiene conjuntos parcialmente ordenados como objetos y funciones que preservan el orden como morfismos.

- *Ch $R$*: Tiene complejos de cadenas de $R$-módulos como objetos y homomorfismos de cadenas como morfismos.

Estas son sólo algunas posibles estructuras, pero también existen otras más abstractas como *signature* o *Htpy*, que no mencionaremos dada su complejidad. Existen ejemplos limitados únicamente por la creatividad humana de cómo aplicar estas estructuras a la economía.